\documentclass[pre,twocolumn,superscriptaddress,longbibliography]{revtex4-2}
\usepackage[english]{babel}
\usepackage[usenames,dvipsnames,svgnames,table]{xcolor}
\usepackage{amsmath, amsfonts,amssymb, mathtools}
\usepackage{hyperref}
\hypersetup{colorlinks=true,  linkcolor=NavyBlue, citecolor=PineGreen,urlcolor=cyan}
\usepackage{physics}
\usepackage{graphicx,color}
\usepackage[utf8]{inputenc}

\usepackage[normalem]{ulem}

\usepackage{algorithm}
\usepackage{algpseudocode}

\graphicspath{{./Figures/}}

\newcommand{\avg}[1]{\left< #1 \right>}

\newcommand{\tvb}[1]{\widetilde{\vb{#1}}}
\newcommand{\tvbg}[1]{\widetilde{\vb*{#1}}}
\newcommand{\uv}[1]{\underline{\vb{#1}}}
\newcommand{\uvg}[1]{\underline{\vb*{#1}}}
\newcommand{\opt}[1]{{#1}^\star }
\newcommand{\ctc}[1]{\ensuremath{\qty[#1]}}

\let\vp\varphi
\newcommand{\lpf}{\Gamma}
\newcommand{\cF}{\mathcal{F}}

\newcommand{\ptar}{\ensuremath{p_{\text{tar}}} }
\newcommand{\pth}{\ensuremath{p^{\text{(th)}}} }
\newcommand{\CAL}{CAL\lowercase{i}PPSO }

\makeatletter
\def\@bibdataout@aps{%
\immediate\write\@bibdataout{%
@CONTROL{%
apsrev41Control%
\longbibliography@sw{%
    ,author="08",editor="1",pages="1",title="0",year="1"%
    }{%
    ,author="08",editor="1",pages="1",title="",year="1"%
    }%
  }%
}%
\if@filesw \immediate \write \@auxout {\string \citation {apsrev41Control}}\fi 
}
\makeatother

\begin{document}

\title{Hard-Sphere Jamming through the Lens of Linear Optimization}

\author{Claudia Artiaco}
\affiliation{Department of Physics, KTH Royal Institute of Technology, Stockholm 106 91, Sweden}

\author{Rafael \surname{Díaz Hernández Rojas} }
\email{rafael.diazhernandezrojas@uniroma1.it}
\affiliation{Dipartimento di Fisica, Sapienza Università di Roma, Piazzale Aldo Moro 5, 00185 Rome, Italy}

\author{Giorgio Parisi}
\affiliation{Dipartimento di Fisica, Sapienza Università di Roma, Piazzale Aldo Moro 5, 00185 Rome, Italy}
\affiliation{INFN, Sezione di Roma1, and CNR-Nanotec, unit\`a di Roma, Piazzale Aldo Moro 5, 00185, Rome, Italy}

\author{Federico Ricci-Tersenghi}
\affiliation{Dipartimento di Fisica, Sapienza Università di Roma, Piazzale Aldo Moro 5, 00185 Rome, Italy}
\affiliation{INFN, Sezione di Roma1, and CNR-Nanotec, unit\`a di Roma, Piazzale Aldo Moro 5, 00185, Rome, Italy}

\begin{abstract}
The jamming transition is ubiquitous. It is present in granular matter, foams, colloids, structural glasses, and many other systems. Yet, it defines a critical point whose properties still need to be fully understood. Recently, a major breakthrough came about when the replica formalism was extended to build a mean-field theory that provides an exact description of the jamming transition of spherical particles in the infinite-dimensional limit. While such theory explains the jamming critical behavior of both soft and hard spheres, investigating the transition in finite-dimensional systems poses very difficult and different problems, in particular from the numerical point of view.
Soft particles are modeled by continuous potentials; thus, their jamming point can be reached through efficient energy minimization algorithms. In contrast, the latter methods are inapplicable to hard-sphere (HS) systems since the interaction energy among the particles is always zero by construction.
To overcome these difficulties, here we recast the jamming of hard spheres as a constrained optimization problem and introduce the \CAL algorithm, capable of readily producing jammed HS packings without including any effective potential. 
This algorithm brings a HS configuration of arbitrary dimensions to its jamming point by solving a chain of linear optimization problems. We show that there is a strict correspondence between the force balance conditions of jammed packings and the properties of the optimal solutions of CALiPPSO, whence we prove analytically that our packings are always isostatic and in mechanical equilibrium. Furthermore, using extensive numerical simulations, we show that our algorithm is able to probe the complex structure of the free-energy landscape, finding qualitative agreement with mean-field predictions. We also characterize the algorithmic complexity of \CAL and provide an open-source implementation of it.
\end{abstract}

\maketitle

\section{Introduction}\label{sec:intro}

Jamming is a pervasive phenomenon: it is present in systems with diverse time and length scales, such as structural glasses, grains, emulsions, foams, and colloids~\cite{liuNonlinearDynamicsJamming1998,liuJammingTransitionMarginally2010,torquatoReviewJammedHardparticlePackings2010,vanheckeReviewJammingSoftParticles2010}. Such ubiquity has been partially understood by recognizing that the jamming point defines a critical point common to all these systems~\cite{ohernJammingZeroTemperature2003,liuNonlinearDynamicsJamming1998,liuJammingTransitionMarginally2010}.
Despite being an out-of-equilibrium transition that brings a system to form a mechanically rigid packing, disordered jammed states 
can be identified as minima of a (properly-defined) free-energy landscape (FEL)~\cite{charbonneauFractalFreeEnergy2014}. In the case of hard-sphere (HS) systems, which are a minimal model for athermal and granular matter~\cite{bauleReviewEdwardsStatisticalMechanics2018a,tigheForceNetworkEnsemble2010}, jamming is reached at infinite pressure, $p=\infty$, and the jammed packings are identified by their packing fraction $\vp_J$. In soft-sphere (SS) systems, the jamming point is still identified by $\vp_J$, but in the limit $p\to0$, at least in the zero temperature limit~\cite{ohernJammingZeroTemperature2003,degiuliTheoryJammingTransition2015}.

Despite its physical relevance, a comprehensive theory of jamming is still far from being formulated. Recently, a mean-field theory has provided an exact description of the jamming transition in infinite-dimensional sphere systems~\cite{charbonneauFractalFreeEnergy2014,parisiMeanfieldTheoryHard2010,charbonneauExactTheoryDense2014,charbonneauGlassJammingTransitions2017,puz_TheorySimpleGlasses2020}. Such theory established that jamming occurs within the so-called Gardner phase~\cite{berthierGardnerPhysicsAmorphous2019,charbonneauExactTheoryDense2014,charbonneauGlassJammingTransitions2017,puz_TheorySimpleGlasses2020}, where all states become marginally stable. This feature implies an abundance of soft modes in jammed or nearly jammed packings. Another consequence is that critical jammed packings are isostatic, meaning that the number of mechanical constraints (i.e., contacts between particles) precisely matches the number of degrees of freedom.
Additionally, the mean-field theory predicted that near jamming, the FEL of HS configurations is a very rough and hierarchically organized hypersurface.

Another important theoretical step came about with the realization that the jamming transition of hard spheres in the high dimensional limit can be thought of as the satisfiability/unsatisfiability threshold of continuous constraint satisfaction problems (CSPs), where the constraints are induced by the requirement that spheres do not overlap. 
From this point of view, jamming criticality defines a universality class encompassing the physical systems mentioned above, as well as CSPs~\cite{franz2016simplest,franzUniversalitySATUNSATJamming2017}, neural networks~\cite{geigerJammingTransitionParadigm2019,spiglerJammingTransitionOverparametrization2019}, and inference problems~\cite{antenucciGlassyNatureHard2019}. In particular, in recent years, the perceptron model has gained a prominent role among the CSPs; it has been employed to investigate special instances of the sphere packing problem~\cite{franzUniversalSpectrumNormal2015,franz2019critical}, and even to analyze their quantum regime~\cite{franz2019impact,artiaco2021quantum}.

From this perspective, it is quite remarkable that the same jamming criticality predicted by mean-field theory (i.e., as $d\to\infty$) has been observed in finite dimensional systems, even down to $d=2$~\cite{charbonneauJammingCriticalityRevealed2015,charbonneauFractalFreeEnergy2014,charbonneauUniversalMicrostructureMechanical2012,dennisJammingEnergyLandscape2020,degiuliForceDistributionAffects2014,lernerLowenergyNonlinearExcitations2013,charbonneauFinitesizeEffectsMicroscopic2021} which is now believed to be the upper critical dimension. Explaining the observed robustness of the jamming phenomenology in HS systems in different dimensions constitutes a challenging open problem, which has been tackled both numerically~\cite{ohernJammingZeroTemperature2003,goodrichFiniteSizeScalingJamming2012,goodrichScalingAnsatzJamming2016,charbonneauJammingCriticalityRevealed2015,charbonneauFractalFreeEnergy2014,charbonneauUniversalMicrostructureMechanical2012,dennisJammingEnergyLandscape2020,degiuliForceDistributionAffects2014,lernerLowenergyNonlinearExcitations2013,charbonneauFinitesizeEffectsMicroscopic2021,artiacoExploratoryStudyGlassy2020,hexnerCanLargePacking2019,hexnerTwoDivergingLength2018,arceriVibrationalPropertiesHard2020,haghBroaderViewJamming2019} and experimentally~\cite{coulaisHowIdealJamming2014,dauchotDynamicalHeterogeneityClose2005,lechenaultCriticalScalingHeterogeneous2008,seguinExperimentalEvidenceGardner2016, wangExperimentalObservationsMarginal2022,asteGeometricalStructureDisordered2005,asteVolumeFluctuationsGeometrical2006}.

A related question is whether finite-dimensional HS and SS systems exhibit the same critical scalings when approaching the jamming transition, i.e., as $\vp\to\vp_J^-$  and $\vp \to \vp_J^+$, respectively. While it is generally believed that HS and SS systems share similar critical behaviors, it has been observed that some critical exponents are markedly different. For instance, near jamming the pressure in hard spheres and harmonic soft spheres scales as $p \sim \abs{\vp-\vp_J}^{\pm 1}$ for $\vp \to \vp_J^\pm$; nonetheless, both critical behaviors can be captured by a single scaling function~\cite{puz_TheorySimpleGlasses2020}. Importantly, it has also been shown that the distributions of forces and inter-particle gaps at the critical point are insensitive to the direction from which jamming is reached. The latter feature has been verified numerically only recently~\cite{charbonneauFinitesizeEffectsMicroscopic2021}, thanks to the algorithm for jamming in hard spheres presented in this article. 

In summary, while HS and SS systems allow a similar theoretical treatment, from the numerical point of view they represent very different types of systems. So far, computer simulations have amply favored SS systems thanks to the efficiency of energy minimization techniques available for interacting systems. 
In particular, the powerful FIRE algorithm~\cite{FIRE} has been successfully employed for studying jammed packings made of several thousands of soft particles, and there is now abundant numerical evidence that the jamming criticality of SS systems agrees with the mean-field predictions~\cite{charbonneauJammingCriticalityRevealed2015,charbonneauFinitesizeEffectsMicroscopic2021,morseGeometricSignaturesJamming2014,charbonneauUniversalNonDebyeScaling2016,morseEchoesGlassTransition2017,arceriVibrationalPropertiesHard2020,charbonneauMemoryFormationJammed2021}.

In contrast, due to the singular interaction potential of HS systems, studies analyzing their critical behavior at jamming are much more scarce. By definition, the interaction energy in HS configurations is either zero (if spheres do not overlap), or infinite (whenever two or more spheres overlap). This makes energy minimization strategies inapplicable. To partially overcome this problem, \emph{nearly} jammed HS configurations have been produced using the Lubachevsky--Stillinger (LS) compression protocol~\cite{lubachevskyGeometricPropertiesRandom1990, md-code}. By properly tuning the compression rate, this protocol can avoid crystallization (whenever $d\geq 3$) and produce highly compressed HS glasses~\cite{torquatoRandomClosePacking2000,zhangConnectionPackingEfficiency2014}. However, since this algorithm relies on simulating the dynamics of HS configurations through elastic collisions, particles do not remain in contact, nor does limit $p=\infty$ is reached. To solve this problem, other molecular dynamics (MD) algorithms have been devised. For instance, the authors of Ref.~\cite{lernerSimulationsDrivenOverdamped2013} considered an overdamped dynamics, in which once two particles collide they are constrained to stay in contact. By cleverly incorporating geometrical information of the configuration into the dynamical equations, they keep a well defined contact network at all times and are able to generate jammed HS packings, coming from this particular dynamical scenario. However, MD approaches that purely rely on elastic collisions ---the most common case--- lack in general the required precision to resolve the full network of contacts that determines a jammed state (as we exemplify in App.~\ref{app:calippso-vs-md} for the LS protocol).

Alternatively, jammed packings of pseudo-hard spheres have been constructed by introducing an effective interaction, and subsequently minimizing the corresponding potential energy. The most common choice has been a logarithmic potential, $v(h) \propto - \log h$, where $h$ is the (dimensionless) distance between spheres~\cite{britoRigidityHardsphereGlass2006,britoGeometricInterpretationPrevitrification2009,henkesExtractingVibrationalModes2012,degiuliForceDistributionAffects2014,arceriVibrationalPropertiesHard2020,charbonneauMemoryFormationJammed2021,altieriJammingTransitionHigh2016}. This potential clearly accounts for the non-overlapping condition, and ample evidence exists that it correctly accounts for the interactions between particles in terms of their \emph{average} position. Nevertheless, it is also important to investigate how to describe HS systems in terms of their \emph{instantaneous} positions. Modeling HS systems by adding an extraneous potential presents two major drawbacks. First, any effective potential introduces fictitious interactions for $h>0$, which become more important as the system approaches its jamming point. Second, the associated energy-minimization algorithms identify jammed packings as equilibrium states at zero temperature, while in true HS jammed states correspond to entropy extrema~\cite{frenkelOrderEntropy2015}. To certify that the logarithmic, or any other potential capture the physics of HS systems, one should carefully study such systems close to their jamming point. The algorithm we propose here is an ideal candidate for such a task: without relying on any effective potentials, it is able to reach the jamming point making use only of the geometrical constraints of HS configurations and on the particle's instantaneous positions. This approach relies on a direct mapping of jamming of hard spheres into a constrained optimization problem.

The purpose of this article is to present in full detail the \textit{CALiPPSO algorithm}, that is able to generate jammed packings of hard spheres in arbitrary dimensions and polydispersity without introducing any interaction potential between particles, thus overcoming the issues mentioned above. It relies on the linear approximation of the original, non-convex optimization problem, and it gradually reaches jamming by iterating over a series of linear programming instances. Hence its name, which stands for \emph{Chain of Approximate Linear Programming for Packing Spherical Objects}. In addition, in this article we show that combining \CAL with the LS compression protocol provides a powerful tool for exploring the physics of jamming in finite-dimensional HS systems. We provide our own implementation as a Julia~\cite{bezansonJuliaFreshApproach2017} package at \cite{code2022github}
% ~\footnote{Our implementation of the CALiPPSO algorithm, \href{https://github.com/rdhr/CALiPPSO.jl}{github.com/rdhr/CALiPPSO}.}

The \CAL algorithm has been successfully employed before for studying the properties of $3d$ packings near the jamming point~\cite{artiacoExploratoryStudyGlassy2020,diazhernandezrojasInferringParticlewiseDynamics2021}. Moreover, in combination with the LS protocol, it was used to confirm the jamming criticality of spheres and other mean-field-like models~\cite{charbonneauFinitesizeEffectsMicroscopic2021}. We refer to those references for extensive discussions on these issues.

The plan of the present article is as follows. In Sec.~\ref{sec:calippso}, we introduce the \CAL algorithm, and analytically show that the configurations that it produces are well-defined jammed states. Specifically: (i) \CAL jammed packings are solutions for the exact optimization problem; (ii) such solutions satisfy the mechanical equilibrium condition, previously derived for soft spheres~\cite{charbonneauJammingCriticalityRevealed2015,degiuliForceDistributionAffects2014}; (iii) they are also always isostatic. Importantly, points (ii) and (iii) imply that the Hessian obtained for hard spheres matches that of soft spheres which has been found to reproduce the marginal stability condition expected from the mean-field solution. In Sec.~\ref{sec:MD+calippso}, we demonstrate, through extensive numerical simulations, that \CAL can be readily coupled with the LS compression protocol to produce a robust and fast algorithm for jamming. We characterize the behavior of the two algorithms combined, and we provide evidence that the LS+\CAL route to jamming allows us to study the hierarchical structure of the FEL of hard spheres with very high accuracy, reproducing the results previously obtained with other methods. In Sec.~\ref{sec:complexity calippso}, we investigate the time complexity of our algorithm and conclude that its running time scales with the cube of the system size. Finally, in Sec.~\ref{sec:conclusion}, we briefly summarize our results and discuss why \CAL should be preferred to other jamming algorithms when an accurate identification of the network of contacts is needed. We finish by providing some examples of other problems that could be tackled using \CAL or similar methods.

Before ending the Introduction, we wish to acknowledge that other works, Refs.~\cite{torquatoRobustAlgorithmGenerate2010,krabbenhoftGranularContactDynamics2012,donevLinearProgrammingAlgorithm2004,hopkinsPhaseDiagramStructural2011,hopkinsDensestBinarySphere2012,hopkinsDisorderedStrictlyJammed2013,jiaoNonuniversalityDensityDisorder2011}, have previously used linear programming methods to produce jammed packings of hard spheres. Our optimization problem is simpler than the ones considered in those studies since it contains fewer optimization variables; thanks to this, we are able to obtain an exact proof of the properties (i)-(iii) mentioned above. To the best of our knowledge, none of the previous works has carried out a similar analysis.
Furthermore, the question of whether linear programming algorithms allow probing the FEL of hard spheres has never been put to test, as we do here. Additionally, CALiPPSO, if initialized with sufficiently highly compressed configurations, produces packings that should coincide with the $p \to \infty$ extrapolation of LS. In contrast, this is not always the case within the other methods~\cite{torquatoRobustAlgorithmGenerate2010}. On the other hand, it should be noted that some of the previous works~\cite{torquatoRobustAlgorithmGenerate2010,hopkinsDensestBinarySphere2012,hopkinsDisorderedStrictlyJammed2013,jiaoNonuniversalityDensityDisorder2011} demonstrated that their algorithms are capable of producing both ordered and disordered packings, within a rather broad range of densities; this feature is not realized by CALiPPSO.

\section{Jammed packings obtained using \CAL}\label{sec:calippso}

\subsection{Jamming as an optimization problem}\label{sec:jamming as OP}

Let us consider $N$ spheres of diameters $\vec{\sigma} = \{\sigma_i\}_{i=1}^N$ inside a cubic box of volume $V=L^d$. We denote the $dN$-dimensional vector of their centers as $\va{r} = \{\vb{r}_i\}_{i=1}^N$, where $\vb{r}_i$ is the $d$-dimensional vector identifying the position of the $i$-th particle. The system's packing fraction is given by $\vp = \frac1V \sum_{i=1}^N v_d(\sigma_i/2)$, where $v_d(R)= \frac{\pi^{d/2}R^d}{\Gamma(1+d/2)}$ is the volume of a hypersphere of radius $R$ in $d$ dimensions. For later use, we define $\sigma_{ij} = \frac{\sigma_i + \sigma_j}{2}$ as the sum of two particles radii.

For hard particles of arbitrary shape, a jammed state must fulfill an excluded volume constraint, as well as a set of mechanical constraints, related to force and torque balance, absence of attractive forces, etc.~\cite{bauleReviewEdwardsStatisticalMechanics2018a}. Restricting to frictionless hard spheres, as we do here, the excluded volume constraint reads $\abs{\vb{r}_i - \vb{r}_j} \geq \sigma_{ij}$, $\forall \, 1 \leq i< j\leq N$. This simply states that spheres cannot overlap. Moreover, among the mechanical requirements, only the force balance condition is relevant, because once it holds the other mechanical constraints are automatically satisfied. In summary, for frictionless hard spheres the excluded volume constraint and the force balance condition for each particle are necessary and sufficient conditions for having jammed configurations~\cite{bauleReviewEdwardsStatisticalMechanics2018a,charbonneauJammingCriticalityRevealed2015}.

Our aim is to bring an initial HS configuration, with initial packing fraction $\vp < \vp_J$, to its jamming point at $\vp_J$. We wish to reach the jamming point by increasing the packing fraction, $\vp \to \vp_J^-$, until the system becomes mechanically rigid, and using a procedure that does not allow any overlap among particles at any time. All of this can be recast as a constrained optimization problem (OP), as we show next. Without loss of generality, we will assume that the system volume is fixed; $\vp$ is then a monotonically increasing function of $\vec{\sigma}$, and maximizing it is equivalent to finding the largest factor by which the particle's diameter can be inflated. Naturally, we look for the optimal value of such inflation factor allowing for particles to be rearranged. Thus, letting $\va{s}=\{\vb{s}_i\}_{i=1}^N$ be the possible particles displacements, $\sqrt{\lpf}$ the inflation factor, and using $^\star$ to denote the optimal value of a quantity, the OP we consider is: find $(\opt{\va{s}}, \opt{\lpf})$ such that if we transform our HS system according to ($\va{r} \leftarrow \va{r}+\opt{\va{s}}$, $\sigma_i \leftarrow  \sqrt{\opt{\lpf}} \sigma_i$), then the packing fraction reaches a (possibly local) maximum value. $\va{s}$ and $\lpf$ must fulfill the non-overlapping constraint between spheres; thus, this is a \emph{constrained} OP. The reason for using the square root of $\lpf$ will become apparent when we will write the linearized version of this OP (see Eq.~\eqref{def:Jamming as LP}). Clearly, by taking an inflation factor equal for all the particles we preserve the degree of polydispersity of the system.

Hence, in a system with periodic boundary conditions and in absence of external forces, finding a jammed HS configuration is equivalent to solving the following constrained OP:
\begin{subequations}
	\label{def:Jamming as optimization problem}
	\begin{align}
		\max & \; \lpf\\
		G_{ij}(\va{s},\lpf) & := \lpf \sigma_{ij}^2 - \abs{ \vb{r}_i + \vb{s}_i - (\vb{r}_j + \vb{s}_j)}^2 \leq 0  \label{seq:exact non-overlap constraints} \\
		\forall & \;   1 \leq \, i < j \leq N\, \notag
	\end{align}
\end{subequations}
where the OP's variables are $(\va{s},\ \lpf)$, while the particles' position and size, $(\va{r}, \va{\sigma})$, play the role of constant parameters. Notice that the constraints in Eq.~\eqref{seq:exact non-overlap constraints} are symmetric upon exchanging $i$ and $j$, making the case $j<i$ redundant. The results presented here can be easily generalized to systems with closed boundaries. In such a case, in the OP \eqref{def:Jamming as optimization problem} one should also require that $\frac{\Gamma \sigma_i^2}{4} \leq (r_{i,\mu}+s_{i,\mu})^2$ and $\frac{\Gamma \sigma_i^2}{4} \leq (L-(r_{i,\mu}+s_{i,\mu}))^2$ for all $i=1,\dots, N$ and $\mu=1,\dots,d$. This would add extra constraints that should be taken into account when counting the effective degrees of freedom and analyzing the stability.

\begin{figure*}
	\includegraphics[width=\linewidth]{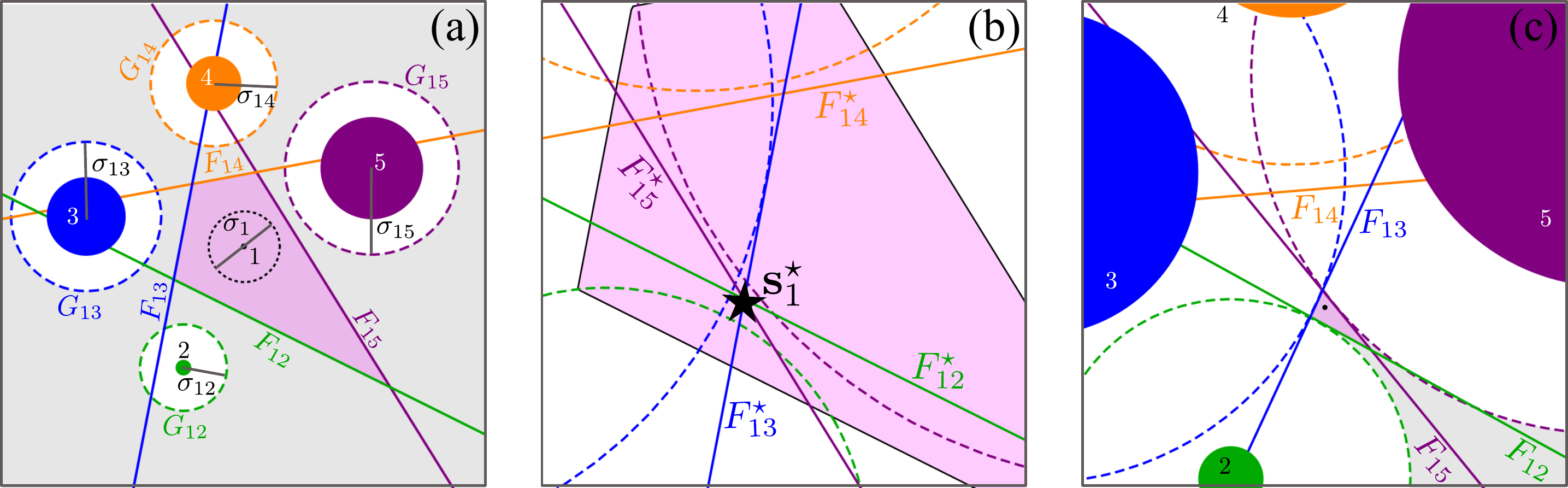}
	\caption{Geometry of \CAL for a problem of a single moving particle (1, black dotted) and four fixed ones (2, green; 3, blue; 4, orange; 5, purple). In all panels, the exact non-overlapping constraints, $\{G_{1j}\}_{j=2}^5$, are drawn with dashed circles, while their linearized version, $\{F_{1j}\}_{j=2}^5$, are identified by solid lines.
	Panel (a) shows the non-convex OP of Eq.~\eqref{def:Jamming as optimization problem} and the associated jamming LOP, Eq.~\eqref{def:Jamming as LP}. The size of particle 1 is indicated by the dotted circle, but only the position of its center (black dot) is relevant since the contribution of $\sigma_1$ has been included in the constraints. $\{G_{1j}\}_{j=2}^5$, induce “holes” in the set of possible displacements (gray region), making it non-convex. In contrast, $\{F_{1j}\}_{j=2}^5$ always define a convex set: a polytope (light pink). Note that such polytope is always strictly contained in the original feasible set.
	Panel (b) depicts the optimal displacement, $\opt{\vb{s}}_1$ (black star), of the jamming LOP instance of (a). As expected for any LOP, $\opt{\vb{s}}_1$ is located at the intersection of the linearized constraints, evaluated at the optimal solution, $\{\opt{F}_{1j} = F_{1j}(\opt{\vb{s}}_1, \opt{\lpf})\}_{j=2}^5$. Note however that $\opt{\vb{s}}_1$ does not saturate the analogous exact constraints.
	In panel (c), we show the new instance of the jamming LOP obtained after updating $\vb{r}_1 \leftarrow \vb{r}_1 + \opt{\vb{s}}_1$, and $\sigma_{i}\leftarrow \sqrt{\opt{\lpf}}\sigma_i$, for $i=1,\dots,5$. This panel shows that \CAL quickly reduces the size of the feasible region, which means that the linear constraints better approximate the exact ones. In this last panel the new size of particle 1 is not shown. Panels (b) and (c) have been magnified by a factor $9.6$ and $4.4$ with respect to panel (a), respectively.}
	\label{fig:diagram-LP}
\end{figure*}

Eq.~\eqref{def:Jamming as optimization problem} represents an exact, albeit non-convex, formulation of jamming as an OP. Indeed, as depicted in Fig.~\ref{fig:diagram-LP}a, the set of points $(\lpf, \va{s})$ that satisfy the non-overlapping constraints (represented as the gray region) is non-convex. Intuitively, the excluded volume constraint between particles creates many “holes” in the set of possible solutions, termed \textit{feasible region}. Thus, as often occurs in non-convex OP's, the problem \eqref{def:Jamming as optimization problem} becomes intractable for large system sizes ($dN \sim \order{10^2}$), and one has to resort to approximations.

A simple, yet powerful one is to assume that the starting HS configuration is already close to the jamming point; therefore, any feasible displacement $\va{s}$ has a negligible magnitude in comparison with the smallest distance between particles centers. Thus, terms of order $\order{\abs{\vb{s}_i}^2}$ can be neglected in each of the constraints $G_{ij}$. This amounts to the so-called approximation of small displacements~\cite{rouxGeometricOriginMechanical2000}, which has been successfully used to analyze the mechanical properties of rigid structures. Within this approach, the exact OP becomes a linear optimization problem (LOP), so it is guaranteed to be convex~\cite{luenbergerLinearNonlinearProgramming2016,boydConvexOptimization2004}.
The resulting LOP reads
\begin{subequations}
\label{def:Jamming as LP}
	\begin{align}
		\max & \; \lpf  \label{seq:objective jamming lp}  \\
		F_{ij}(\va{s},\lpf) & := -2 \vb{r}_{ij}\cdot \vb{s}_{ij} + \lpf \sigma_{ij}^2  - \abs{\vb{r}_{ij}}^2  \leq 0,  \label{seq:linear non-overlap constraints} \\
		\forall & \;   1 \leq \, i < j \leq N\, \notag
	\end{align}
\end{subequations}
with $\vb{r}_{ij} := \vb{r}_i - \vb{r}_j$, and $\vb{s}_{ij} := \vb{s}_i - \vb{s}_j$. 
Despite the fact that $F_{ij}$ depends on $\va{s}$ only through $\vb{s}_{ij}$, to simplify the notation, we will consider the general case where it depends on the full displacement vector. Notice that by increasing the diameters by a factor $\sqrt{\lpf}$ we are able to keep both the objective function ($\lpf$) and the constraints strictly linear. In the following, we will refer to \eqref{def:Jamming as LP} as the jamming LOP. For future use, we introduce the Lagrange multiplier, $\lambda_{ij} \geq 0$, associated to the constraint $F_{ij}(\va{s},\Gamma)$. The Lagrange multipliers play a fundamental role in our algorithm since they determine the set of contact forces at jamming, as we will show below when analysing the dual of the jamming LOP (Eq.~\eqref{eqs:dual jamming lop}).

The main advantage of using the LOP in \eqref{def:Jamming as LP} with respect to the exact OP is that the optimal solution of the former, $(\opt{\va{s}},\, \opt{\lpf})$, can be found by means of several linear programming methods.
Moreover, it is straightforward to show that any point that satisfies the set of linear constraints $\{F_{ij}(\va{s},\lpf)\}_{1\leq i<j\leq N}$ satisfies also the exact ones $\{G_{ij}(\va{s},\lpf)\}_{1\leq i<j\leq N}$; indeed, $G_{ij}(\va{s},\lpf)=F_{ij}(\va{s},\lpf) -\abs{\vb{s}_{ij}}^2$. This feature is illustrated by Fig.~\ref{fig:diagram-LP}a, where it is shown that the feasible region of the LOP (pink) is contained in the analogous set of the original OP (gray). 

On the other hand, a significant drawback is that the optimal solution of the jamming LOP is, in general, a sub-optimal solution for the exact OP. That is, solving \eqref{def:Jamming as LP} \emph{once}, does not necessarily yield a jammed packing because, even if the linear constraints \eqref{seq:linear non-overlap constraints} are saturated, the exact constraints in \eqref{seq:exact non-overlap constraints} might not be so. 
More precisely, sub-optimality is a consequence of the fact that an optimal solution of the LOP is always located in the boundary of the polytope defined by the linear constraints~\cite{boydConvexOptimization2004,luenbergerLinearNonlinearProgramming2016}, and that such point, in general, does not minimize the separation between inflated particles. Therefore, some of the exact constraints $\{G_{ij}\}$ remain unsaturated.
Geometrically, this means that $\opt{\va{s}}$ of the jamming LOP is determined by the intersection of the set of constraints $\{F_{ij}\}$ that are saturated once evaluated with $\opt{\lpf}$. But if the intersection point does not coincide with the point where $F_{ij}$ are tangent to $G_{ij}$, then $(\opt{\va{s}}, \opt{\lpf})$ will be sub-optimal with respect to the exact OP. This is depicted in Fig.~\ref{fig:diagram-LP}b for a single particle.

Yet, importantly the configuration obtained after solving the LOP, $(\va{r}, \va{\sigma}) \leftarrow (\va{r}+\opt{\va{s}}, \sqrt{\opt{\lpf}}\va{\sigma})$, will have no overlaps and a larger packing fraction. Consequently, it will be closer to jamming and we can use it to generate a new instance of the LOP \eqref{def:Jamming as LP}, that better approximates the exact OP. This is shown in Fig.~\ref{fig:diagram-LP}c.

The above considerations suggest that, if $x(t)$ denotes the value of the quantity $x$ in the $t$-th instance of the jamming LOP \eqref{def:Jamming as LP}, we can use the optimal solutions as initialization points, and proceed iteratively as $(\va{r}(t+1), \va{\sigma}(t+1) ) = (\va{r}(t)+\opt{\va{s}}(t), \sqrt{\opt{\lpf}(t)}\va{\sigma}(t))$ to reach the jamming point.
Jamming is realized when the particles cannot be further inflated nor moved, i.e., $(\opt{\va{s}}(n), \ \opt{\lpf}(n)) = (\va{0}, 1)$; we henceforth refer to this optimal solution as the \textit{convergence condition} of the \CAL algorithm and $n$ as the number of linear optimizations required to reach it.

We can intuitively understand the functioning of CALiPPSO in the following way. During the first few linear optimizations and the corresponding configuration updates, the space of possible solutions is quickly reduced, as depicted in Figs.~\ref{fig:diagram-LP}a-c for a single particle. This reduction is mainly caused by the tighter bounds imposed by the linear constraints $F_{ij}(\va{s},\lpf)$, in comparison with the exact non-overlapping constraints of Eq.~\eqref{seq:exact non-overlap constraints} (cf.\ Fig.~\ref{fig:diagram-LP}a). Importantly, the polytope defined by the constraints $F_{ij}(\va{s},\lpf)$ suppresses rearrangements that would allow particles to escape from the cages formed by their neighbors, thus efficiently preventing hopping over (entropic) barriers (cf.\ Fig.~\ref{fig:diagram-LP}c). This feature is particularly relevant to prevent crystallization in monodisperse configurations. 
In this way, the linear constraints of the \CAL algorithm are responsible for efficiently trapping particles, a geometric metaphor for the enchantments used by Calypso to keep Odysseus captive in her island for several years.

As a final remark, we highlight that the convergence condition of \CAL corresponds also to an optimal solution of the exact OP. That is, if $(\va{r}^{(J)}, \va{\sigma}^{(J)})$ defines a jammed packing obtained using \CAL and we use these values to generate another instance of the OP~\eqref{def:Jamming as optimization problem}, then $(\opt{\va{s}}, \opt{\lpf}) = (\va{0},1)$ and $(\va{r}^{(J)}, \va{\sigma}^{(J)})$ constitutes a local optimum of the exact OP as well. At least in $d=2$ and $d=3$ the global optimum corresponds instead to a crystalline structure. For instance, the densest packing of $3d$ monodisperse spheres corresponds to an FCC structure with $\vp_J^{(FCC)}=\pi/\sqrt{18} \approx 0.74$. Similarly, in $d<10$, the densest known monodisperse packings have an ordered structure~\cite{conwaySpherePackingsLattices2013,charbonneauThreeSimpleScenarios2021}. Since our aim is to study amorphous packings, we want to avoid precisely such ordered or partially crystallized solutions. It is therefore convenient that the \CAL algorithm prevents particles from performing large displacements: in such a way only a local optimum, very likely corresponding to a disordered configuration, can be obtained. The results we report below confirm that we can avoid producing ordered monodisperse packings whenever a random initial configuration is chosen for $d\geq 3$. When $d=2$, instead, (partial) crystallization always ensues in monodisperse packings. This is a consequence of the Euler criterion applied to the planar graph formed by the network of contacts~\cite{blumenfeldDisorderCriterionExplicit2021,hinrichsenRandomPackingDisks1990} (see App.~\ref{app:further-2d} for more details). As we show below, adding bidispersity or polydispersity helps to prevent crystallization in bidimensional systems.

While these arguments show that \CAL is able to generate a maximally dense disordered configuration of hard spheres, a more rigorous analysis is needed to show that such configuration is a mechanically stable packing. This is the issue we address next.

\subsection{\CAL produces well-defined, globally stable jammed states}\label{sec:properties calippso packings}

In this Section, we analytically prove that a HS packing obtained once the \CAL convergence condition is reached corresponds to a \textit{valid jammed state}. This means that: (a) such HS configuration satisfies the excluded volume and force balance constraints for each particle (local property); (b) it is a collectively stable packing (global property). The stability property follows from a relation between the number of contacts ($N_c$) and the number of degrees of freedom in a configuration ($N_{dof}$), specifically, $N_c \geq N_{dof}$~\cite{donevPairCorrelationFunction2005,goodrichFiniteSizeScalingJamming2012,donevCommentJammingZero2004,ohernReplyCommentJamming2004,torquatoReviewJammedHardparticlePackings2010,bauleReviewEdwardsStatisticalMechanics2018a}. Counting $N_{dof}$ requires some care due to the presence of symmetries and unstable particles. Fortunately, \CAL also provides a univocal way to determine $N_{dof}$; see Eq.~\eqref{eq:Ndof} below. Before continuing, we recall some common terminology. When the latter inequality does not hold ($N_c < N_{dof}$), a packing is said to be \textit{hypostatic}; if the equality is verified ($N_c = N_{dof}$), the packing is \textit{isostatic}; finally, a packing is \textit{hyperstatic} when the strict inequality is fulfilled ($N_c > N_{dof}$). Note that the condition (b) does not follow necessarily from (a) since there can be hypostatic packings where force balance holds for each particle and no overlaps are present~\cite{rouxGeometricOriginMechanical2000,torquatoReviewJammedHardparticlePackings2010}. 

It is known that critical jamming emerges together with isostaticity. Remarkably, isostaticity is verified in packings produced via \CAL once all the relevant degrees of freedom are considered, as we will show here. For our proof, we will adopt a pedagogical approach, so several logical steps will be explicitly made, and we will emphasize the connection between a geometrical analysis of jamming and the optimization approach developed here. Our proof will go along the following lines: to show (a) we will first notice that the excluded volume constraint is verified by construction. Then, we will use general results of convex optimization theory~\cite{boydConvexOptimization2004} to show rigorously that optimality of the jamming LOP \eqref{def:Jamming as LP} implies mechanical equilibrium. The property (b) will be derived from the existence of the solution of the force balance equations. These equations are encoded in the constraints of the dual optimization problem associated to the jamming LOP; see Eq.~\eqref{eqs:dual jamming lop} below. Leveraging on the property that such dual problem is itself a LOP, we will show that the solution of the force balance equations is unique. Since this can only happen when $N_c=N_{dof}$, the packings produced must be isostatic.

Let us begin by pointing out that, since $G_{ij}(\va{s},\lpf) \leq F_{ij}(\va{s},\lpf)$, $\forall \, 1\leq i < j \leq N$, it is guaranteed that the solutions of the LOP \eqref{def:Jamming as LP} always satisfy the non-overlapping constraints of HS systems, Eq.~\eqref{seq:exact non-overlap constraints}. 
To continue the proof, we will need few results of linear optimization, Eqs.~\eqref{eqs:jamming standard LP}-\eqref{eqs:KKT} below, particularized for the jamming LOP. We first introduce some notation. Let $M:=\frac{N(N-1)}{2}$  and $\tvbg{\lambda} = \{\lambda_{ij} \geq 0\}_{i<j}$ be the set of $M$ non-negative dual variables or Lagrange multipliers associated with the constraints $\{ F_{ij} \}_{i<j}$. In the following, we will show that we only need to consider $\order{N}$ constraints in order for \CAL to work. For the remaining part of this Section, we will use the simplified notation $\va{x}:=(\va{s}, \lpf)$, and define the ($dN+1$)-dimensional vector $\va{y}:=(\va{0}, -1)$ ; the zero vector of such space will be denoted by $\vec{\emptyset} = (\va{0},0)$. In terms of this new variables, the objective function \eqref{seq:objective jamming lp} is equal to $\va{y}\cdot \va{x}$. We rewrite the jamming LOP \eqref{def:Jamming as LP} in the more conventional form:
\begin{subequations}\label{eqs:jamming standard LP}
	\begin{align}
		\min & \; \va{y}\cdot \va{x} \\
	\text{s.t. }	\cF \va{x} & \ \preceq \tvbg{\rho} \; ; \label{seq:ilp matrix constraints}
	\end{align}
\end{subequations}
where $\preceq$ denotes element-wise comparison; $\cF$ is a $M\times (dN+1)$ matrix with entries
\begin{subequations}\label{eqs:entries ILP constraints matrix}
	\begin{align}
		&\cF_{ij}^{\mu k} = -2 r_{ij,\mu} (\delta_{ik} - \delta_{jk}) \, ,\\
		&\cF_{ij}^{dN+1} = \sigma_{ij}^2 \, ,
	\end{align}
\end{subequations}
for $1\leq k\leq N$, $1\leq \mu \leq d$, and $1\leq i < j \leq N$; while $\tvbg{\rho}$ is an $M$-dimensional vector with components $\rho_{ij}=\abs{\vb{r}_{ij}}^2,\ i<j$. Thus, expression \eqref{seq:ilp matrix constraints} is nothing else than the full set of non-overlapping constraints, emphasizing their affine character. In addition, for later use we introduce the \emph{dual} OP of \eqref{eqs:jamming standard LP}:
\begin{subequations}\label{eqs:dual jamming lop}
	\begin{align}
		\max &\  - \tvbg{\rho}\cdot \tvbg{\lambda} \\
		\text{s.t. } \cF^T \tvbg{\lambda} + \va{y} & = \vec{\emptyset}  \, ; \label{seq:equality constraints dual vars} \\
		\tvbg{\lambda} & \succeq  \tvb{0} \, . \label{seq:non-negative dual vars}
	\end{align}
\end{subequations}
Notice that the dual OP is also a LOP, but one in which the inequality constraints, Eq.~\eqref{seq:ilp matrix constraints} (or \eqref{seq:linear non-overlap constraints}) have been replaced by equalities, Eq.~\eqref{seq:equality constraints dual vars}. This is a standard property of LOPs~\cite{boydConvexOptimization2004,luenbergerLinearNonlinearProgramming2016}, and will be fundamental to prove the isostaticity of CALiPPSO packings.

Readers familiar with a geometrical interpretation of the mechanical properties of granular packings~\cite{rouxGeometricOriginMechanical2000} might recognize that Eq.~\eqref{seq:equality constraints dual vars} is equivalent to the mechanical equilibrium condition, provided that the Lagrange multipliers are identified with the contact forces. We expect that the details of our proof will make clear the correspondence between the optimization and the geometrical approach. In fact, the CALiPPSO algorithm works by exploiting the deep connection between the mechanical equilibrium condition to be fulfilled by jammed packings (captured by the jamming LOP's dual \eqref{eqs:dual jamming lop}) and the constraints that such condition imposes on the density and configurational degrees of freedom of the system (represented in the primal LOP \eqref{eqs:jamming standard LP}). In particular, one can rely on the  \emph{strong duality} property of LOPs~\cite{boydConvexOptimization2004,luenbergerLinearNonlinearProgramming2016}, which implies that if either the primal or the dual LOP has a finite optimal solution, so does the other. When both LOPs can be solved to optimality, their optimal values are equal (or, in other words, there is no duality gap). In our case, this means that if $\opt{\va{x}}$ is an optimal solution of \eqref{eqs:jamming standard LP}, then $\opt{\tvbg{\lambda}}$ ---the optimal solution of the dual \eqref{eqs:dual jamming lop}--- is such that $\tvbg{\rho}\cdot \opt{\tvbg{\lambda}}= - \va{y}\cdot \opt{\va{x}}= \opt{\lpf}$. Moreover, as we will show next, the equality constraints \eqref{seq:equality constraints dual vars} imposed on $\opt{\tvbg{\lambda}}$ are equivalent to the force balance requirement for each particle.

An explicit derivation of the force balance equations for  $\tvbg{\lambda}$ can be obtained using the Lagrangian, defined as
\begin{equation}
	\mathcal{L}(\va{x}, \tvbg{\lambda}) = \va{y}\cdot \va{x} + \sum_{\mathclap{1\leq i<j \leq N}} \lambda_{ij} F_{ij}(\va{x}) \, . \label{eq:lagrangian}
\end{equation}
Because of the strong duality between the LOP \eqref{eqs:jamming standard LP} and \eqref{eqs:dual jamming lop}, the Karush--Kuhn--Tucker (KKT) conditions~\cite{boydConvexOptimization2004} imply the following results for any primal and dual optimal points $(\opt{\va{x}}, \opt{\tvbg{\lambda}}):= (\opt{\va{s}}, \opt{\lpf}, \opt{\tvbg{\lambda}})$:
\begin{subequations}\label{eqs:KKT}
	\begin{align}
		& \opt{\lambda}_{ij} F_{ij}(\opt{\va{x}})  = 0,  \label{seq:kkt-complimentary-slackness} \\
	\nabla \mathcal{L}(\opt{\va{x}}, \opt{\tvbg{\lambda}})=	& \va{y} + \sum_{\mathclap{1\leq i<j \leq N}} \opt{\lambda}_{ij} \nabla F_{ij}(\opt{\va{x}}) = \vec{\emptyset}\, , \label{seq:kkt-zero-grad}
	\end{align}
\end{subequations}
where $\nabla =  (\pdv{\va{s}},\ \pdv{\lpf})$, and $\pdv{\va{s}}$ has been used as a shorthand notation of $(\pdv{s_{1,1}}, \pdv{s_{1,2}}, \dots, \pdv{s_{1,d}}, \pdv{s_{2,1}}, \dots, \pdv{s_{N,d}})$. Importantly, equality \eqref{seq:kkt-complimentary-slackness}, termed \emph{complementary slackness}, implies that if the constraints are inactive, i.e., $F_{ij}(\opt{\va{x}})<0$, then $\lambda_{ij}=0$. Conversely, a Lagrange multiplier will only be positive, $\lambda_{ij}>0$, when the associated constraint is active, $F_{ij}(\opt{\va{s}}, \opt{\lpf})=0$.

Such active Lagrange multipliers play a major role in our algorithm because they can be used to obtain the physical contact forces between particles. To emphasize that our analysis of the mechanical equilibrium of the packing is based on the linear constraints \eqref{seq:linear non-overlap constraints} that are saturated in an optimal solution, we will use $\ctc{ij}$, with $i<j$, to indicate that there is an active linear constraint between particles $i$ and $j$. So, we define $\mathcal{C} = \{\ctc{ij}\}$, the set of linear contacts, whose cardinality is $\abs{\mathcal{C}}=N_c \ll M $. The last inequality follows from the fact that not all particles are in contact with each other; therefore, the amount of positive Lagrange multipliers is much smaller than $M$ as long as $d\ll N$, as we assume here. It is useful to consider the $N_c$-dimensional vector of only positive dual variables, $\uvg{\lambda}= \{\lambda_{ij}\}_{\ctc{ij}\in \mathcal{C}}$.

Proving the mechanical equilibrium condition is now straightforward: the linearity of the LOP \eqref{eqs:jamming standard LP} implies that each of the derivatives of Eq.~\eqref{seq:kkt-zero-grad} results in one entry of $\cF$ (for fixed $i<j$). Thus, plugging \eqref{eqs:entries ILP constraints matrix} into Eq.~\eqref{seq:kkt-zero-grad} leads to the following equation for the components associated to the $i$-th particle: 
\begin{equation}\label{eq:sum-dual-vars-positions}
	\sum_{\substack{j\\j\neq i}}^{1,N} \opt{\lambda}_{ij} \vb{r}_{ij} =
	\sum_{\substack{j \in \partial i\\
	[ij] \in \mathcal{C}}} \opt{\lambda}_{ij} \vb{r}_{ij} = \vb{0} \, .
\end{equation}
where, in the second equality, $\partial i$ is the set of all linear contacts of particle $i$. That is, because of the complementary slackness condition \eqref{seq:kkt-complimentary-slackness}, the sum over $N-1$ terms is reduced to one that only contains active dual variables, usually of order $d$. On the other hand, the component associated to $\pdv{\lpf}$ of Eq.~\eqref{seq:kkt-zero-grad} reads
\begin{equation}\label{eq:sum-dual-vars-radii}
	 \sum_{1\leq i<j\leq N} \opt{\lambda}_{ij} \sigma_{ij}^2 = \sum_{ \ctc{ij}\in \mathcal{C} } \opt{\lambda}_{{ij}}  \sigma_{ij}^2 = 1 \, .
\end{equation}
We note in passing that from the last equation we can estimate how far from the jamming point a given $\opt{\va{x}}$ is. Indeed, given that $\opt{\lpf}\geq 1$, from the absence of duality gap mentioned above, it follows that $\tvbg{\rho}\cdot \opt{\tvbg{\lambda}} \geq 1 $. Whence, from \eqref{eq:sum-dual-vars-radii} and the complementary slackness property, it is straightforward to obtain
\begin{equation}\label{eq:saturation gaps}
    \sum_{ \ctc{ij}\in \mathcal{C} } \opt{\lambda}_{ij} (\abs{\vb{r}_{ij}}^2 - \sigma_{ij}^2) \geq 0 \, .
\end{equation}

These are conditions to be fulfilled by any optimal solution $\opt{\va{x}}$ of the jamming LOP. However, when the convergence criterion $\opt{\va{x}}=(\va{0},1)$ is reached, we have $F_{\ctc{ij}} (\va{0}, 1)=0$, and equality holds in \eqref{eq:saturation gaps}. This means that $\abs{\vb{r}_\ctc{ij}} = \sigma_\ctc{ij}$, i.e., linear contacts become physical contacts, and $\mathcal{C}$ then determines the full network of contacts at jamming. Once we know that the norm of $\vb{r}_\ctc{ij}$ can be fixed, we can rescale the corresponding dual variables $\opt{\lambda}_{ij} = f_{ij}/\sigma_{ij}$ and introduce the unit vector $\vb{n}_{ij} = \frac{\vb{r}_{ij}}{\abs{\vb{r}_{ij}}} $ to rewrite Eq.~\eqref{eq:sum-dual-vars-positions} as
\begin{equation}
    \sum_{j\in \partial i} f_{ij} \vb{n}_{ij}=\vb{0}.
\end{equation}
This is the force balance equation for the $i$-th sphere. We can write analogous equations for the full configuration as
\begin{equation}\label{eq:force balance}
	\mathcal{S} \uv{f} = \va{0}
\end{equation}
where $\uv{f}=\{f_{ij}\}_{\ctc{ij} \in \mathcal{C} }$ is the vector containing the contact forces magnitudes, and $\mathcal{S}$ is a $dN \times N_c$ matrix, whose entries are given by $\mathcal{S}_{k,\mu}^{\ctc{ij}} = (\delta_{ik} - \delta_{jk}) n_{ij, \mu}$. Crucially, this expression is identical to the one derived in previous works~\cite{charbonneauJammingCriticalityRevealed2015,degiuliForceDistributionAffects2014,puz_TheorySimpleGlasses2020,rouxGeometricOriginMechanical2000}, and it determines the force balance condition for jammed packings. Notice that $\uv{f}$ represents \emph{contact} forces, which are finite despite the singular potential of hard spheres. The connection between contact forces and dual variables is completely analogous to the one between generalized forces and constraint conditions in Lagrangian mechanics; namely, $\vb{f}_{ij}$ can be obtained from the Lagrange multiplier $\opt{\lambda}_{ij}$ times the derivative of the constraints involved (cf.\ Eq.~\eqref{seq:kkt-zero-grad} from the KKT conditions). This completes the proof that, upon convergence, \CAL generates packings in mechanical equilibrium and with physical contacts, even if in the intermediate steps this latter feature is not necessarily true.

The global stability property or, equivalently, the requirement that a jammed state produced with our algorithm is not hypostatic, does not follow from the force balance condition, Eq.~\eqref{eq:force balance}, alone. 
To prove that \CAL always produces isostatic packings (property (b) above), we note that once convergence has been reached, it follows that $\mathcal{S}_{k,\mu}^{\ctc{ij}} = - \frac1{2\sigma_{ij}} (\cF^T)_{k, \mu}^{i,j}$ if particles $i$ and $j$ are in contact. This means that Eq.~\eqref{eq:force balance} is equivalent to the equality constraints of the dual LOP \eqref{seq:equality constraints dual vars} but with a reduced matrix which only involves particles in contact. Letting $\mathcal{R}$ be such a $(dN+1)\times N_c$ matrix, Eq.~\eqref{seq:equality constraints dual vars} can be rewritten as 
\begin{equation} \label{eq:linear system dual vars}
	\mathcal{R} \uvg{\lambda} + \va{y} = \vec{\emptyset} \, .
\end{equation}
This expression forms a system of $dN+1$ equations in $N_c$ unknowns, in which the first $dN$ equations form a homogeneous system, equivalent to Eq.~\eqref{eq:force balance} (once rescaled by $\frac1{\sigma_{ij}}$). Eq.~\eqref{eq:linear system dual vars} represents the link between the geometric interpretation of HS packings, and the optimization perspective given by the KKT conditions. Indeed, the first $dN$ rows of \eqref{eq:linear system dual vars} are nothing else but the rightmost equality of Eq.~\eqref{eq:sum-dual-vars-positions} for each particle, while the last row corresponds to Eq.~\eqref{eq:sum-dual-vars-radii}.

Proving the isostaticity of \CAL packings requires to accurately count the degrees of freedom. We first notice that, because of our assumption of periodic boundaries, there are $d$ uniform translations that leave relative distances and displacements among particles invariant; thus, the number of degrees of freedom is decreased by $d$. This is nicely reflected by the fact that out of the $dN$ homogeneous equations in \eqref{eq:linear system dual vars}, $d$ are linearly dependent. Another important consideration is the presence of rattlers, which are particles with $d$ or fewer contacts. Due to the low number of contacts, these particles are unstable, and in most cases they do not belong to the backbone of the network of contacts. Therefore, they do not contribute to the rigidity of the configuration. The only exception are monodisperse systems in $2d$, where rattlers are subject to forces; this case is discussed in Appendix~\ref{app:further-2d}. Rattlers should be excluded when counting the number of contacts and degrees of freedom, and only spheres with at least $d+1$ contacts ---henceforth termed stable particles--- should be considered. Of course, identifying whether a particle is stable or not can only be done a posteriori, when $\mathcal{C}$ is constructed after \CAL has converged. If $N_s$ is the number of stable particles, in the absence of any other symmetries or external constraints, we have
\begin{equation} \label{eq:Ndof}
	N_{dof} = d(N_s-1) + 1 \qc
\end{equation}
where the extra degree of freedom is a consequence of the fact that $\lpf$ is also a variable of the jamming LOP. Equivalently, one can think of density as an additional degree of freedom. Thus, when we say that the isostatic condition $N_c=N_{dof}$ is verified, we mean that all the degrees of freedom of the jamming LOP (i.e., the displacements of stable particles and inflation factor) have been considered.

In contrast, if one considers only configurational degrees of freedom, i.e., $N_{dof}'=d(N_s-1)$, we have $N_c = N_{dof}' + 1$, which makes the packings produce by \CAL hyperstatic.
This amounts to say that requiring \CAL configurations to have a finite bulk modulus (or being rigid), imposes an extra constraint~\cite{donevPairCorrelationFunction2005,goodrichFiniteSizeScalingJamming2012}.
To highlight our optimization approach, we will use \textit{isostatic} when the degrees of freedom are
counted as in Eq.~\eqref{eq:Ndof}.

As mentioned above, the $d$ uniform translations reduce the number of linearly independent equations in~\eqref{eq:linear system dual vars}. Naturally, this feature is also present in $\mathcal{S}$.
Therefore, considering Eq.~\eqref{eq:force balance}, it is easy to see that to have a consistent system of equations, it must happen that $N_c \geq N_{dof}'$. However, if $N_c =N_{dof}'$ only the homogeneous solution exists. Yet, Eq.~\eqref{eq:sum-dual-vars-radii} prevents such scenario, whence $N_c \geq N_{dof}'+1 = N_{dof}$. This shows that \CAL packings can never be hypostatic, and thus are always collectively stable.

Finally, to show that the packings obtained with our method are isostatic we make use of the fact that there is zero duality gap between the jamming primal and dual LOPs, Eqs.~\eqref{eqs:jamming standard LP} and~\eqref{eqs:dual jamming lop}, respectively.
The absence of duality gap implies that the solution of the linear system \eqref{eq:linear system dual vars} is unique~\cite{rouxGeometricOriginMechanical2000}. Given that $\uvg{\lambda} \succ \uv{0}$, and that Eq.~\eqref{eq:linear system dual vars} is not a homogeneous system (recall $y_{dN+1}=-1$), the uniqueness of the solution implies that the number of independent equations matches the number of unknowns, $N_{dof} = N_c$. Therefore, the resulting packings are isostatic. This completes the proof that packings produced with the \CAL algorithm are valid jammed states.

Let us note that the result of two paragraphs above, i.e., $N_c \geq N_{dof}$, only relies on the KKT conditions (Eq.~\eqref{eqs:KKT}), and thus is more general than the isostaticity property. In some “pathological” cases, it may happen that $N_c > N_{dof}$. In our experience, hyperstaticity only occurres in $2d$ monodisperse packings, where also large crystalline domains are formed. We comment further on this particular case in Appendix~\ref{app:further-2d}. Importantly, all the tests we performed in $d\geq 3$ (see Secs.~\ref{sec:MD+calippso}, \ref{sec:complexity calippso} and App.~\ref{app:further-characterization}) have never produced such hyperstatic configurations, even with monodisperse systems. Similarly, in bidimensional, polydisperse systems isostaticity is recovered.

It should now be clear that the dual jamming LOP (in particular Eq.~\eqref{seq:equality constraints dual vars}), together with the complementary slackness property (Eq.~\eqref{seq:kkt-complimentary-slackness}), contain all the requirements to guarantee that \CAL produces valid jammed states. We also showed that an equivalent condition is that any \CAL optimal dual solution $\opt{\uvg{\lambda}}$ must fulfill Eq.~\eqref{eq:linear system dual vars}. We emphasize that the matrices $\mathcal{S}$, $\cF$, and $\mathcal{R}$ are determined entirely by the geometrical features of the configuration. Notably however, we derived them not from geometric considerations, but following results of optimization theory; specifically the KKT conditions, Eq.~\eqref{eqs:KKT}. The \CAL algorithm works by exploiting the correspondence between optimization and geometric descriptions of jammed packings. This can be considered the heart of our proof: any optimal solution $\opt{\va{x}}$ must fulfill the KKT conditions, as written in Eqs.~\eqref{eq:sum-dual-vars-positions} and \eqref{eq:sum-dual-vars-radii} (due to the convex nature of the jamming LOP), and at convergence these conditions are equivalent to Eq.~\eqref{eq:linear system dual vars} (a geometric property).
Mathematically, we can understand that the CALiPPSO algorithm succeeds at producing isostatic configurations because the dual optimization problem associated to the jamming LOP (see Eq.~\eqref{eqs:dual jamming lop}) is itself a LOP. Given that, once an optimal solution is found the two systems of linear equations associated to their constraints are satisfied simultaneously (because there can be no duality gap), the only possibility is that $N_c=N_{dof}$.

Before closing this part, we comment on an additional property that can be derived from Eq.~\eqref{eq:sum-dual-vars-radii}, for a packing in arbitrary dimensions, $d$. Notice that for monodisperse systems, or whenever the distribution of diameters is significantly peaked around its mean value $\overline{\sigma}$  (i.e., whenever $\text{Var}[\va{\sigma}] \ll \overline{\sigma}^2$), it is easy to see that $\vp \sim N {\overline{\sigma}}^d$. In this case, the scaling of Eq.~\eqref{eq:sum-dual-vars-radii} with the system size implies that the mean force $\overline{f}$ is such that $N \overline{f} \sim 1/\overline{\sigma}$, whence we obtain $\overline{f} \sim N^{1/d -1}$.

\subsection{Algorithmic implementation of \CAL}
\label{sec:calippso implementation}

The \CAL algorithm is rather simple; it consists of a single loop that iterates over successive LOP instances in order to reach the jamming point. 

Its performance can be easily enhanced by using neighbor lists to reduce the number of constraints of the LOP. Indeed, when two particles are far apart, their associated constraint in Eq.~\eqref{seq:linear non-overlap constraints} becomes irrelevant. Hence, for sufficiently distant particles, the evaluation of the inequality~\eqref{seq:linear non-overlap constraints} can be omitted without affecting the optimization procedure.
By implementing the neighbor-list approach, instead of including the $N(N-1)/2$ possible constraints, we only consider $M'\sim c_d N$ of them, where $c_d$ is a prefactor that depends on the dimensionality and should be, at most, of the order of the kissing number (i.e., the maximum number of non-overlapping spheres such that each of them touch a common sphere) in the corresponding dimension. Clearly, this reduces the size of the constraints matrix $\cF$ of Eq.~\eqref{eqs:entries ILP constraints matrix} to be $M'\times (dN+1)$.

The neighbor list is constructed utilizing a cutoff distance, $\ell(\vp)$, which (possibly) depends on the system's packing fraction, as explained below. For each particle $i$ we define its list of neighbors, $\tilde{\partial} i := \{j \ | \ \abs{\vb{r}_{ij}} \leq \ell(\vp) \}$, where the distance between particles $i$ and $j$ is computed following the nearest image convention~\cite{donevNeighborListCollisiondriven2005}. Therefore, the constraint $F_{ij}(\va{s}, \lpf)$ is included only if $j\in \tilde{\partial} i$ (and $i<j$). That is, for a given packing fraction, the full set of constraints becomes $\tvb{F}(\va{s},\lpf) = \{F_{ij}(\va{s}, \lpf) \ |\ i<j,\ j \in \tilde{\partial} i \}$ 

It is useful to consider a cutoff distance dependent on $\vp$ since at low packing fraction, when a single update can result in particles being displaced over large distances, $\ell(\vp)$ should be kept large enough to ensure that no overlaps occur even after such large displacements. In contrast, for $\vp \lesssim \vp_J$, when each linear optimization iteration generates only very small rearrangements, $\ell(\vp)$ can be set to a small value, keeping track only of the nearest neighbors for each particle. Notice that we have assumed that $\ell$ is the same for each particle, but it is straightforward to generalize our algorithm to the case where each particle has a different cutoff distance, $\ell_i(\vp)$. This situation could be useful, for instance, with highly polydisperse packings, in which the smallest particles might perform larger displacements and therefore more neighbors need to be taken into account to avoid overlaps. In such case, the list of neighbors for the $i$-th particle could be defined as  $\tilde{\partial} i := \{j \ | \  \abs{\vb{r}_{ij}} \leq \max \{\ell_i(\vp), \ell_j(\vp)\} \}$.

\begin{figure}[!htb]
	\begin{algorithm}[H]
		\caption{\CAL algorithm\\for jamming hard spheres}\label{alg:ILP algorithm}
		\begin{algorithmic}[1]
			\Require A HS configuration $(\va{r}, \vec{\sigma})$, without overlaps; tolerance for convergence criterion $(\text{tol}_{\vb{s}}, \text{tol}_\lpf)$.
			\Statex
			\Procedure{CAL\lowercase{i}PPSO}{$\va{r},\vec{\sigma};\  \text{tol}_{\vb{s}}, \text{tol}_\lpf$}
			\State Compute initial density, $\vp$, and cutoff $\ell(\vp)$
			\Statex 
			
	        \Repeat
				\For{$1=1,\dots,N$} \Comment{Construct neighbor lists}
					\State $\tilde{\partial i} \gets \{j |\  \abs{\vb{r}_{ij}} \leq\ell(\vp) \}$ 
				\EndFor
				\Statex
				\State $\tvb{F} := \{F_{ij}(\va{s}, \lpf) \ | \, i<j\qc j \in \tilde{\partial} i \}$ 
				\Comment{Define set of relevant constraints, $F_{ij}$, from Eq.~\eqref{seq:linear non-overlap constraints}.}
				\State Solve the jamming LOP \eqref{def:Jamming as LP}, with constraints $\tvb{F}$.
				\State $\va{s}\gets \opt{\va{s}}$
				\State $\lpf \gets \opt{\lpf}$
				\State Store $\mathcal{C} :=\{ \ctc{ij} \ \vert \ \lambda_{ij}>0 \}$ \Comment{Define contacts indices from active dual variables}
				\State Store active dual variables, $\uvg{\lambda}$.
				\State $(\va{r}, \vec{\sigma}) \gets (\va{r}+\va{s}, \sqrt{\lpf}\vec{\sigma})$ \Comment{Update the configuration}
				\State Recompute $\vp$ and $\ell(\vp)$
			\Until{$\max_{i} \abs{\vb{s}_i}< \text{tol}_{\vb{s}}$ \textbf{and} $\sqrt{\lpf}-1<\text{tol}_\lpf$ }
	
			\Statex 
			\State $(\va{r}_J,\ \va{\sigma}_J)  \gets (\va{r},\ \va{\sigma})$ \Comment{Define jammed configuration}
			\For{$\ctc{ij} \in \mathcal{C}$} \Comment{Construct network of contacts}
			\State $\vb{n}_{ij} = \frac{\vb{r}_{ij}}{\sigma_{ij}}$ \Comment{Store contact vectors}
			\State $f_{ij} = \frac{\lambda_{ij}}{\sigma_{ij}}$ \Comment{Store forces magnitudes}
			\EndFor
	
			\Statex
			\State \textbf{return} $\ \va{r}_J,\ \vec{\sigma},\ \{\vb{n}_{ij}\},\ \{f_{ij}\}$ \Comment{Output}
		\EndProcedure
		\end{algorithmic}
	\end{algorithm}
\end{figure}

The \CAL algorithm is reported as pseudo-code in Algorithm~\ref{alg:ILP algorithm}. Before further comments, we anticipate that more details on how to initialize the algorithm will be addressed in the next Section~\ref{sec:MD+calippso}, while the analysis of the convergence time of \CAL is postponed to Sec.~\ref{sec:complexity calippso}.

As a first remark, we highlight that \CAL has no free parameters, except for $\ell(\vp)$. Working with monodisperse configurations, we found that, if the initial packing fraction is not too small, setting $\ell/\sigma \in [3,4]$ produces good results, while when the system is very close to jamming (i.e., during the last linear optimizations), further reducing the cutoff distance to $\ell =1.4\sigma$ suffices.
However, since $\vp_J$ is not known \emph{a priori} for a given configuration, we choose $\ell$ according to the optimal inflation factor from the previous linear optimization, $\opt{\lpf}_0$. For instance, in our tests we used $\ell=3.5\sigma$ when $\sqrt{\opt{\lpf}}-1 \geq 10^{-5}$, and $\ell=1.4\sigma$ otherwise. We verified that the results are insensitive to specific value of $\ell\in [1.4, 5]\sigma$.

Second, if the initial packing fraction is not close to $\vp_{J}$, say $\vp/\vp_J < 0.5$ imposing bounds on the particles' displacements, $\abs{\vb{s}_i}< s_{\text{bound}}$ for $i=1,\dots,N$, is convenient. In this way, one can avoid both large rearrangements (that could lead to crystallization in $3d$ monodisperse systems), and the need to make $\ell(\vp)$ too large. We found that the naive bound, $s_{\text{bound}}= \frac{1}{2\sqrt{d}}[ \ell(\vp)- \sqrt{\opt{\lpf}_{0}}\sigma]$, was enough to avoid overlaps in all the cases we tested. However, if the degree of polydispersity is very broad, tighter bounds might be needed. Additionally, bounding $\abs{\vb{s}_i}$ might be useful also at later stages of the chain of linear optimizations to effectively reduce the feasible region of the jamming LOP and speed up the optimization. This should be done with some care because, if these constraints become active, they would play the role of external forces. Therefore, they must be taken into account when assessing the mechanical equilibrium of the configuration. If ignored, active displacement bounds might cause a packing to be non-isostatic (in the sense defined above i.e., $N_c=N_{dof}$ with $N_{dof}$ as in Eq.~\eqref{eq:Ndof}), given that $N_c$ only considers contact forces. A practical solution to guarantee isostaticity and that only real contacts are included, is to perform the last linear optimization without bounds on any $\abs{\vb{s}_i}$.

In the pseudo-code of Algorithm~\ref{alg:ILP algorithm} we have implicitly assumed, without loss of generality, that $\opt{\tvbg{\lambda}}$ is obtained \emph{simultaneously} when solving the jamming LOP (line 8). This is certainly the case when the jamming LOP \eqref{def:Jamming as LP} is solved using interior-point methods~\cite{boydConvexOptimization2004,luenbergerLinearNonlinearProgramming2016} as we do here (see below). However, if the optimal solution is obtained using, e.g., the primal simplex method~\cite{luenbergerLinearNonlinearProgramming2016}, the Lagrange multipliers would be computed \emph{after} such solution is found. Conversely, if the dual simplex method was employed (i.e., if the LOP \eqref{eqs:dual jamming lop} is solved instead of the primal, original jamming LOP), $(\opt{\va{s}}, \opt{\lpf})$ would be obtained from $\opt{\tvbg{\lambda}}$. As already remarked, these arguments rely on the strong duality theorem: once the primal or dual optimal solution is available, the other one can be accessed straightforwardly. 

When the \CAL convergence criterion has been met within a given tolerance $(\text{tol}_{\vb{s}}, \text{tol}_\lpf)$, we directly obtain the jamming packing fraction $\vp_J$, the particles' position and size $(\va{r}^{(J)}, \va{\sigma}^{(J)})$, as well as the set of contact forces $\uv{f}=\{f_{\ctc{ij}}\}$. From their knowledge, it is possible to investigate all the properties of the jamming transition of hard spheres, such as the jamming critical exponents~\cite{charbonneauFractalFreeEnergy2014,charbonneauExactTheoryDense2014,charbonneauFinitesizeEffectsMicroscopic2021,charbonneauGlassJammingTransitions2017}, and the structure the free-energy landscape (FEL)~\cite{artiacoExploratoryStudyGlassy2020,dennisJammingEnergyLandscape2020} (see also Sec.~\ref{sec:LS-parameters}).

\begin{figure}[!htb]
	\includegraphics[width=\columnwidth]{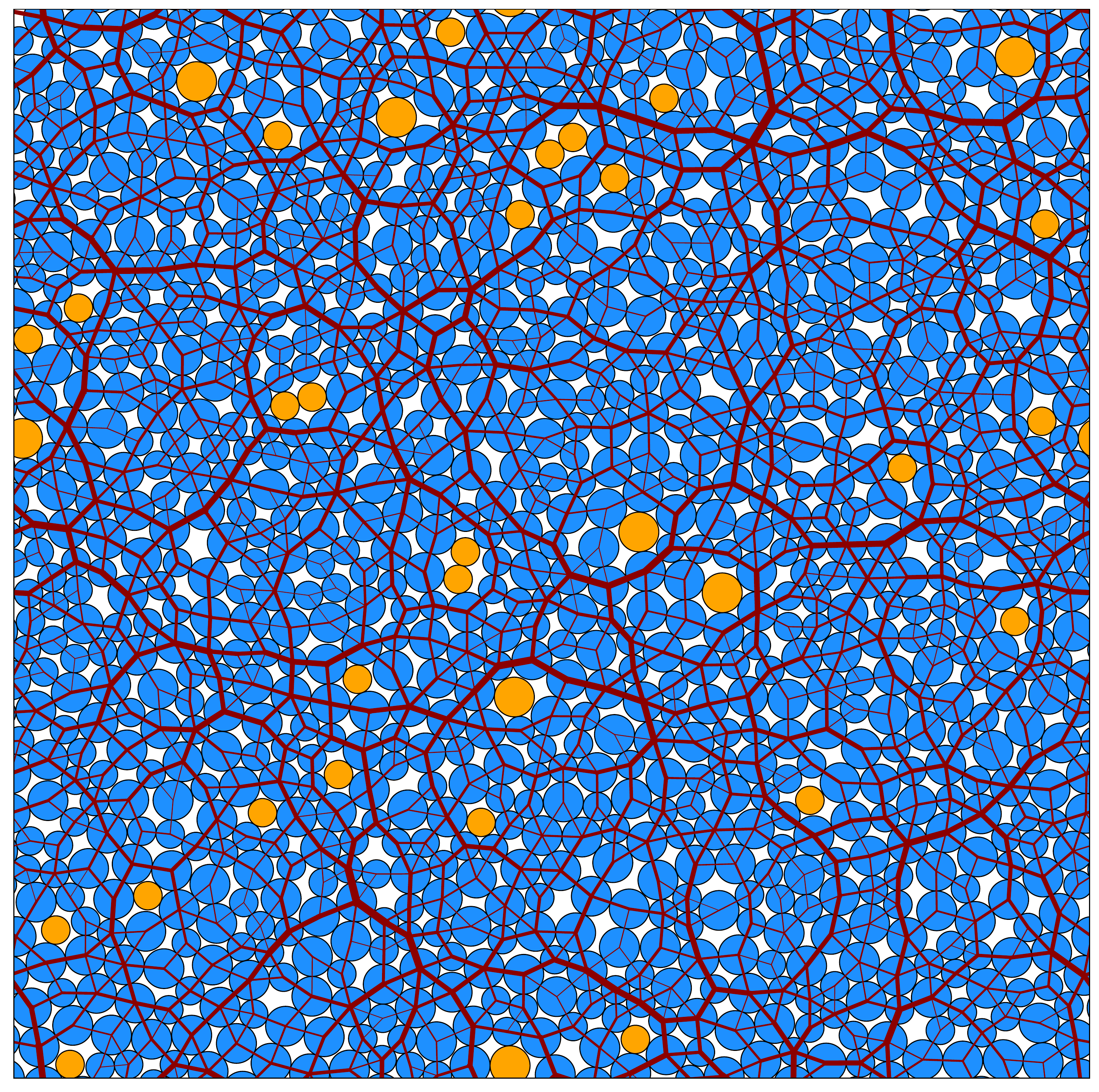}
	\includegraphics[width=\columnwidth]{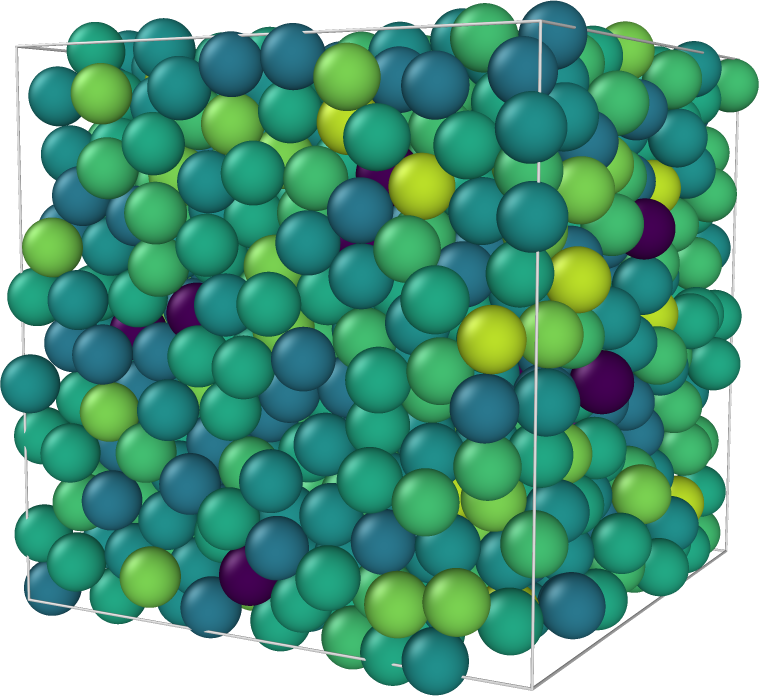}
	\caption{Upper panel: jammed packing in a two-dimensional, bidisperse system (with diameter ratio $1:1.4$). Rattlers are coloured in orange and the full network of contacts is shown, with the magnitude of contact forces represented by the thickness of the lines. Bottom panel: monodisperse packing in three dimensions. The contact network is not shown to avoid cluttering but particles are coloured according to the number of their contacts: lighter particles have more contacts, while the darkest ones are rattlers (zero contacts). Both packings were obtained initializing \CAL with low density configurations: $\vp_{0}=0.4$ in $2d$ ($\vp_{J}=0.839$), and $\vp_{0}=0.2$ in $3d$ ($\vp_{J}=0.635$).}
	\label{fig:jammed configs}
\end{figure}

In \cite{code2022github}, we provide our own implementation of the \CAL algorithm, written in the Julia programming language~\cite{bezansonJuliaFreshApproach2017} and making use of the JuMP~\cite{jump} modeling package. To solve all the jamming LOP instances, we used the Gurobi Solver~\cite{gurobi}. We tested our code also with other free, open-source optimizers such as HiGHS~\cite{highs} and GLPK~\cite{glpk}. We found that even though all these solvers have a relatively low precision (about $10^{-9}$), \CAL is able to produce valid jammed packings systematically. This is a remarkable feature that contrasts with other algorithms, such as FIRE~\cite{FIRE}, which requires quad-precision computations to avoid over-shooting the jamming critical point~\cite{charbonneauJammingCriticalityRevealed2015,charbonneauUniversalNonDebyeScaling2016}. 

The precision of the solver determines the tolerance for satisfying the constraints and, consequently, the precision with which the dual variables are computed. Using Gurobi with the highest overall accuracy available ($10^{-9}$), overlaps larger than $10^{-8}$ never occur, and the force balance condition is satisfied with higher precision by several orders of magnitude (about $10^{-13}$). Moreover, using such solver, active and non-active dual variables are distinguished with double precision in most of the systems we examined. This implies that true contacts are identified with high precision and mechanical equilibrium is also guaranteed within a reasonably small tolerance.

Using our implementation~\cite{code2022github} of Alg.~\ref{alg:ILP algorithm} we are able to produce packings as the ones shown in Fig.~\ref{fig:jammed configs}. In the top panel, we illustrate the network of contacts in a $2d$ bidisperse configuration, while in the bottom panel spheres are colored according to the number of their contacts (visualization done using \cite{ovito}). These packings were crunched from an initial random configuration with $\vp=0.4$ (in $2d$) and $\vp=0.2$ (in $3d$). We mention that, compared with their corresponding jamming densities, these values amount to $\vp/\vp_J < 0.5$ and $\vp/\vp_J < 0.3$, respectively. This means that, even if the justification to transform the exact jamming OP \eqref{def:Jamming as optimization problem} into the LOP \eqref{def:Jamming as LP} was based on the assumption that $\vp \lesssim \vp_J$, so $\{\abs{\vb{s}_i}\}_{i=1}^N\;\forall i$ are small, the \CAL algorithm is sufficiently robust to produce valid packings even if initialized relatively far from jamming. We tested our algorithm in $d=2-5$ dimensions (see App.~\ref{app:further-characterization}), and we verified that in all cases the jammed packings thus produced satisfy the mechanical equilibrium condition and are isostatic. This shows convincingly that our algorithm generates typical jammed configurations, and can be employed to carefully explore the jamming transition of HS systems.

\section{The LS+\CAL route\\to jamming: probing the free energy landscape} \label{sec:MD+calippso}

In this Section, we present results obtained when combining \CAL with the Lubachevsky--Stillinger (LS) compression protocol~\cite{lubachevskyGeometricPropertiesRandom1990}. Specifically, the idea is to use configurations compressed with LS as initial conditions of the \CAL algorithm. This scheme will allow us to improve the performance of \CAL and to study in detail the influence of the initial condition on the jammed packings it produces. We argue that the jammed states we obtained following such combined approach, referred as LS+CALiPPSO, reflect the hierarchical structure of the free energy landscape (FEL)~\cite{charbonneauFractalFreeEnergy2014}.
Our numerical results suggest that, if \CAL is initialized from a configuration at very high pressure (as specified below), the packings it produces likely coincide with the ones that would be obtained extrapolating the LS compression to the infinite pressure limit (see Sec.~\ref{sec:phase diagram}). Hence, using LS+\CAL we can generate jammed packings~\cite{donevCommentJammingZero2004,ohernJammingZeroTemperature2003,ohernReplyCommentJamming2004}, reproducing many of the properties observed previously in the literature~\cite{charbonneauFractalFreeEnergy2014,artiacoExploratoryStudyGlassy2020,charbonneauUniversalMicrostructureMechanical2012}.

As we mentioned in the Introduction, the LS protocol allows us to compress hard spheres to very high pressures. However, it is not able to strictly reach the jamming condition, $1/p=0$. Nevertheless, the LS compression protocol is an excellent tool to \emph{approach} the jamming point: besides being fast, it has been amply verified that it closely reproduces the (phenomenological) equation of state~\footnote{Note that Eq.~\eqref{eq:hs-glass-eos} has been derived from a free volume analysis~\cite{salsburgEquationStateClassical1962}. Therefore, it does not correspond to the true thermodynamic equation of state~\cite{puz_TheorySimpleGlasses2020,charbonneauGlassJammingTransitions2017} (see also Fig.~\ref{fig:hs-eos-liquid-and-glass} in App.~\ref{app:further-md} for more details).} of HS glasses~\cite{md-code,berthierGrowingTimescalesLengthscales2016,parisiMeanfieldTheoryHard2010,salsburgEquationStateClassical1962}:
\begin{equation}\label{eq:hs-glass-eos}
	p = \frac{d}{1- \vp/\vp_J}\,,
\end{equation}
with $p=\beta P V/N$ the reduced pressure of the system. 

The LS compression protocol increases the particle diameters with a uniform growth rate, $\dot{\sigma}(t)=\kappa$. This compression is performed simultaneously to the dynamical evolution of the configuration. HS dynamics can be efficiently simulated using event-driven MD in arbitrary dimensions \cite{md-code,charbonneauFractalFreeEnergy2014,berthierGrowingTimescalesLengthscales2016,charbonneauUniversalMicrostructureMechanical2012}, or even in some mean-field models \cite{charbonneauNumericalDetectionGardner2015}. We used the implementation of Ref.~\cite{md-code}, so we limit our analysis to finite-dimensional, monodisperse systems. Before proceeding, let us note that depending on $\kappa$ the LS protocol can produce monodisperse configurations that possess some degree of crystallization~\cite{torquatoRandomClosePacking2000,hopkinsDisorderedStrictlyJammed2013,md-code}. As stated above, we want to produce only disordered packings; thus, we took care in choosing $\kappa$ to avoid any ordering in our configurations, as explained in the following.

In this work, we will focus on three-dimensional systems because of their special role as a minimal model displaying jamming phenomenology~\footnote{Jamming criticality can also be clearly observed in two dimensional systems, provided that bidisperse or polydisperse packings are used. In that sense, $3d$ monodisperse systems are simpler because the random particles' positions are the only source of disorder.}. Nevertheless, our approach is valid in any dimensionality $d>1$, (in $d=1$ the resultant packings are inevitably ordered chains). Monodisperse systems in $d\leq 3$ are prone to crystallize if slow compression rates are used. To avoid the formation of ordered domains in three-dimensional systems, we initially perform a fast compression with $\kappa^{(0)}=5\times10^{-3}$ (see App.~\ref{app:further-md} for more details); instead, there is no need to include this initial fast compression in higher dimensions.

Our LS+\CAL protocol to reach the jamming point of HS systems is composed of the following steps:
\begin{enumerate}
	\item We generate a random HS configuration, i.e., we draw from a uniform distribution the spheres' positions, at low packing fraction $\vp_0$ and without overlaps, and initialize the LS compression protocol with it.
	\item \label{lsilp:fast-compression} To avoid crystallization, we perform a fast compression, with a compression rate $\kappa^{(0)}=5\times10^{-3}$, until $\ptar^{(0)}=500$. Note that this step can be safely avoided in $d\geq4$ (see, e.g. Fig.~\ref{fig:results-4d-5d}).
	\item \label{lsilp:ls} We initialize a new LS compression with the HS configuration obtained at $\ptar^{(0)}=500$, and further compress it until a target pressure $\ptar \gg1 $ (see below for detailed values) is reached. This second compression is performed with a smaller compression rate $\kappa$. 
	\item \label{lsilp:ilp} We use the HS configuration at $\ptar$ from the previous step to initialize the \CAL algorithm. Following the implementation of Algorithm~\ref{alg:ILP algorithm}, \CAL is executed until the convergence condition $(\opt{\lpf}, \opt{\va{s}})= (1, \va{0})$ is verified, within a given tolerance. In this way, we obtain an HS configuration at the jamming point. All the results reported below have been obtained at fixed tolerance: $\text{tol}_{\vb{s}}=10^{-9}, \text{tol}_\lpf=10^{-12}$. 
\end{enumerate}

The two protocols involved, LS and CALiPPSO, compress HS configurations at very different speeds. In particular, the LS compression is a finite-time protocol. Thus, while it is unable to produce the states given by a quasi-static construction, such as state-following or adiabatic compression~\cite{charbonneauGlassJammingTransitions2017,rainoneFollowingEvolutionGlassy2016}, it is relatively slow when compared with the \CAL instantaneous inflation of spheres. The \CAL algorithm instead is, for all purposes, a quenched compression, or crunching. Consequently, LS+\CAL is an out-of-equilibrium procedure that brings an HS configuration to its jamming point without following the thermodynamic equation of state (see also Sec.~\ref{sec:phase diagram}).

\subsection{The role of the initial condition and the FEL structure}\label{sec:LS-parameters}

The most relevant parameters of the LS+\CAL protocol are the LS compression rate, $\kappa$ (see step~\ref{lsilp:ls} above), and the target pressure from which \CAL is initialized, $\ptar$ (see step~\ref{lsilp:ilp} above). In addition, we can change the number of particles in the system, $N$. Notice that, due to memory effects in hard spheres~\cite{charbonneauMemoryFormationJammed2021}, in principle the value $\vp_{\text{tar}}$ from which \CAL is initialized, i.e., the packing fraction such that $p(\vp_{\text{tar}}) =\ptar$, should be considered as a parameter of the protocol. However, for simplicity we will assume that, in the glassy regime, the exact value of $\vp$ is uniquely determined by $p$, and the influence of $\vp_{\text{tar}}$ is considered to be implicitly captured by $\ptar$. Therefore, in the following, we will investigate only the influence of $\kappa$, $\ptar$, and $N$ on the final jammed states.

For fixed $\kappa$ and $N$, we construct a \emph{sample} by compressing via LS the same initial HS configuration to different values of $\ptar$ in the range $[10^3, 10^{11}]$. In such a way, all the jammed states of a sample belong to the same glassy state. This feature can be exploited to explore the FEL's structure at jamming. Since our compression protocol stops as soon as $p\geq \ptar$, the precise value of $p$ obtained for a target $\ptar$ can slightly vary. For a given $\ptar$, there is a small sample variability (usually less than $1\%$) in the actual value of the pressure.

\begin{figure}[!htb]
	\includegraphics[width=\columnwidth]{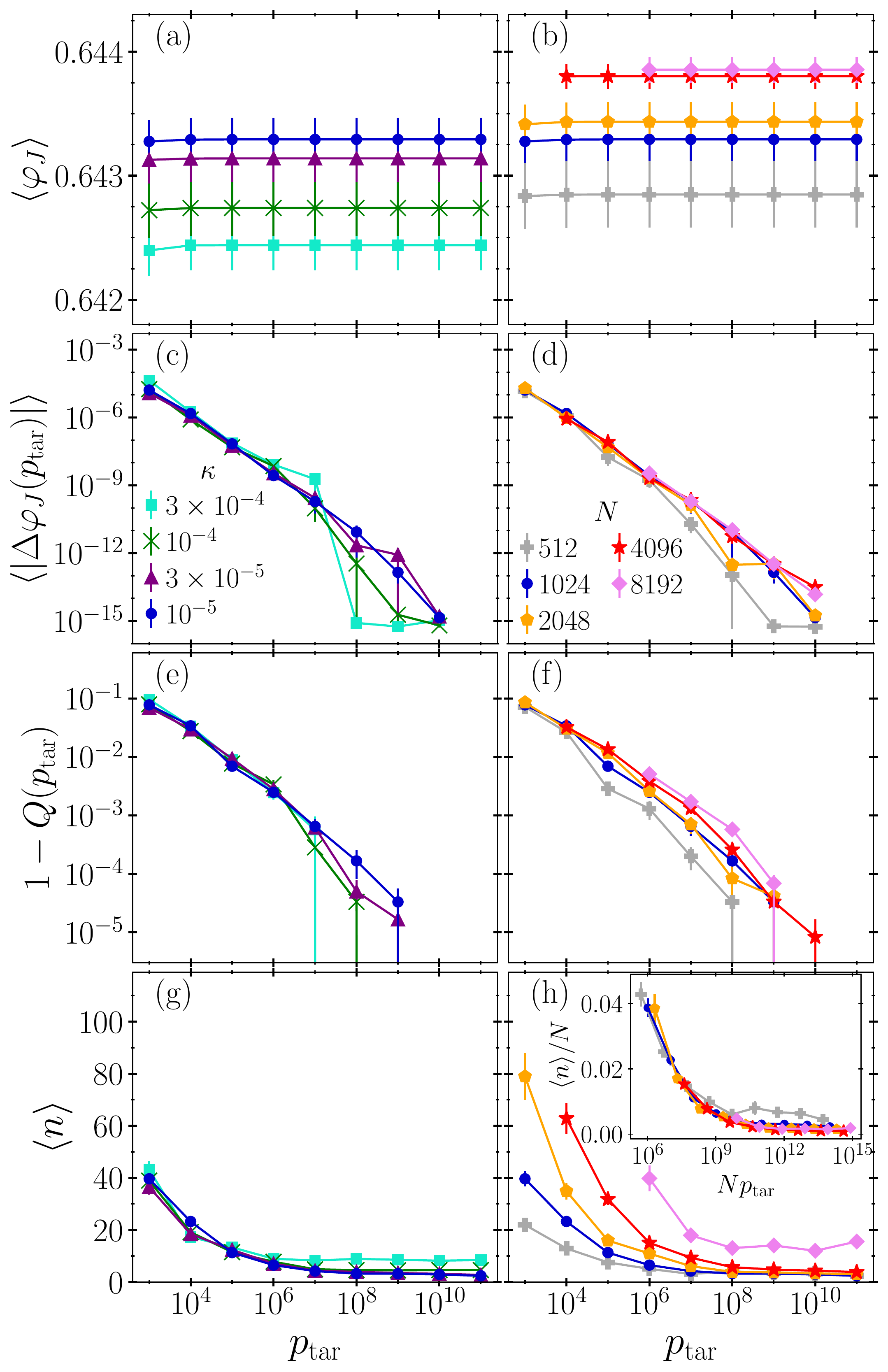}
	\caption{Dependence of various properties of the jamming configurations obtained via LS+\CAL as a function of $\ptar$, for $N=1024$ (left) and $\kappa=10^{-5}$ (right); note that plots on a same row share the scale of their vertical axes. We present results for several values of the compression rate $\kappa$ (left column) and the system size $N$ (right column). All values reported are the average and standard error over 20 samples. (a)-(b) Jamming packing fraction, $\avg{\vp_J}$. (c)-(d) Difference of the jamming packing fraction obtained at $\ptar$ from the one obtained at $p_{\max}=10^{11}$ in the same sample. (e)-(f) Similarity measure of the contact networks within the same sample, $1-Q(\ptar)$ (defined in the main text). (g)-(h) Number of linear optimizations, $n$, needed for \CAL to reach the jamming convergence condition. Inset of (h): size scaling collapse showing that $n/N$ is only a function of the thermodynamic pressure $P\sim Np$, as explained in the main text. Panels (c-f) support the presence of a hierarchical structure of the jamming landscape, as explained in the text.}
	\label{fig:dependence-ptar-kappa}
\end{figure}

Let us begin by analyzing the dependence of $\vp_J$ on the LS+\CAL parameters. In particular, let us consider its sample average, $\avg{\vp_J}$. All the averages reported in the following are computed over 20 samples, unless otherwise stated. We consider $\ptar\in [10^3, 10^{11}]$, and several values of $\kappa$ and $N$. Fig.~\ref{fig:dependence-ptar-kappa}a shows the dependence of $\avg{\vp_J}$ on $\ptar$, for various $\kappa$, at $N=1024$. Fig.~\ref{fig:dependence-ptar-kappa}b illustrates analogous results for different system sizes, at fixed $\kappa=10^{-5}$. In agreement with previous studies~\cite{ohernJammingZeroTemperature2003,torquatoReviewJammedHardparticlePackings2010,parisiMeanfieldTheoryHard2010}, we find that $\avg{\vp_J}$ is always close to $0.64$. We see that changing the target pressure of the LS compression protocol has a very small effect in the pressure range we considered. In contrast, slowing down the compression or increasing the system size increases $\vp_J$, albeit always within a narrow interval. These features are not surprising. Lowering $\kappa$, the particles have more time to rearrange; therefore, the system can reach deeper minima of the FEL, yielding higher values of $\vp_J$~\cite{artiacoExploratoryStudyGlassy2020}. Instead, border or periodic effects are reduced in larger systems, so the particles are less constrained, and higher values of $\vp_J$ can be achieved. Nevertheless, the data suggest a rather quick convergence to the thermodynamic limit value.

To further investigate the properties of the jamming configurations obtained via the LS+\CAL algorithm, we define the quantity $| \Delta \vp_J(\ptar) | := | \vp_J(p_{\max}) - \vp_J(\ptar) |$, where $p_{\max} = 10^{11}$ is the largest target pressure considered, and is used as a reference for comparison. $| \Delta \vp_J |$ quantifies to what extent the packing fraction of a jammed configuration changes if the \CAL crunching begins from a smaller $\ptar$. We seldom find that $\Delta \vp_J$ is a non-monotonic function of $\ptar$, so in Figs.~\ref{fig:dependence-ptar-kappa}c-d we consider its absolute value. We report $ \avg{| \Delta \vp_J (\ptar) |}$, averaged over 20 samples, for several values of $\kappa$ at $N=1024$ in Fig.~\ref{fig:dependence-ptar-kappa}c, and for different system sizes $N$ at $\kappa=10^{-5}$ in Fig.~\ref{fig:dependence-ptar-kappa}d. We observe that in general $\avg{| \Delta \vp_J |}$ is a decreasing function of $\ptar$. Moreover, from Fig.~\ref{fig:dependence-ptar-kappa}c we see that $\avg{\abs{\Delta \vp_J(\ptar)}}$ is independent of the compression rate only for $\ptar \leq 10^6$.
For larger pressures, the effects of $\kappa$ become relevant. For instance, using the fastest compression ($\kappa=3\times 10^{-4}$), $\avg{\abs{\Delta \vp_J(\ptar)}} = 0$ within the numerical precision for $\ptar \geq 10^8$. This means that, in each sample, the LS+\CAL algorithm finds the same $\vp_J$. Instead, smaller compression rates yield $\avg{\abs{\Delta \vp_J(\ptar)}} > 0$ for higher pressure, with $\Delta \vp_J$ remaining finite within an interval that grows as $\kappa$ decreases. A cleaner signature of this behavior is discussed below considering the similarity of the network of contacts (see Figs.~\ref{fig:dependence-ptar-kappa}e-f). On the other hand, Fig.~\ref{fig:dependence-ptar-kappa}d shows that size effects are negligible, at least for $N\geq 1024$.

As explained before, employing the LS+\CAL protocol, we can easily extract the network of contacts of a jammed configuration. Notice that this network univocally defines a minimum of the FEL~\cite{charbonneauFractalFreeEnergy2014}. Thus, a more refined measure of similarity between jammed configurations can be obtained comparing the contact networks rather than the packing fractions of two configurations. In particular, we are interested in quantifying the similarity between two configurations of the same sample at different target pressures. Considering again the configuration produced from $p_{\max}$ as reference, we define $Q(\ptar) := \abs{\mathcal{C}(\ptar)\cap \mathcal{C}(p_{\max})}/N_c(p_{\max})$, where $\mathcal{C}(\ptar)\cap \mathcal{C}(p_{\max})$ is the intersection between the contact networks of the two configurations, while $N_c(p_{\max})$ is the number of contacts of the jammed configuration at $p_{max}$. Therefore, $Q(\ptar)$ measures the number of common contacts between the packings obtained initializing \CAL with two configurations in the same sample, one at $\ptar$ and the other at $p_{\max}$. $Q(\ptar) = 1$ only when the contact networks exactly coincide. Similar observables have been used elsewhere~\cite{charbonneauFractalFreeEnergy2014,charbonneauMemoryFormationJammed2021,dennisJammingEnergyLandscape2020} to investigate the FEL's structure near and at jamming. Our findings are reported in Figs.~\ref{fig:dependence-ptar-kappa}e (resp.\ \ref{fig:dependence-ptar-kappa}f) for fixed $N$ and different $\kappa$ (resp.\ fixed $\kappa$ and different $N$). For the values of $\ptar, \kappa$, and $N$ considered here, we see that the main structure of the contact network is shared by jammed packings in the same sample, i.e., within a given meta-basin~\cite{charbonneauGlassJammingTransitions2017,charbonneauFractalFreeEnergy2014}. Yet, a small fraction of the contacts ($\sim 10 \%$) are only determined gradually, as the \CAL input configuration goes down in the landscape as $\ptar$ increases. 
Thus, the behavior of $1-Q(\ptar)$ reported in these figures is in accordance with that of $\abs{\Delta \vp_J(\ptar)}$ discussed above.	
Using $Q(\ptar)$ the influence of changing the compression speed, as well as the system size, are distinguished more cleanly. We observe that for a given $\kappa$ or small $N$, there is a threshold pressure above which we obtain $Q(\ptar)=1$ for all samples.

These findings are in agreement with the rough and hierarchical structure of the FEL predicted for hard spheres in infinite dimensions (see Sec.~\ref{sec:intro}). Assuming such mean-field picture to be valid also in finite-dimensional systems, as suggested by recent numerical evidence~\cite{charbonneauFractalFreeEnergy2014,artiacoExploratoryStudyGlassy2020,dennisJammingEnergyLandscape2020,charbonneauGlassJammingTransitions2017}, our results indicate that the FEL's structure could only be fully resolved with an infinitesimally slow compression, $\kappa \to 0$ (assuming that crystallization can be avoided in this limit; see the discussion at the beginning of Sec.~\ref{sec:phase diagram}). Instead, when $\kappa$ is finite, there will be a threshold pressure $\pth$, such that if $\ptar > \pth$ the jamming packings obtained within the same sample will be inevitably the same: the system is trapped in one minimum. The fact that $\pth$ increases with $N$ can be understood considering that the number of minima in the FEL increases tremendously with the system size~\cite{stillinger1999exponential,charbonneauGlassJammingTransitions2017,puz_TheorySimpleGlasses2020} which suggests that for large systems finding diverse minima is much likelier, even for very high values of $\ptar$. The role of $\pth$ will be discussed in more detail in the next Section. 

The dependence of the final packing on $\ptar$ is reflected not only in physical quantities, such as $\vp_{J}$ and the network of contacts but also in algorithmic properties. In Figs.~\ref{fig:dependence-ptar-kappa}g-h we consider the number of linear optimizations needed to reach the convergence criterion, $n$, as a function of $\ptar$. In agreement with the rest of the panels, these curves illustrate that, for sufficiently large $\ptar$, the value of $n$ remains essentially unchanged, despite the fact that the distance to $\vp_{J}$ decreases by several orders of magnitude. Moreover, from Fig~\ref{fig:dependence-ptar-kappa}h we see a clear dependence of $n$ on the system size $N$. Interestingly, such dependence can be teased out by assuming $n = N \mathcal{G}(N \ptar)$, where $\mathcal{G}(x)$ is a scaling function such that $\mathcal{G}(x)\sim $ constant, for $x\to\infty$. We put to test this scaling in the inset of Fig~\ref{fig:dependence-ptar-kappa}h, obtaining a very good collapse of the curves at different $N$. The small deviations observed in $N=512$ systems at very large target pressure are likely explained from the fact that, in such cases, $n$ matches the minimal number of iterations, $n_0=2$ (see Sec.~\ref{sec:complexity calippso}). We can rationalize the scaling variable $N\ptar$ by noting that, from the definition of the reduce pressure~\cite{santosStructuralThermodynamicProperties2020}, we have that $Np = \beta P V$, where $P$ is the usual thermodynamic pressure, $\beta$ is the inverse temperature, and $V$ is the system's volume. Since in all our MD simulations we fix $\beta=10$ and $V=1$, $N\ptar \sim P$ provides the natural variable to measure how far from jamming a configuration is, independently of the number of particles. Our results hence indicate that $n = N \mathcal{G}(P)$; for fixed $P$, this linear relation suggests that \CAL works by blocking few degrees of freedom at each linear optimization (see also the discussion in Sec.~\ref{sec:complexity calippso}).

Finally, we compare the properties of the packings obtained via the LS+\CAL protocol to the ones that would be obtained using only the LS compression protocol in the limit $p\to\infty$. Unfortunately, it is very difficult to access the microscopic details of the jammed packings, such as contacts and interparticle gaps, purely from an extrapolation of the data at finite pressure (see, for instance, Fig.~\ref{fig:gaps-after-md} in App.~\ref{app:calippso-vs-md} and the corresponding discussion). However, recalling that an HS glass in $d$ dimensions is well described empirically by the equation of state~\eqref{eq:hs-glass-eos}, we can use this expression to fit the values of $p$ and $\vp$ computed during the MD simulations of the LS protocol, and then estimate the jamming packing fraction $\vp_J^{(LS)}$ that would correspond to $1/p=0$.

Fixing $N=1024$, we estimate $\vp_J^{(\text{LS})}$ using the LS data in the range $p\in[10^5, 10^{11}]$, and we perform the sample average of such values. Then, we compare it to the sample average of $\vp_J$ obtained via the LS+\CAL protocol. The results (mean and standard error) for various compression rates $\kappa$ are reported in Table~\ref{tab:phiJ-MD}. We only include the LS+\CAL results at $\ptar = 10^3$ and $10^5$ because with larger pressures the sample average of $\vp_J$ is identical, within the statistical error, to the one at $\ptar = 10^5$. The results in Tab~\ref{tab:phiJ-MD} imply that, at fixed $\kappa$, the jammed state obtained via the LS+\CAL protocol for sufficiently large $\ptar$ (presumably smaller than $\pth$) coincides with $\vp_J^{(\text{LS})}$. Similar results have been obtained performing a comparison of $\vp_J$ and $\vp_J^{(\text{LS})}$ within each sample. However, given that for $\ptar \geq 10^5$ the uncertainty in the estimation of $\vp_J^{(\text{LS})}$ (about $10^{-5}$) is much larger than the values of $\vp_J - \vp_J^{(\text{LS})}$ (about $10^{-7}$) and $\abs{\Delta \vp(\ptar)}$, a more precise comparison for larger $\ptar$ is unfeasible.

\begin{table}[!htb]
	\centering
	\caption{Average values of the jamming density of the LS compression protocol (second column), and of LS+\CAL (third and fourth columns). Each row corresponds to a different growth rate (first column). $\vp_J^{(\text{LS})}$ is the $p\to\infty$ limit packing fraction within the LS protocol, estimated via a fit as explained in the main text. The values and uncertainties reported are obtained averaging over 20 samples of $N=1024$ particles.} 
	\begin{tabular}{|c|c| c | c |}
		\hline
		$\kappa$            & $\displaystyle \avg{\vp_J^{(LS)}}$ & $\avg{\vp_J(\ptar=10^{3})}$ & $\avg{\vp_J(\ptar=10^{5})}$ \\[2mm]
		\hline
		$ 3\times 10^{-4} $ & 0.64244(20)        & 0.64240(21)      & 0.64244(20)  \\
		$ 10 ^{-4} $        & 0.64274(23)        & 0.64272(22)      & 0.64274(23)  \\
		$ 3\times 10^{-5}$  & 0.64314(19)        & 0.64313(19)      & 0.64314(19)  \\
		$ 10^{-5} $         & 0.64329(18)        & 0.64328(17)      & 0.64329(18)  \\
		\hline
	\end{tabular}
	\label{tab:phiJ-MD}
\end{table}

\subsection{The LS+\CAL phase diagram}\label{sec:phase diagram}

\begin{figure}[!htb]
    \centering
	\includegraphics[width=\linewidth]{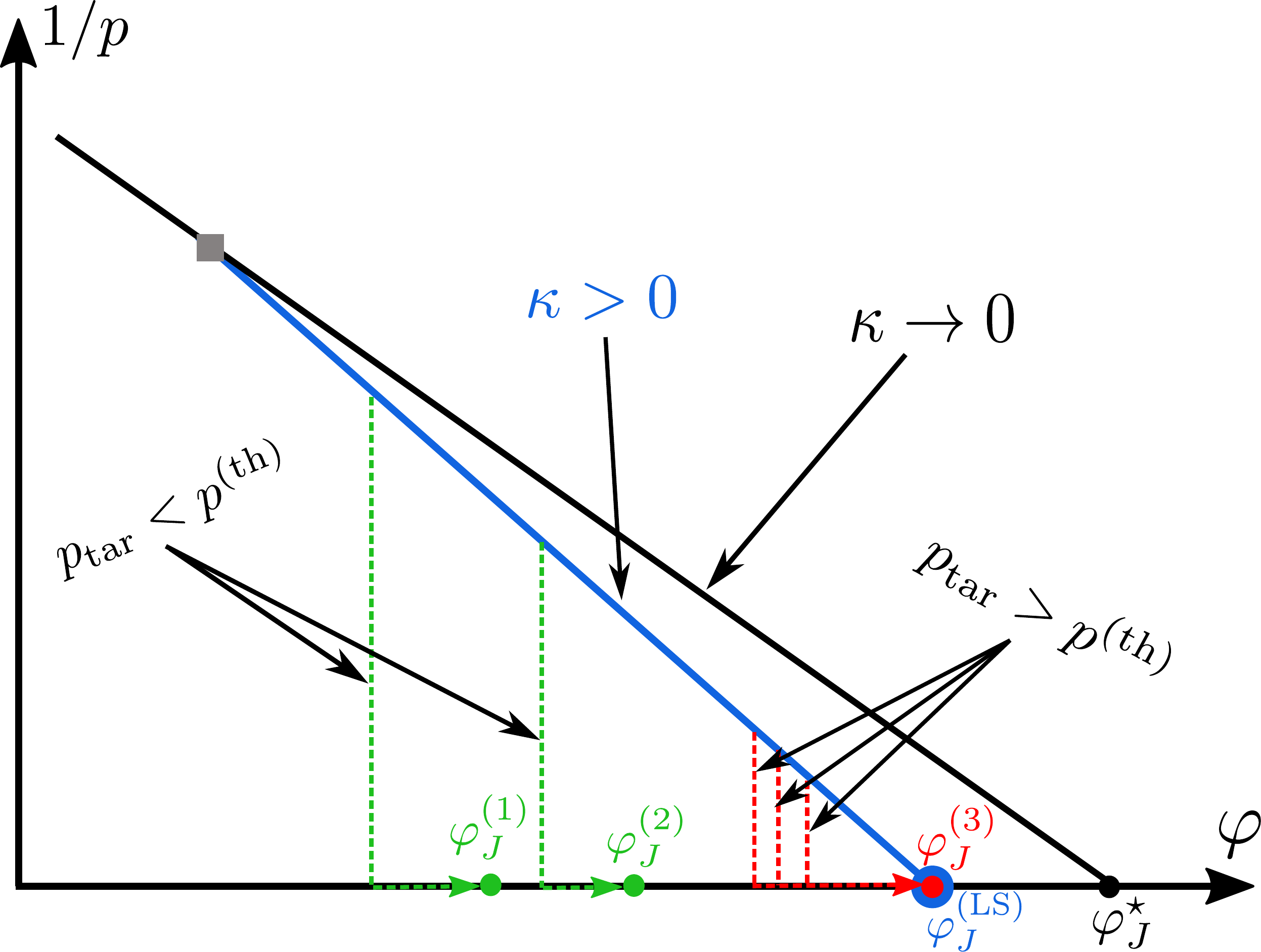}
	\caption{Sketch of LS+\CAL route to jamming in the $(\vp,1/p)$ plane. We include a schematic comparison of the LS protocol (blue line, $\kappa>0$) and the quasi-static one (black line, $\kappa \to 0$) obtained by, e.g., adiabatic compression, that ends in $\opt{\vp}_J$. These two protocols coincide up to a certain pressure, identified by the gray square, but at larger pressure the finite compression rate of LS makes it detach from the idealized thermodynamic line.
	$\vp_J^{(\text{LS})}$ is the jamming packing fraction that would be obtained by LS when $1/\ptar=0$, and is identified with the big blue circle to represent the large uncertainty in its estimation with respect to $\vp_J$ from \CAL (see Sec.~\ref{sec:LS-parameters}).
	\CAL crunching is represented by dashed lines, initialized at a given $\ptar<\infty$.
	If $\ptar \lesssim \pth$, \CAL leads to packings of different densities, e.g. $\vp_J^{(1)}< \vp_J^{(2)}$ (green lines). Conversely, when $\ptar \gtrsim \pth$, different initial conditions lead to the same jammed state, $\vp_J^{(3)}$ (red lines) that very likely coincides with $\vp_{J}^{(\text{LS})}$. 
	Note that the interval of different values of $\vp_J$ has been drastically magnified, and that the $\kappa\to0$ path is used only as a schematic reference since no finite-time numerical algorithm could follow such a path.
	}
	\label{fig:route-MD-ILP-to-jamming}
\end{figure}

In view of the results presented in Sec.~\ref{sec:LS-parameters}, we can sketch the path followed by HS configurations brought to jamming via the LS+\CAL algorithm. Such path is illustrated in Fig.~\ref{fig:route-MD-ILP-to-jamming} in the $(\vp, 1/p)$ plane, where it is compared with the path one would obtain considering an infinitesimally slow compression (black solid line), i.e., the hypothetical thermodynamic equation of state of the glass. Let us stress that in practice no algorithm can reach the $\kappa \to 0$ limit. Moreover, in $3d$, monodisperse systems decreasing $\kappa$ below a critical compression rate inevitably leads to partial crystallization~\cite{zhangConnectionPackingEfficiency2014} (accompanied by significantly larger values of $\varphi_J$). Since our results have been obtained at finite $\kappa$, and we have checked the absence of any sign of crystallization (see Fig.~\ref{fig:rdf-vs-p}), in our discussion we will neglect the possibility of forming a crystal during the LS compression, and use the $\kappa\to0$ limit just as an idealized reference.

In Fig.~\ref{fig:route-MD-ILP-to-jamming}, the first half of our protocol, i.e., the LS compression, is identified by the $\kappa>0$ line which, for small $p$, is virtually identical to the thermodynamic one. However, for any finite $\kappa>0$, the glass obtained from the LS protocol inevitably detaches from the path associated with the quasi-static limit. The \CAL crunching is identified by the dashed, red lines, highlighting the fact that no value of $p$ can be assigned during this process. Fig.~\ref{fig:route-MD-ILP-to-jamming} illustrates also the threshold pressure introduced before, such that, if $\ptar< \pth$, \CAL produces packings with different jamming density and network of contacts. Conversely, when $\ptar > \pth$, \CAL converges to the same jammed microstate.
As discussed above, such a state is presumably identical to the $p\to\infty$ limit of the LS protocol. From the FEL perspective, this implies that in the regime $\ptar < \pth$ the hierarchical structure of the landscape affects the final packings realized by the LS+\CAL algorithm, while for $\ptar > \pth$ such structure is not longer detected by the algorithm and it always reaches the same minimum. In addition, our data show that $\pth$ is not universal: $\pth$ increases as $\kappa$ decreases or $N$ increases. Nevertheless, further studies are needed to better characterize such dependence.
Finally, we note that the values of $\vp_J$ obtained using either \CAL or the $p\to\infty$ limit of LS are expected to be smaller than the quasi-static jamming density $\opt{\vp}_{J}$. This discrepancy has already been observed in polydisperse systems~\cite{berthierEquilibriumSamplingHard2016}, and mean-field models~\cite{hwangForceBalanceControls2020}, although in $3d$ monodisperse HS systems such difference should be very small~\cite{charbonneauUniversalMicrostructureMechanical2012}.

\section{Time complexity of \CAL}\label{sec:complexity calippso}

In this Section, we explore the performance of the \CAL algorithm as a function of the system size $N$. In particular, we will analyse the time required by the algorithm to converge ($\tau$), the number of linear optimizations ($n$), and the mean time required per linear optimization in a given system ($t_{\rm LOP}$), as a function of $N$. 
The results presented in this Section are obtained for $3d$, monodisperse systems, fixing $\ptar=10^{7}$, and within two different LS compression protocols. In the first one, we set $\kappa_f=10^{-5}$ for all system sizes; in the second one, we introduce a size-dependent growth rate, $\kappa_N$, such that $\dot{\vp}$ is constant for all values of $N$. More precisely, given that $\dot{\vp} \sim N \sigma^2 \dot{\sigma} \sim (N\vp^2)^{1/3} \kappa$, if all configurations must be subject to the same compression rate for a given value of $\vp$, then $\kappa_N \sim N^{-1/3}$ (more generally, in $d$ dimensions we would have $\kappa_N \sim N^{-1/d}$). We consider system sizes up to $N_{\max}=16384$. For the second compression protocol, we fix $\kappa_N = (N_{\max}/N)^{1/3} \kappa_f$. Notice that, since the smallest system size is $N_{\min}=256$, the ratio between the inflation rates of the two protocols is $\kappa_{N_{\min}}/\kappa_{f}=4$ at most. Comparing the two scenarios is useful to guarantee that the scalings we obtain are intrinsic to CALiPPSO.

Since the performance of the LS protocol has been analyzed before~\cite{torquatoRobustAlgorithmGenerate2010}, we will not discuss it here. However, in App.~\ref{app:calippso-vs-md} we show a comparison of the \CAL and LS times and their dependence on $\ptar$ (see Fig.~\ref{fig:times LS and ILP}). 

We test our implementation of the \CAL algorithm in a 6 cores computer, with processor Intel Core i7-8700 at 3.2 GHz. Each jamming LOP is solved using the Gurobi Solver~\cite{gurobi} (version 9.1) along with the JuMP package~\cite{jump} of the Julia programming language~\cite{bezansonJuliaFreshApproach2017}. All the jamming LOP instances are solved using the Gurobi's barrier method, with 6 threads running concurrently, and setting the feasibility and optimality tolerance to their most stringent values, $10^{-9}$. The rest of the solver parameters are used with their default values~\cite{gurobi}.

\begin{figure}[!htb]
	\includegraphics[width=\linewidth]{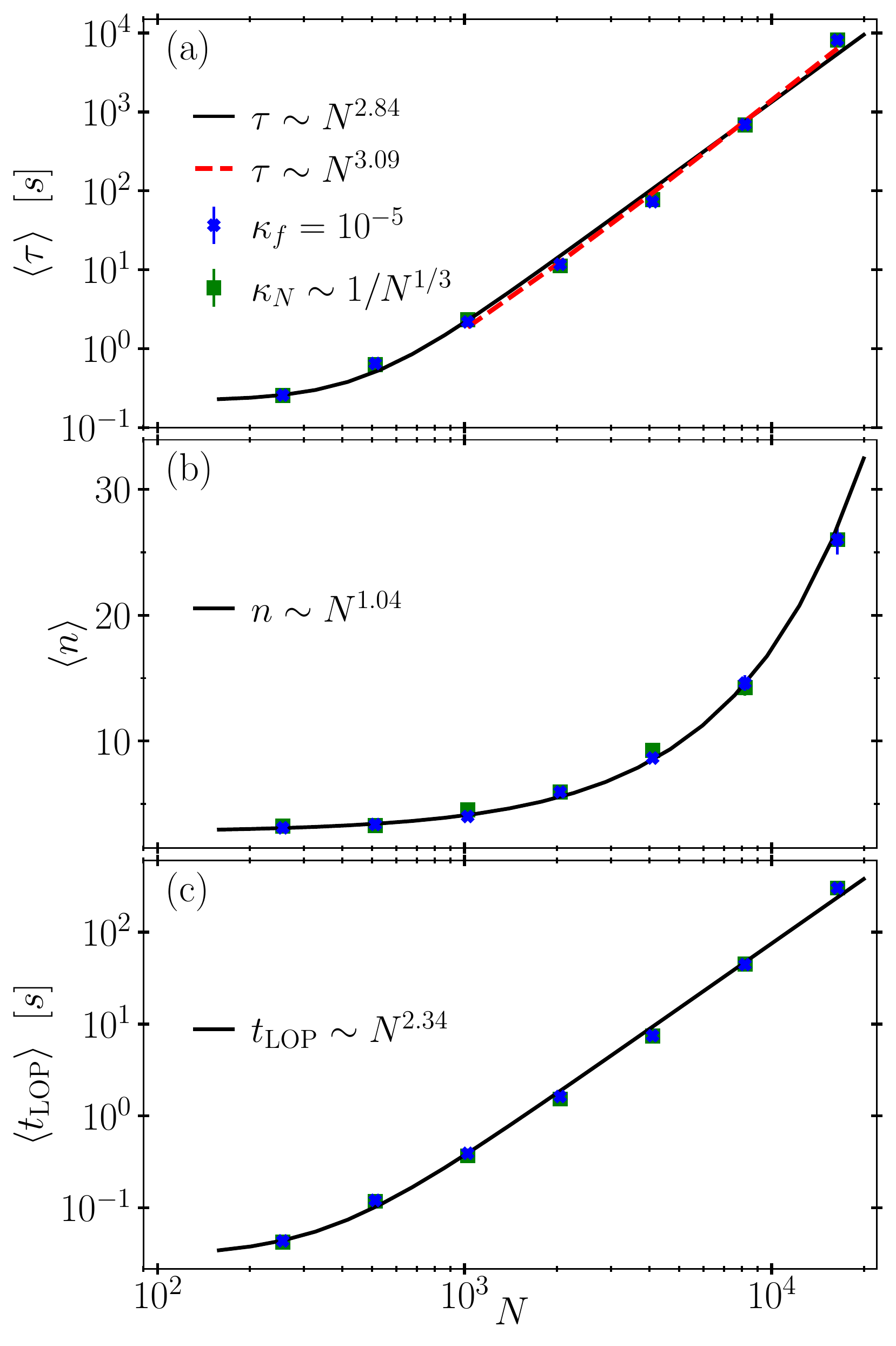}
	\caption{Complexity of \CAL as a function of system size $N$, measured by the convergence time, $\tau$ (panel (a)), number of linear optimizations used to reach the convergence condition, $n$ (panel (b)), and the mean time required in solving the jamming LOPs of a system, $t_{\rm LOP}$ (panel (c)). We set $d=3$, $\ptar=10^7$ and used two LS compression protocols: a fixed growth rate $\kappa_f=10^{-5}$ (blue crosses) and a size scaled compression $\kappa_N$ (green squares). Data reported correspond to the average over $M=100$ samples for each value of $N$ and compression strategy. Solid lines show the size scaling obtained by least-squares fits to the data obtained using $\kappa_f$; the red, dashed line in panel (a) is an analogous fit obtained considering only $N\geq 1024$ (see main text for more details). The fits obtained with the $\kappa_N$ protocol are virtually identical.}
	\label{fig:scaling-with-size}
\end{figure}

In Fig.~\ref{fig:scaling-with-size}, we report the results obtained after averaging over $100$ samples for each value of $N$ and compression protocol, along with the corresponding standard error. Notice that in all three panels, the mean values obtained at fixed compression, $\kappa_f$ (blue crosses), are very similar to the ones obtained with the size-scaled compression rate, $\kappa_N$ (green squares). Therefore, here we only present the data analysis made within the former protocol. In panel~(a), we illustrate the size dependence of $\tau$, as well as the result of a least-square fit of the form $\tau = \tau_0 + c N^\alpha$ (black line). We chose the offset time, $\tau_0$, such that residuals of the linear model $\ln(\tau-\tau_0) = \ln c + \alpha \ln N$ were minimized, and obtained $\alpha = 2.84\pm 0.09$. We can see that data at small $N$ deviate from the large size trend. Performing again the fit, but including only data for $N\geq 1024$, we find that $\alpha=3.09\pm0.13$ (red, dashed line). Similar values are obtained for the compression protocol with $\kappa_N$, whence we argue that this is the intrinsic size dependence of $\tau$ of CALiPPSO. We can conclude that, for high enough $\ptar$ (see Sec.~\ref{sec:phase diagram}), the complexity of \CAL scales approximately as $N^3$.

We can also investigate which components of the \CAL algorithm are more sensitive to the system size. In other words, does $\tau$ increase with $N$ due to a much larger number of LOP instances required for convergence? Or, instead, is it the time spent for each linear optimization that gives the largest contribution to $\tau$? To answer this questions, in panel (b) (resp.\ (c)) we plot the behavior of $\avg{n}$ (resp.\ $\avg{t_{\rm LOP}}$) for different values of $N$ and the two protocols considered. Our results clearly show that the increasing computational cost for jamming larger systems is due to the time required to solve each LOP instance. Indeed, analogous fits to panel~(a) yield that $t_{\rm LOP} \sim N^{2.34\pm 0.05}$, while $n \sim N^{1.04\pm 0.03}$. Hence, the main contribution to the growth of $\tau$ with $N$ comes from the increasing difficulty in solving a single instance of the jamming LOP, while the number of steps required for convergence grows only linearly with the system size. Notice that the scaling exponent obtained for $\avg{ t_{\rm LOP}}$ vs.\ $N$ is significantly smaller than the one from the worst-case scenario analysis that would yield a dependence $N^{3.5}$, given the complexity of the interior point method itself~\cite{boydConvexOptimization2004}.

Our data show also that the minimal number of iterations required by \CAL is approximately $n_0=2$, a value confirmed by analogous results for very large $\ptar$ (cf.\ Fig.~\ref{fig:dependence-ptar-kappa}h). 
This finding can be intuitively explained by assuming that, even if the initial condition is very close to jamming and a single linear optimization would make a system reach its final density, $\opt{\va{s}}$ would be determined by saturating the \emph{linear} constraints. Therefore, an extra iteration would be needed to make linear contacts precisely match physical ones.

It is worth recalling here that from the scaling presented in Sec.~\ref{sec:LS-parameters}, $\avg{n}\sim N \mathcal{G}(P)$, the linear dependence of the number of iterations with  $N$ is expected to be valid even for smaller values of $\ptar$. In other words, even if \CAL was initialized from a smaller pressure, the proportionality relation $\avg{n}\sim N$ would be valid for a fixed $P$.

Importantly, our tests have also shown that the convergence condition for $\lpf$ is achieved noticeably faster than the analogous condition on $\va{s}$, especially for $N>1024$. That is, the last iterations of \CAL are employed to fine-tune the particles' positions in order to attain stability (by matching linear with real contacts), and not in increasing the system density. This suggests that our algorithm might be improved by implementing some sort of “relaxation” during the configuration update. For instance, writing the optimal inflation factor of the jamming LOP as $\lpf^\star=1+\gamma$, with $\gamma>0$, we can update the particles' diameter as $\sigma \to \sqrt{1 + c \gamma} \sigma$, for some $0<c<1$, instead of the rule considered so far. By doing so, it is likely that both $\lpf$ and $\va{s}$ would converge at a much similar rate because the slightly larger amount of free volume would allow obtaining larger optimal displacements when needed. This goes beyond the scope of the present work, and such alternatives will be explored in further studies.

Another relevant point emerging from  Fig.~\ref{fig:scaling-with-size} is that the simple assumption $\avg{\tau} \sim \avg{n} \cdot \avg{t_{LOP}}$ would give a slightly larger exponent for the scaling of \(\tau\) with $N$ than the one obtained from the direct fit. There are two mechanisms that might be responsible for this mismatch. The first and most obvious one is the contribution of the other steps described in Algorithm~\ref{alg:ILP algorithm}. Clearly, besides the linear optimization, \CAL also requires computing and updating the neighbor lists, identifying rattlers, modifying and storing arrays, etc. Because each of these operations is less demanding than the jamming LOP, their combined effect is to reduce the overall exponent. However, a second factor that should be considered is that $t_{\rm LOP}$ and $n$ are not necessarily independent variables, thus $\avg{n} \cdot \avg{t_{LOP}} \neq \avg{n t_{\rm LOP} } \sim \avg{\tau}$. The fact that the first of these quantities yields an $N$ dependence with a larger exponent implies that $n$ and $t_{\rm LOP}$ are anti-correlated. In other words, it is likely that our algorithm converges by performing a relatively small number of expensive linear optimizations.

As we mentioned in Sec.~\ref{sec:calippso}, the \CAL algorithm is suitable for systems in any dimensions $d$, as we evince in Fig.~\ref{fig:results-4d-5d} in Appendix~\ref{app:further-characterization} for $d=4$ and $d=5$. Here, we anticipate that its performance is affected by the dimensionality of the system because, as $d$ increases, each particle is surrounded by more neighbors. That is, even if at jamming each particle is in contact with an average of $2d$ other spheres, the amount of spheres within $\ell(\vp)$ that induce a constraint could grow much faster. More precisely, solving each jamming LOP has a polynomial complexity~\cite{boydConvexOptimization2004,nocedalNumericalOptimization2006} on the $\max (N_{dof}, M')$, following the notation of Sec.~\ref{sec:calippso}. Now, $N_{dof} \sim d N$, while $M' = \tilde{z}_d N$, where $\tilde{z}_d$ denotes the average number of near contacts in a $d$-dimensional system. Isostaticity and geometrical constraints imply $N_{dof} \leq M' \leq k_d N$, with $k_d$ the kissing number in $d$ dimensions. Thus, as long as $k_d $ is not much larger than $d$, we can expect our complexity analysis to hold. Unfortunately, $k_d$ increases exponentially in $d$, and having tighter bounds on $\tilde{z}_d$ is not trivial. Previous data in $d=4-6$~\cite{torquatoReviewJammedHardparticlePackings2010,md-code} suggest that the abundance of near contacts increases rapidly. Nevertheless, the constraint matrix $\mathcal{F}$ of the jamming LOP~\eqref{eqs:jamming standard LP} will remain rather sparse as long as $\tilde{z}_d \ll N$; this should avoid $t_{LOP}$ from reaching the worst-case complexity mentioned above. Finally, let us notice that $\tilde{z}_d$ is also influenced by the \CAL initial configuration parameters (e.g., $\ptar$ if the LS compression protocol is used). Nonetheless, our tests have shown that \CAL is an efficient algorithm even for moderately high values of $d$, as discussed in App.~\ref{app:further-characterization}. A more quantitative analysis is beyond the scope of the present work, and is left for future studies.

\section{Conclusions}\label{sec:conclusion}

In this work, we introduced the \CAL algorithm to produce disordered jammed packings of hard spheres with very high accuracy. In contrast with most of the existing algorithms, \CAL does not require introducing any effective potentials between particles. Instead, it is based on formulating the packing problem of hard spheres as a non-convex optimization problem, which is then solved through a series of more tractable linear optimization problems. 

Section~\ref{sec:calippso} contains our main results. We showed that, even if the linear problems are only approximations of the original problem, once convergence is attained, \CAL produces hard-sphere (HS) configurations that are optimal also with respect to the original non-convex problem. Importantly, using results of optimization theory, we analytically proved that \CAL packings are globally stable, in mechanical equilibrium, and isostatic.

From the analysis of the complexity of \CAL in $3d$ systems in Section~\ref{sec:complexity calippso}, we showed that its convergence time likely scales as $N^3$, where $N$ is the system size, making \CAL a very efficient algorithm. We provide our own implementation of \CAL in \cite{code2022github}.

Notably, from the isostatic packings produced by the \CAL algorithm, one can easily extract all the relevant information on the microstructure of the jammed HS configurations. Achieving the same precision with techniques based on molecular dynamics (MD) is certainly not as straightforward. When employing MD-based algorithms few gaps might be misclassified as contacts, potentially leading to unstable packings. A detailed discussion on this is provided in Appendix~\ref{app:calippso-vs-md}. \CAL solves these issues thanks to the fact that the complete microstructure of a jammed packing is obtained purely from static quantities. In fact, contact forces are identified from active Lagrange multipliers. Thus, forces are computed independently from inter-particle gaps.

To improve the performance of CALiPPSO, in Section~\ref{sec:MD+calippso} we combined it with the Lubachevsky-Stillinger (LS) compression protocol and verified that using these two algorithms together we can readily produce typical jammed packings. By means of extensive numerical simulations, we showed that the LS+\CAL protocol is capable of probing the hierarchical structure of the free-energy landscape with unprecedented accuracy. Studying the landscape structure in finite-dimensional HS systems is relevant for several reasons; for example, it provides a direct test of the recent mean-field theory of glasses and jammed systems~\cite{charbonneauFractalFreeEnergy2014,charbonneauGlassJammingTransitions2017,puz_TheorySimpleGlasses2020}. LS+\CAL represents an optimal candidate to accurately test whether the landscape of HS configurations possesses the ultrametric structure predicted by the theory, and recently confirmed in soft-sphere packings~\cite{dennisJammingEnergyLandscape2020}. 

Our method opens up numerous research directions towards the characterization of the jamming critical properties in finite-dimensional HS systems. First, exploring how the LS+\CAL protocol navigates the landscape upon reaching jamming could inform on the existence of the Gardner phase~\cite{berthierGardnerPhysicsAmorphous2019,kurchanExactTheoryDense2013,charbonneauGlassJammingTransitions2017} in finite-dimensional models. Moreover, our algorithm can be profitably used to confirm whether the Gardner-like algorithmic transition found in Ref.~\cite{charbonneauMemoryFormationJammed2021} is a generic feature of packing algorithms. Furthermore, by defining a cost function for hard spheres at jamming, \CAL can be employed to explore the stability of the jammed packings, as well as their spectral properties. Finally, another promising extension would be to adapt this algorithm to tackle constraint satisfaction problems. For instance, one could apply it to the spherical perceptron model~\cite{franz2016simplest,franzUniversalitySATUNSATJamming2017} in finite dimensions and study the properties of the SAT/UNSAT transition when it is approached from the SAT phase. This would provide a numerical validation of the corresponding mean-field results~\cite{altieriJammingTransitionHigh2016,franzUniversalitySATUNSATJamming2017}. We defer the study of these and other topics to future works.

%%%%%%%%%%%%%%%%%%%% Acknowledgements %%%%%%%%%%%%%%%%%%%%
\acknowledgements
We thank Edan Lerner for his very valuable help to understand other algorithms dealing with hard-sphere packings, and two anonymous referees whose comments helped us to improve the manuscript.
This work was supported by the Simons Foundation (Grant No. 454949 (G.P.)). C.A. acknowledges financial support from the European Research Council (ERC) under the European Union's Horizon 2020 research and innovation program (grant No. 101001902).

%%%%%%%%%%%%%%%%%%%% APPENDIX %%%%%%%%%%%%%%%%%%%%

\appendix

\section{Overconstrained systems: The case of $2d$ monodisperse packings}\label{app:further-2d}

As we showed in Sec.~\ref{sec:properties calippso packings}, our protocol produces configurations that satisfy the stability condition, $N_c \geq N_{dof}$. Nevertheless, even though the vast majority of packings produced by \CAL are isostatic, in the case of \emph{monodisperse} disks additional topological and geometrical constraints lead to hyperstatic configurations (i.e., the strict inequality is verified). As we explain here, in such overcontrained packings the excess of contacts only occur between pairs of rattlers.

\begin{figure}[!htb]
    \centering
    \includegraphics[width=\linewidth]{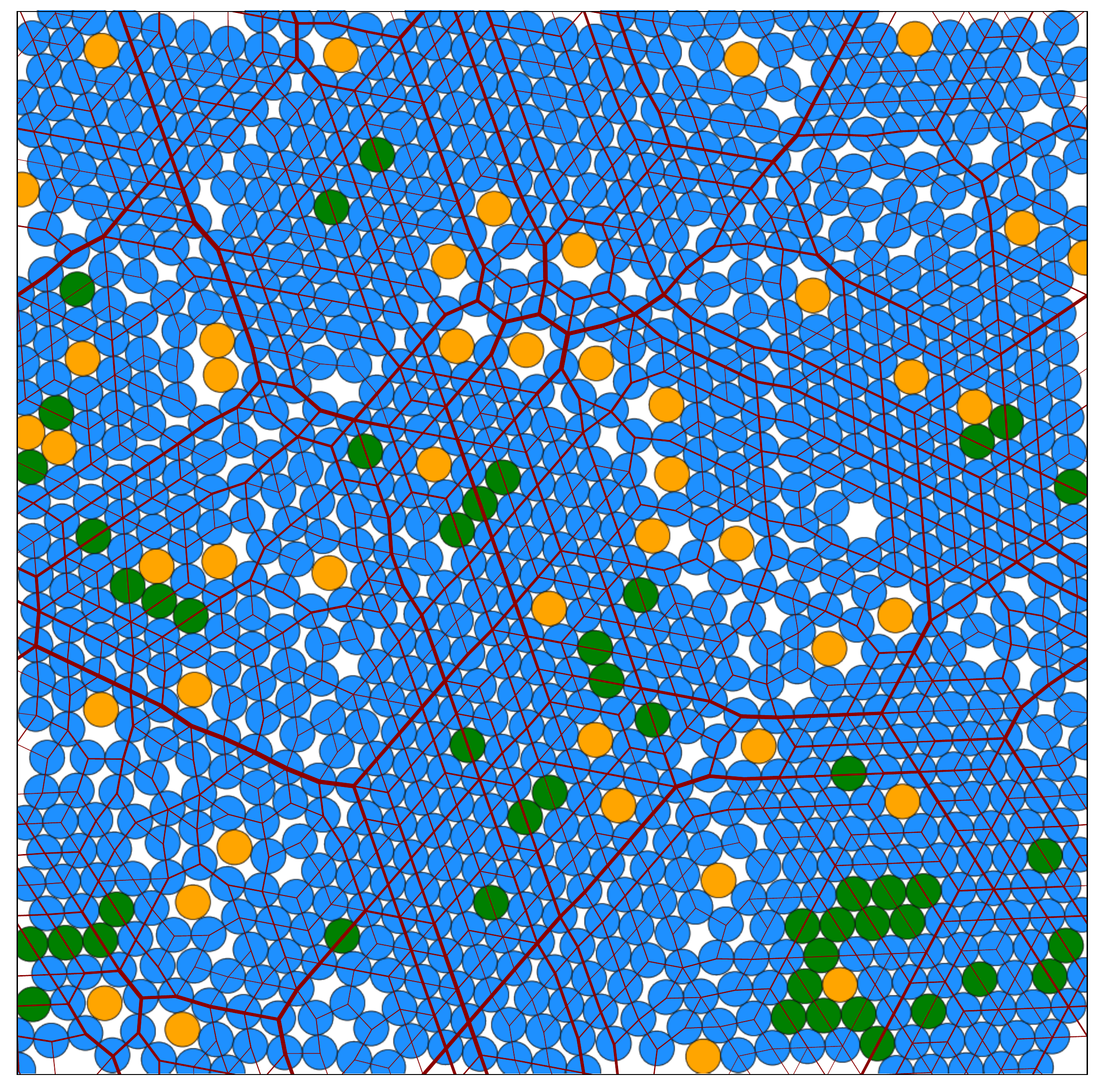}
    \caption{Unstable, jammed packing of $N=1024$ monodisperse hard disks. The presence of large crystalline domains is apparent. Particles coloured in orange are rattlers upon which no contact force is acting, while green disks are rattlers with finite contact forces. Particles of this latter type cannot be removed without breaking the force balance condition.}
    \label{fig:2d-mono}
\end{figure}

When dealing with monodisperse configurations in $2d$, CALiPPSO constructs a solution that is mathematically valid (i.e., $(\opt{\va{s}}, \opt{\lpf})$ optimize the jamming LOP, and $\opt{\uvg{\lambda}}$ fulfils the force balance conditions) but \emph{physically unstable}. The reason is that some rattlers have two parallel force bearing contacts. Hence, such particles could move in the direction perpendicular to the line defined by the contacts, without affecting any mechanical constraint. Moreover, force bearing rattlers are part of the backbone of the system and, if they were removed, mechanical equilibrium would be broken across the system. We observed that such instabilities are invariably accompanied by the formation of large crystalline domains. These features are exemplified in Fig.~\ref{fig:2d-mono}.

This atypical case can be understood by first noting that in $2d$ monodisperse systems partial crystallization is practically unavoidable. This is due to the fact that the Euler criterion (applied to planar graphs), and the requirement that the faces of such a graph are regular polygons add extra constraints to the network of contacts in monodisperse disks~\cite{blumenfeldDisorderCriterionExplicit2021,hinrichsenRandomPackingDisks1990}. Now, in a region with partial crystallization, the average coordination number is usually larger than the isostaticity requirement, $z=2d- \order{1/N}$. For instance, in a triangular lattice $z=6$. However, as we have showed in Sec.~\ref{sec:properties calippso packings}, the CALiPPSO algorithm always generates isostatic packings with respect to the stable particles. To reconcile these two opposing conditions, in $2d$ monodisperse systems \CAL produces packings where some rattlers exert finite forces upon stable particles. From the point of view of linear optimization, the excess of active constraints causes the jamming dual LOP \eqref{eqs:dual jamming lop} to have degenerate solutions~\cite{luenbergerLinearNonlinearProgramming2016}.

Formally, such configurations are hyperstatic since there are more contacts than the number of degrees of freedom. Nevertheless, if the contacts exerted by rattlers on stable particles are taken as “external” forces acting individually on such stable particles, isostaticity is recovered. In other words, if $N_c$ exclusively counts contacts between stable-stable and rattler-stable particles, and all rattlers (even the ones with force bearing contacts) are excluded from $N_{dof}$, then the isostatic condition $N_c=N_{dof}$ is once again verified.  We stress that, even in this uncommon scenario, the force balance condition is always satisfied for both rattlers and stable particles. 

From these considerations, we argue that whenever the solution to the jamming LOP leads to finite forces between rattlers and stable particles, the resultant system must be hypterstatic. Such a hyperstaticity in \CAL packings signals the presence of some structural ordering.

\section{Further characterization of \CAL jammed packings} \label{app:further-characterization}

\begin{figure}
	\includegraphics[width=\linewidth]{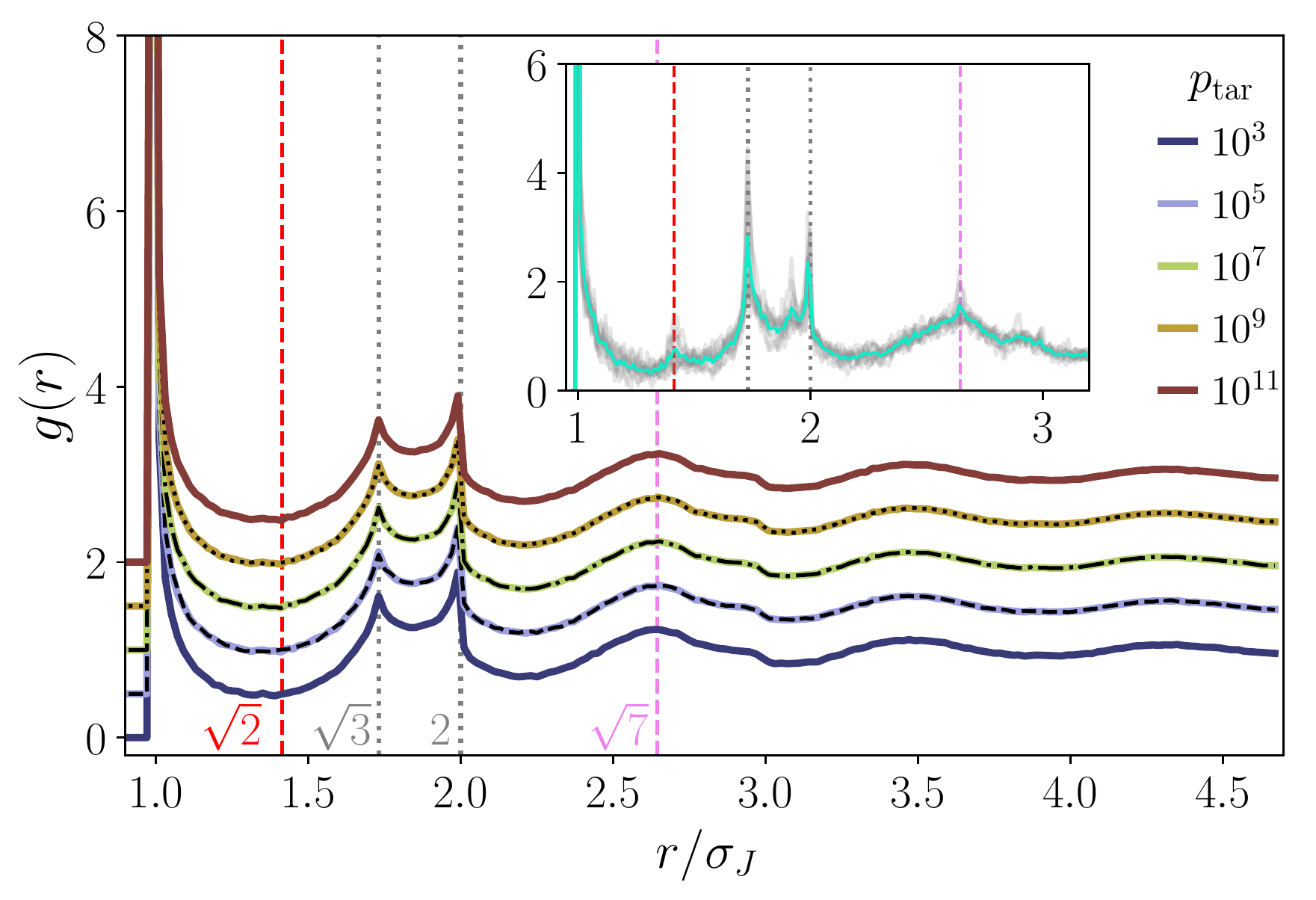}
	\caption{Radial distribution function of jammed packings obtained initializing \CAL from different target pressures (displaced vertically for clarity), as indicated in the legend, using $\kappa=3\times10^{-4}$ for the LS compression. Each curve is the average over $M=20$ configurations of $N=1024$ spheres. Changing the compression rate does not alter $g(r)$, as illustrated by the black lines corresponding to $\kappa=10^{-4}$ (dashed),  $\kappa=3\times10^{-5}$ (dash-dotted), and $\kappa=10^{-5}$ (dotted). 
	Inset: We show, for comparison, $g(r)$ for 10 jammed configurations with partial crystallization (gray lines), and their average (cyan curve). Such kind of configurations have been excluded from all the results presented in this study. The peaks at $\sqrt{2}$ and $\sqrt{7}$ (vertical dashed lines) are not present in the amorphous packings, confirming the absence of crystallization. Besides, the ones at $\sqrt{3}$ and $2$ are considerably smaller and broadened.
	}
	\label{fig:rdf-vs-p}
\end{figure}

First, we show that our LS+\CAL protocol produces packings without any crystallization. To do so, we compute the radial distribution function $g(r)$ of the jammed packings with $N=1024$ monodisperse particles in $3d$. The curves we obtained initializing \CAL from different values of $\ptar$ are reported in Fig.~\ref{fig:rdf-vs-p} and are fully consistent with the analogous results obtained with other methods~\cite{md-code,rissoneLongRangeAnomalousDecay2021}. In particular, the absence of a peak at $r=\sqrt{2}\sigma_J$ indicates that no crystalline order is present in our packings. To exemplify that even a small degree of crystallization yields a noticeable peak at such location, in the inset of the same figure, we plot $g(r)$ of 10 slightly denser configurations, obtained using a smaller $\kappa^{(0)}$ during the LS compression which have partially crystallized. The densest of these 10 extra configurations has a $\vp_J$ that is only $5\%$ larger than the average of the packings without crystallization. In addition, we see that the very sharp peaks at $r=\sqrt{3}\sigma_J$ and $r=2\sigma_J$ broadens considerably in the disordered configurations, while the one at $r=\sqrt{7}\sigma_J$ disappears and only the underlying shoulder remains.

\begin{figure}
	\includegraphics[width=\linewidth]{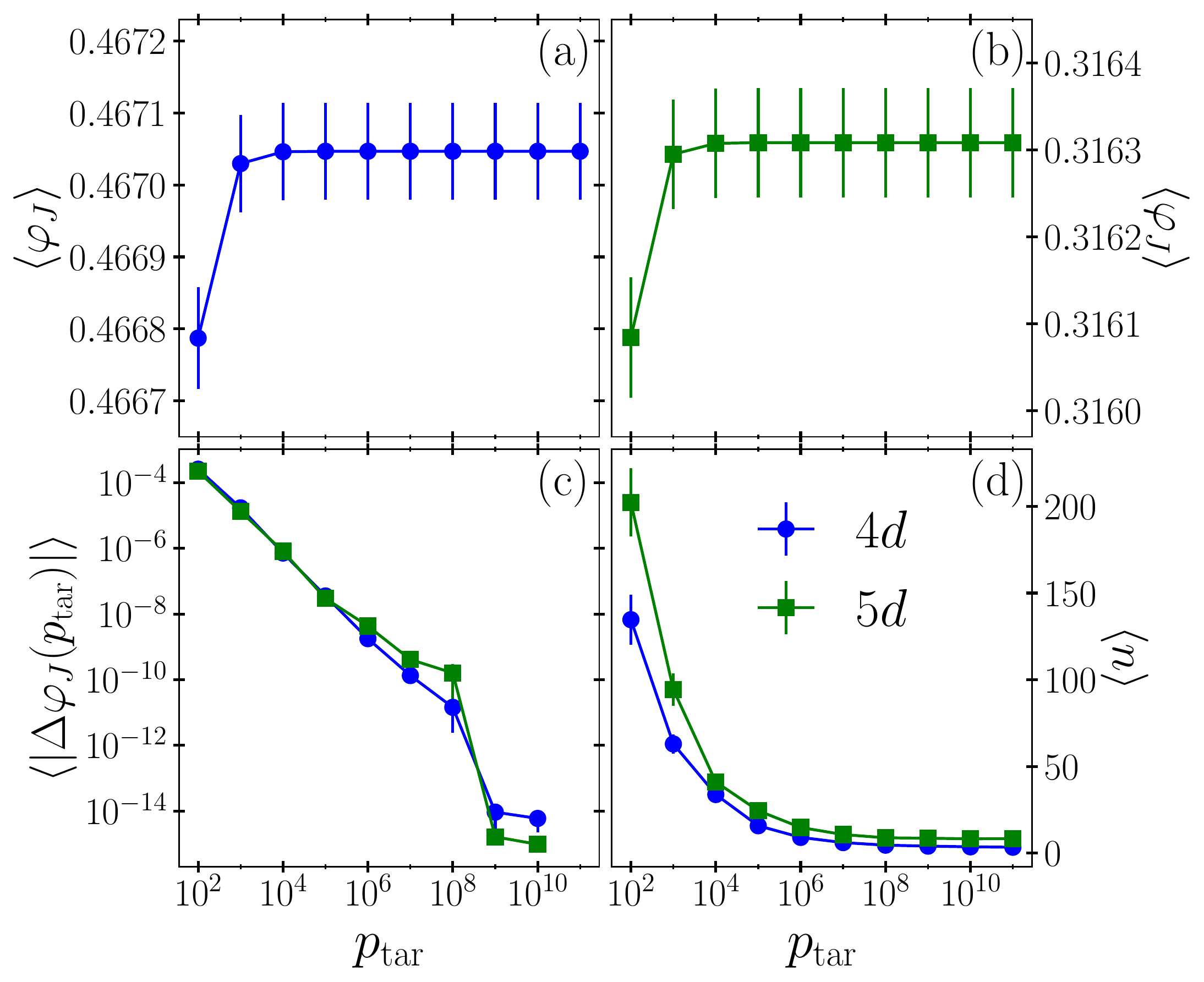}
	\caption{Results with the LS+\CAL protocol in higher dimensions: $d=4$ (blue circles) and $d=5$ (green squares). In analogy with Fig.~\ref{fig:dependence-ptar-kappa}, we explore the effect of $\ptar$ on the jamming packing fraction ($d=4$ in panel (a) and $d=5$ in panel (b)), as well as $\Delta \vp_J(\ptar)$ (panel (c)), and the number of linear optimizations, $n$ (panel (d)). The values reported correspond to the average over 20 samples together with the corresponding standard error.}
	\label{fig:results-4d-5d}
\end{figure}

Finally, we characterize the jammed packings obtained via the LS+\CAL protocol in $d>3$. In particular, we analyse monodisperse configurations of $N=1024$ particles in $d=4$ and $d=5$, for $\kappa=10^{-5}$. As mentioned at the beginning of Sec.~\ref{sec:MD+calippso}, when producing these packings, the fast compression with $\kappa^{(0)}$ (point~\ref{lsilp:fast-compression} of our LS+\CAL protocol) has been omitted. In analogy with the analysis of Sec.~\ref{sec:LS-parameters}, we compute $\vp_J$, $\Delta \vp_J(\ptar)$, and $n$ when \CAL is initialized from different values of the target pressure. In Fig.~\ref{fig:results-4d-5d}, we report the average over 20 samples of these quantities, which agree with the behavior found in $d=3$ systems and the hierarchical FEL structure we described above. We verified that all the obtained configurations at jamming are isostatic and in mechanical equilibrium. These results demonstrate that \CAL can readily produce valid jammed packings in higher dimensions.

\section{Further details on the LS compression protocol}\label{app:further-md}

\begin{figure}
	\includegraphics[width=\linewidth]{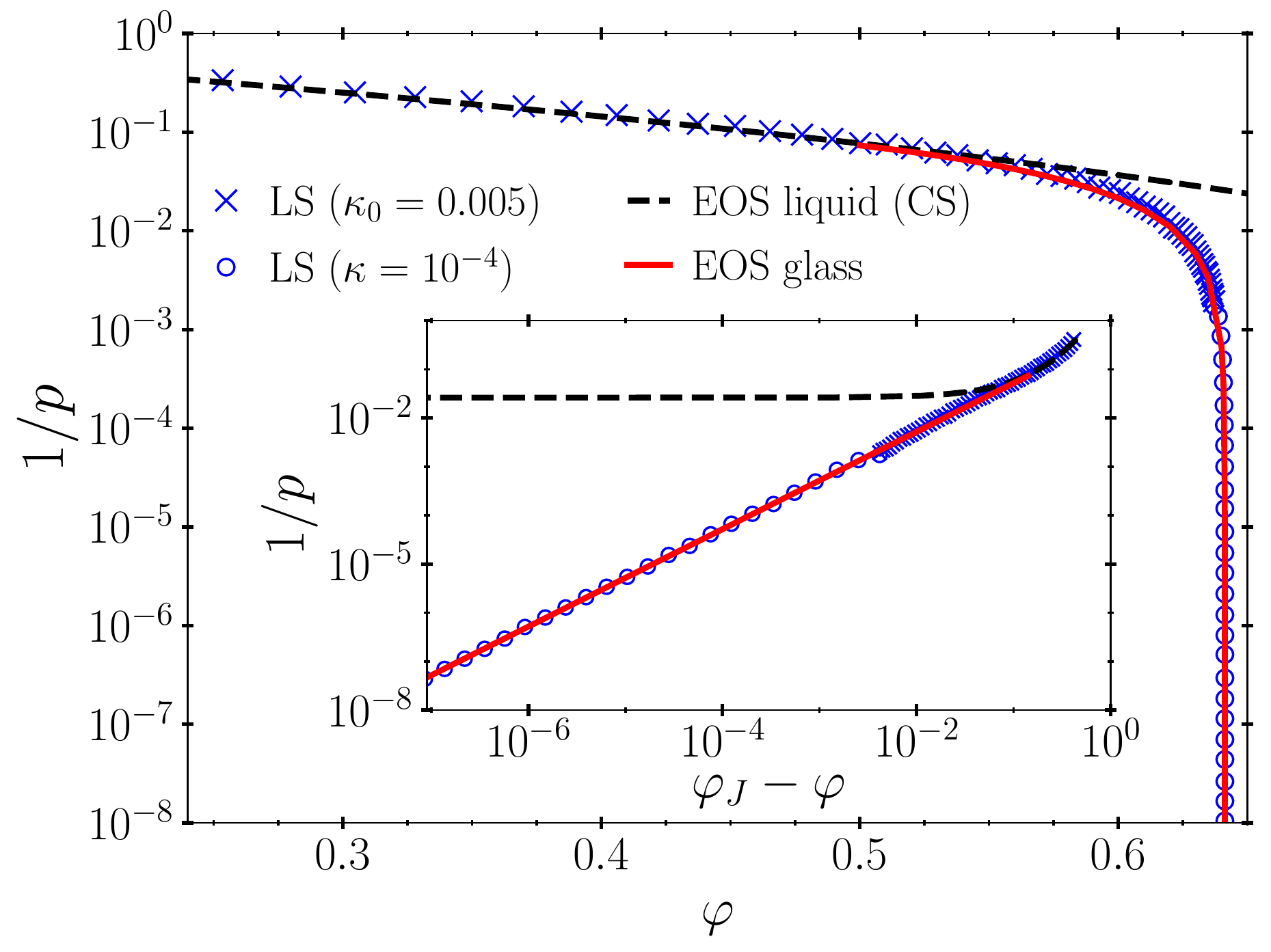}
	\caption{Comparison of the pressure obtained during the LS compression (symbols) with the Carnahan--Starling equation of state \cite{santosStructuralThermodynamicProperties2020} for the liquid (black, dashed line), and Eq.~\eqref{eq:hs-glass-eos} for the glass phase (solid, red line). The fast compression is represented by the blue crosses, while the circles correspond to the slower compression in the glass phase. The inset shows that the pressure diverges as predicted by Eq.~\eqref{eq:hs-glass-eos} in the main text.}
	\label{fig:hs-eos-liquid-and-glass}
\end{figure}

In Fig.~\ref{fig:hs-eos-liquid-and-glass}, we exemplify the LS compression part of our LS+\CAL algorithm (i.e., the first three steps described at the beginning of Sec.~\ref{sec:MD+calippso}). From this Figure it is clear that using the fast compression the liquid (crosses) undergoes a smooth transition to a glass (circles); thus, there are no signatures of the crystalline phase. On the other hand, it should be noted that having a finite $\kappa$, the LS fails to equilibrate the liquid all the way up to the glass transition density~\cite{santosStructuralThermodynamicProperties2020,parisiMeanfieldTheoryHard2010}, \(\vp \simeq 0.58\). Yet, the glass phase is well described by the free-volume equation of state, as shown by the excellent agreement between the numerical data and the plot of Eq.~\eqref{eq:hs-glass-eos} (solid red line). As mentioned above, this is not the true thermodynamic equation of state. The inset of Fig.~\ref{fig:hs-eos-liquid-and-glass} confirms the divergent behavior of $p$ as $\vp \to \vp_J$. In short, even if the LS is a rather quick compression protocol, it efficiently produces glassy configurations. Naturally, a more complex compression method can be used in order to attain better thermalized systems before initializing the \CAL crunching part. However, as we argued in Sec.~\ref{sec:LS-parameters}, this protocol suffices to probe the most salient features of jammed packings.

\section{Comparing \CAL to MD based algorithms}\label{app:calippso-vs-md}

\begin{figure}[t]
	\includegraphics[width=\linewidth]{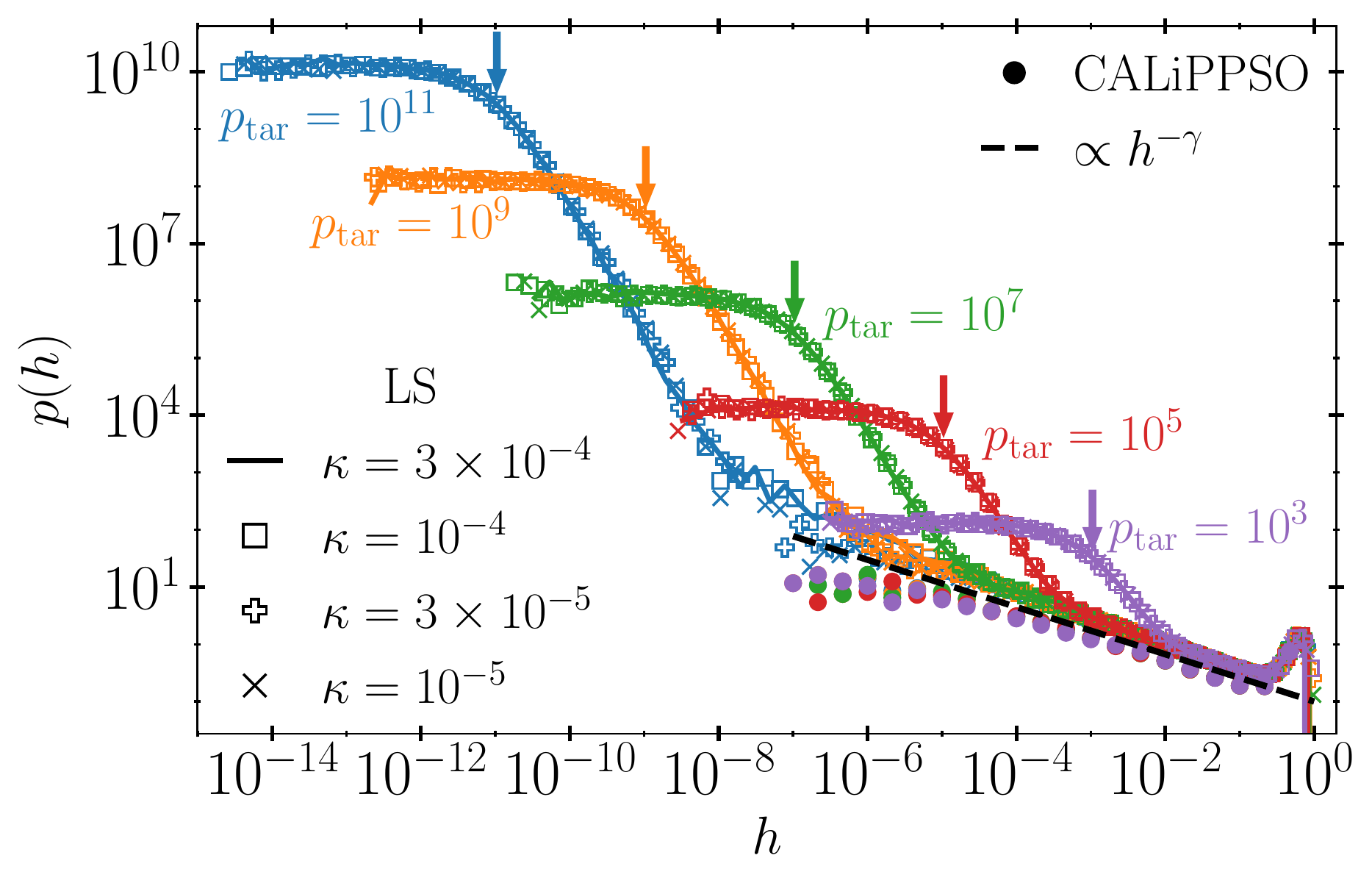}
	\caption{Probability distribution function (pdf) of the inter-particle gaps, $h$, after terminating the LS compression at different target pressures and with different compression rates (as indicated by colors and legend in the lower left). We show also the analogous pdf's obtained with \CAL for the same values of $\ptar$, with configurations compressed using $\kappa=3\times10^{-5}$ (other values yield virtually identical distributions). These curves have been displayed downwards by a small fraction for clarity. For each value of $\kappa$ and $\ptar$, we report the distributions obtained averaging over 20 configurations of $N=1024$ particles. Down arrows indicate the intersection of the distributions with $h=1/\ptar$ and correspond to the beginning of the plateau. Thus, for a given value of $\ptar$,  LS “contacts” can be identified as the gaps smaller than $1/\ptar$. We observe the presence of the regime $p(h)\sim h^{-\gamma}$ (black, dashed line) which is in good agreement with the distribution of gaps predicted by the mean-field theory. As we argue in the main text, since intermediate values of $h$ cannot be associated unequivocally to either of these two classes, using exclusively LS-type algorithms we cannot recover the detailed structure of jammed packings.}
	\label{fig:gaps-after-md}
\end{figure}

\begin{figure}[t]
	\includegraphics[width=\linewidth]{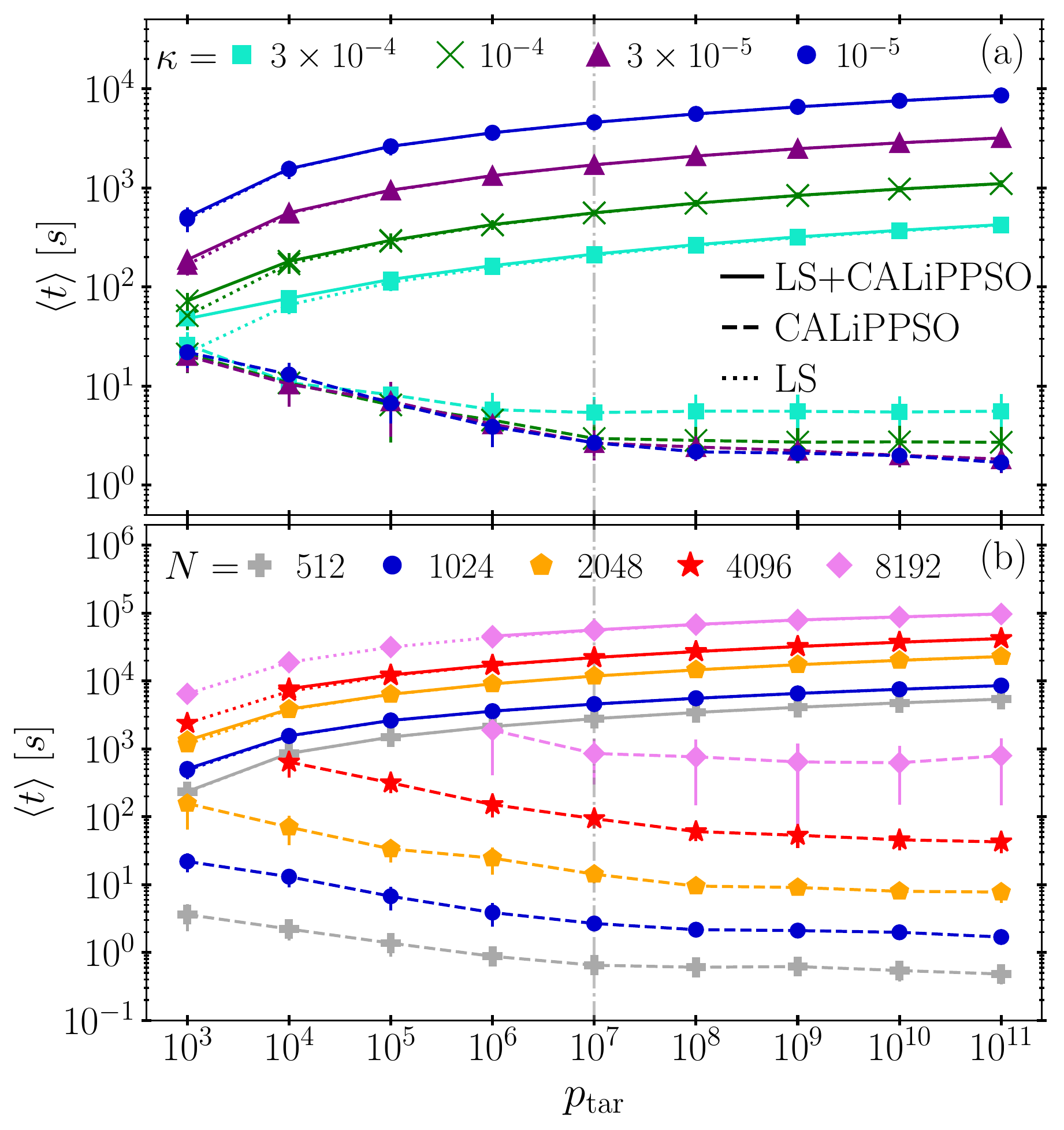}
	\caption{Running time of LS (dotted lines) and \CAL algorithms (dashed), and their sum (solid), as a function of $\ptar$. We explore the effect of using different compression rates with fixed $N=1024$ (upper panel), as well as changing the system size but a constant $\kappa=10^{-5}$ (lower one), as indicated in the legends. For the analysis of Sec.~\ref{sec:complexity calippso}, we used $\ptar=10^7$ (highlighted by the vertical line).}
	\label{fig:times LS and ILP}
\end{figure}

In the main text, we mentioned that using only the LS compression it is impossible to extract the full structural information of a jammed state. To support this claim, in Fig.~\ref{fig:gaps-after-md} we present the probability distribution function (pdf) of the interparticle gaps, defined as $h_{ij} = \frac{r_{ij}}{\sigma_{ij}} -1$, obtained at different values of $\ptar$. That is, the gaps' pdf once the MD simulations have reached $\ptar$, but without using CALiPPSO.
For a given value of target pressure, the associated pdf has three different regimes: (i)~for very small values of the gaps, \(h \lesssim 1/\ptar\), a plateau forms, whose height increases (linearly) with $\ptar$; (ii)~for $1/\ptar \ll h$, the pdf follows the scaling predicted by mean-field theory~\cite{charbonneauGlassJammingTransitions2017,puz_TheorySimpleGlasses2020,charbonneauFractalFreeEnergy2014}, $p(h)\sim h^{-\gamma}$, with $\gamma=0.4163$ (dashed, black line); (iii)~for  $ 1/\ptar \lesssim h $ a third, intermediate regime sets in which cannot be identified with any well-known property of the configurations. The first two regimes can be identified, respectively, as the nascent contact-singularity and the critical gap distribution of the radial distribution function characteristic of spheres packings~\cite{md-code,donevPairCorrelationFunction2005,santosStructuralThermodynamicProperties2020}.
We verified that if only the gaps smaller than $1/\ptar$ are counted as contacts, the resulting system is highly hypostatic. Notably, these features are independent of the value of $\kappa$ employed.

Fig.~\ref{fig:gaps-after-md} illustrates also the gap distributions obtained at convergence of \CAL (circular markers), using the LS configurations as seeds. 
Notice that using CALiPPSO, $p(h)$ is rather insensitive to the value of $\ptar$ from which a configuration is crunched. Specifically, we also recover the mean-field prediction, but in a consistent range spanning several decades, independently of $\ptar$. Moreover, deviations from $p(h) \sim h^{-\gamma}$ observed in the left tail of $p(h)$ can be ascribed to finite-size effects~\cite{charbonneauFinitesizeEffectsMicroscopic2021}. The plateau proportional to $\ptar$ present in the LS distribution, completely disappears after the crunching: it becomes $\delta(h)$, i.e., particles are in contact. Thus, the intermediate regime (iii) for $ 1/\ptar \lesssim h $ disappears as well because using \CAL all the interparticle distances can be classified either as force bearing contacts, or as true gaps.

To better understand the difference between the results of our approach and highly compressed configurations obtained from MD simulations, let us assume a fixed value of $\ptar$. From the discussion of the previous paragraphs, by comparing $p(h)$ obtained within LS with the corresponding distribution within CALiPPSO, we can conclude that a significant fraction of gaps in the intermediate domain of the LS distribution will form valid contacts once $\vp_J$ is reached. Yet, for any $\vp<\vp_J$, there is no clear criterion for distinguishing this type of “pre-contact” gaps from true ones (i.e., those that remain finite at $\vp_J$). To overcome this difficulty, previous studies have computed a time-averaged network of contacts. In a nutshell, this technique identifies the contacts of a particle by averaging the collisions it undergoes with its neighbors over a sufficiently large time window. Naturally, the plateau of $h<1/\ptar$ is a consistent contribution to such average because such very small gaps constitute a rather constant background of collisions associated with potential contacts. The remaining contacts needed to achieve stability are therefore obtained from the collisions with particles whose typical distance lies in the intermediate regime. Instead, as we have shown here, identifying forces with strictly positive Lagrange multipliers of the final jamming LOP offers a clear-cut distinction between gaps and contacts. This makes CALiPPSO a more precise packing algorithm than dynamical or time-average approaches coming from the near jamming regime. We emphasize that precisely identifying real contacts and gaps is of utmost importance because of the marginal stability of jammed packings and the long-range correlations of their networks of contacts.

To conclude, we compare (i) the time required by LS to reach a given $\ptar$, (ii) the time \CAL takes to successively jam the configuration from the target pressure, and (iii) their sum, i.e., the total time of our LS+\CAL method. For the first one, it is clear that the larger the desired $\ptar$ the longer the LS protocol takes (dotted lines in Fig.~\ref{fig:times LS and ILP}). Besides, intuition suggests the duration of the \CAL crunching process is reduced the larger $\ptar$ gets, leading to a trade-off of the optimal total time of LS+CALiPPSO. However, from the results of Figs.~\ref{fig:dependence-ptar-kappa}(g-h), we know that if $\ptar >\pth$ this is not necessarily the case. Indeed, those figures show that $\avg{n}$ remains essentially unchanged in this very high-pressure regime. Hence, it is expected that the convergence time of \CAL remains roughly constant, as confirmed in Fig.~\ref{fig:times LS and ILP} (dashed lines). Thus, at least for moderately large $N$, increasing $\ptar$ is actually detrimental to the performance of LS+CALiPPSO. Recall however that in order to properly sample the free energy landscape (see Sec.~\ref{sec:LS-parameters}) a large target pressure, $\pth > \ptar \gg 1$ is useful. Therefore, even if choosing a very high value of $\ptar$ leads to a longer LS compression, it should be favored in order to better model the thermodynamic route to the jamming line. This is the reason why we fixed $\ptar=10^{7}$ in Sec.~\ref{sec:complexity calippso} to explore the \CAL algorithmic complexity.

From a more practical point of view, if many independent configurations have to be compressed simultaneously, a long LS compression does not necessarily hinders the performance of LS+CALiPPSO, provided $\ptar<\pth$. That is, given that LS~\cite{md-code} relies on event-driven MD that can be efficiently implemented using a serial algorithm, as many configurations as available threads can be compressed. However, some optimizers, as Gurobi~\cite{gurobi}, benefit from being executed with parallelized algorithms, which hinders the possibility of concurrently executing \CAL on several systems. Therefore, when selecting $\ptar$ to optimize the total time of LS+\CAL these two different behaviors should be considered.

As a final remark, we mention that a relatively large $\kappa$ can produce a non-monotonic behavior of time of the full LS+\CAL protocol. In this case, a longer LS compression to reach a larger pressure is convenient up to a certain $\ptar$, that depends on $\kappa$. Beyond such pressure, however, further increasing the target pressure leads to a larger time because the \CAL crunching is not substantially accelerated. In other words, the small speed gain of \CAL does not compensate for the longer LS compression.

%%%%%%%%%%%%%%%%%%%% REFERENCES %%%%%%%%%%%%%%%%%%%%
% \bibliographystyle{abbrv}
\bibliography{refs-ILP.bib}

\begin{thebibliography}{97}%
\makeatletter
\providecommand \@ifxundefined [1]{%
 \@ifx{#1\undefined}
}%
\providecommand \@ifnum [1]{%
 \ifnum #1\expandafter \@firstoftwo
 \else \expandafter \@secondoftwo
 \fi
}%
\providecommand \@ifx [1]{%
 \ifx #1\expandafter \@firstoftwo
 \else \expandafter \@secondoftwo
 \fi
}%
\providecommand \natexlab [1]{#1}%
\providecommand \enquote  [1]{``#1''}%
\providecommand \bibnamefont  [1]{#1}%
\providecommand \bibfnamefont [1]{#1}%
\providecommand \citenamefont [1]{#1}%
\providecommand \href@noop [0]{\@secondoftwo}%
\providecommand \href [0]{\begingroup \@sanitize@url \@href}%
\providecommand \@href[1]{\@@startlink{#1}\@@href}%
\providecommand \@@href[1]{\endgroup#1\@@endlink}%
\providecommand \@sanitize@url [0]{\catcode `\\12\catcode `\$12\catcode
  `\&12\catcode `\#12\catcode `\^12\catcode `\_12\catcode `\%12\relax}%
\providecommand \@@startlink[1]{}%
\providecommand \@@endlink[0]{}%
\providecommand \url  [0]{\begingroup\@sanitize@url \@url }%
\providecommand \@url [1]{\endgroup\@href {#1}{\urlprefix }}%
\providecommand \urlprefix  [0]{URL }%
\providecommand \Eprint [0]{\href }%
\providecommand \doibase [0]{https://doi.org/}%
\providecommand \selectlanguage [0]{\@gobble}%
\providecommand \bibinfo  [0]{\@secondoftwo}%
\providecommand \bibfield  [0]{\@secondoftwo}%
\providecommand \translation [1]{[#1]}%
\providecommand \BibitemOpen [0]{}%
\providecommand \bibitemStop [0]{}%
\providecommand \bibitemNoStop [0]{.\EOS\space}%
\providecommand \EOS [0]{\spacefactor3000\relax}%
\providecommand \BibitemShut  [1]{\csname bibitem#1\endcsname}%
\let\auto@bib@innerbib\@empty
%</preamble>
\bibitem [{\citenamefont {Liu}\ and\ \citenamefont
  {Nagel}(1998)}]{liuNonlinearDynamicsJamming1998}%
  \BibitemOpen
  \bibfield  {author} {\bibinfo {author} {\bibfnamefont {A.~J.}\ \bibnamefont
  {Liu}}\ and\ \bibinfo {author} {\bibfnamefont {S.~R.}\ \bibnamefont
  {Nagel}},\ }\bibfield  {title} {\bibinfo {title} {Nonlinear dynamics: Jamming
  is not just cool any more},\ }\href {https://doi.org/10.1038/23819}
  {\bibfield  {journal} {\bibinfo  {journal} {Nature}\ }\textbf {\bibinfo
  {volume} {396}},\ \bibinfo {pages} {21--22} (\bibinfo {year}
  {1998})}\BibitemShut {NoStop}%
\bibitem [{\citenamefont {Liu}\ and\ \citenamefont
  {Nagel}(2010)}]{liuJammingTransitionMarginally2010}%
  \BibitemOpen
  \bibfield  {author} {\bibinfo {author} {\bibfnamefont {A.~J.}\ \bibnamefont
  {Liu}}\ and\ \bibinfo {author} {\bibfnamefont {S.~R.}\ \bibnamefont
  {Nagel}},\ }\bibfield  {title} {\bibinfo {title} {The {{Jamming Transition}}
  and the {{Marginally Jammed Solid}}},\ }\href
  {https://doi.org/10.1146/annurev-conmatphys-070909-104045} {\bibfield
  {journal} {\bibinfo  {journal} {Annual Review of Condensed Matter Physics}\
  }\textbf {\bibinfo {volume} {1}},\ \bibinfo {pages} {347--369} (\bibinfo
  {year} {2010})}\BibitemShut {NoStop}%
\bibitem [{\citenamefont {Torquato}\ and\ \citenamefont
  {Stillinger}(2010)}]{torquatoReviewJammedHardparticlePackings2010}%
  \BibitemOpen
  \bibfield  {author} {\bibinfo {author} {\bibfnamefont {S.}~\bibnamefont
  {Torquato}}\ and\ \bibinfo {author} {\bibfnamefont {F.~H.}\ \bibnamefont
  {Stillinger}},\ }\bibfield  {title} {\bibinfo {title} {Jammed hard-particle
  packings: {{From Kepler}} to {{Bernal}} and beyond},\ }\href
  {https://doi.org/10.1103/RevModPhys.82.2633} {\bibfield  {journal} {\bibinfo
  {journal} {Reviews of Modern Physics}\ }\textbf {\bibinfo {volume} {82}},\
  \bibinfo {pages} {2633--2672} (\bibinfo {year} {2010})}\BibitemShut {NoStop}%
\bibitem [{\citenamefont {{van
  Hecke}}(2010)}]{vanheckeReviewJammingSoftParticles2010}%
  \BibitemOpen
  \bibfield  {author} {\bibinfo {author} {\bibfnamefont {M.}~\bibnamefont {{van
  Hecke}}},\ }\bibfield  {title} {\bibinfo {title} {Jamming of soft particles:
  Geometry, mechanics, scaling and isostaticity},\ }\href
  {https://doi.org/10.1088/0953-8984/22/3/033101} {\bibfield  {journal}
  {\bibinfo  {journal} {Journal of Physics: Condensed Matter}\ }\textbf
  {\bibinfo {volume} {22}},\ \bibinfo {pages} {033101} (\bibinfo {year}
  {2010})}\BibitemShut {NoStop}%
\bibitem [{\citenamefont {O'Hern}\ \emph {et~al.}(2003)\citenamefont {O'Hern},
  \citenamefont {Silbert}, \citenamefont {Liu},\ and\ \citenamefont
  {Nagel}}]{ohernJammingZeroTemperature2003}%
  \BibitemOpen
  \bibfield  {author} {\bibinfo {author} {\bibfnamefont {C.~S.}\ \bibnamefont
  {O'Hern}}, \bibinfo {author} {\bibfnamefont {L.~E.}\ \bibnamefont {Silbert}},
  \bibinfo {author} {\bibfnamefont {A.~J.}\ \bibnamefont {Liu}},\ and\ \bibinfo
  {author} {\bibfnamefont {S.~R.}\ \bibnamefont {Nagel}},\ }\bibfield  {title}
  {\bibinfo {title} {Jamming at zero temperature and zero applied stress:
  {{The}} epitome of disorder},\ }\href
  {https://doi.org/10.1103/PhysRevE.68.011306} {\bibfield  {journal} {\bibinfo
  {journal} {Physical Review E}\ }\textbf {\bibinfo {volume} {68}},\ \bibinfo
  {pages} {011306} (\bibinfo {year} {2003})}\BibitemShut {NoStop}%
\bibitem [{\citenamefont {Charbonneau}\ \emph
  {et~al.}(2014{\natexlab{a}})\citenamefont {Charbonneau}, \citenamefont
  {Kurchan}, \citenamefont {Parisi}, \citenamefont {Urbani},\ and\
  \citenamefont {Zamponi}}]{charbonneauFractalFreeEnergy2014}%
  \BibitemOpen
  \bibfield  {author} {\bibinfo {author} {\bibfnamefont {P.}~\bibnamefont
  {Charbonneau}}, \bibinfo {author} {\bibfnamefont {J.}~\bibnamefont
  {Kurchan}}, \bibinfo {author} {\bibfnamefont {G.}~\bibnamefont {Parisi}},
  \bibinfo {author} {\bibfnamefont {P.}~\bibnamefont {Urbani}},\ and\ \bibinfo
  {author} {\bibfnamefont {F.}~\bibnamefont {Zamponi}},\ }\bibfield  {title}
  {\bibinfo {title} {Fractal free energy landscapes in structural glasses},\
  }\href {https://doi.org/10.1038/ncomms4725} {\bibfield  {journal} {\bibinfo
  {journal} {Nature Communications}\ }\textbf {\bibinfo {volume} {5}},\
  \bibinfo {pages} {3725} (\bibinfo {year} {2014}{\natexlab{a}})}\BibitemShut
  {NoStop}%
\bibitem [{\citenamefont {Baule}\ \emph {et~al.}(2018)\citenamefont {Baule},
  \citenamefont {Morone}, \citenamefont {Herrmann},\ and\ \citenamefont
  {Makse}}]{bauleReviewEdwardsStatisticalMechanics2018a}%
  \BibitemOpen
  \bibfield  {author} {\bibinfo {author} {\bibfnamefont {A.}~\bibnamefont
  {Baule}}, \bibinfo {author} {\bibfnamefont {F.}~\bibnamefont {Morone}},
  \bibinfo {author} {\bibfnamefont {H.~J.}\ \bibnamefont {Herrmann}},\ and\
  \bibinfo {author} {\bibfnamefont {H.~A.}\ \bibnamefont {Makse}},\ }\bibfield
  {title} {\bibinfo {title} {Edwards statistical mechanics for jammed granular
  matter},\ }\href {https://doi.org/10.1103/RevModPhys.90.015006} {\bibfield
  {journal} {\bibinfo  {journal} {Reviews of Modern Physics}\ }\textbf
  {\bibinfo {volume} {90}},\ \bibinfo {pages} {015006} (\bibinfo {year}
  {2018})}\BibitemShut {NoStop}%
\bibitem [{\citenamefont {Tighe}\ \emph {et~al.}(2010)\citenamefont {Tighe},
  \citenamefont {Snoeijer}, \citenamefont {Vlugt},\ and\ \citenamefont {van
  Hecke}}]{tigheForceNetworkEnsemble2010}%
  \BibitemOpen
  \bibfield  {author} {\bibinfo {author} {\bibfnamefont {B.~P.}\ \bibnamefont
  {Tighe}}, \bibinfo {author} {\bibfnamefont {J.~H.}\ \bibnamefont {Snoeijer}},
  \bibinfo {author} {\bibfnamefont {T.~J.~H.}\ \bibnamefont {Vlugt}},\ and\
  \bibinfo {author} {\bibfnamefont {M.}~\bibnamefont {van Hecke}},\ }\bibfield
  {title} {\bibinfo {title} {The force network ensemble for granular
  packings},\ }\href {https://doi.org/10.1039/B926592A} {\bibfield  {journal}
  {\bibinfo  {journal} {Soft Matter}\ }\textbf {\bibinfo {volume} {6}},\
  \bibinfo {pages} {2908--2917} (\bibinfo {year} {2010})}\BibitemShut {NoStop}%
\bibitem [{\citenamefont {DeGiuli}\ \emph {et~al.}(2015)\citenamefont
  {DeGiuli}, \citenamefont {Lerner},\ and\ \citenamefont
  {Wyart}}]{degiuliTheoryJammingTransition2015}%
  \BibitemOpen
  \bibfield  {author} {\bibinfo {author} {\bibfnamefont {E.}~\bibnamefont
  {DeGiuli}}, \bibinfo {author} {\bibfnamefont {E.}~\bibnamefont {Lerner}},\
  and\ \bibinfo {author} {\bibfnamefont {M.}~\bibnamefont {Wyart}},\ }\bibfield
   {title} {\bibinfo {title} {Theory of the jamming transition at finite
  temperature},\ }\href {https://doi.org/10.1063/1.4918737} {\bibfield
  {journal} {\bibinfo  {journal} {The Journal of Chemical Physics}\ }\textbf
  {\bibinfo {volume} {142}},\ \bibinfo {pages} {164503} (\bibinfo {year}
  {2015})}\BibitemShut {NoStop}%
\bibitem [{\citenamefont {Parisi}\ and\ \citenamefont
  {Zamponi}(2010)}]{parisiMeanfieldTheoryHard2010}%
  \BibitemOpen
  \bibfield  {author} {\bibinfo {author} {\bibfnamefont {G.}~\bibnamefont
  {Parisi}}\ and\ \bibinfo {author} {\bibfnamefont {F.}~\bibnamefont
  {Zamponi}},\ }\bibfield  {title} {\bibinfo {title} {Mean-field theory of hard
  sphere glasses and jamming},\ }\href
  {https://doi.org/10.1103/RevModPhys.82.789} {\bibfield  {journal} {\bibinfo
  {journal} {Reviews of Modern Physics}\ }\textbf {\bibinfo {volume} {82}},\
  \bibinfo {pages} {789--845} (\bibinfo {year} {2010})}\BibitemShut {NoStop}%
\bibitem [{\citenamefont {Charbonneau}\ \emph
  {et~al.}(2014{\natexlab{b}})\citenamefont {Charbonneau}, \citenamefont
  {Kurchan}, \citenamefont {Parisi}, \citenamefont {Urbani},\ and\
  \citenamefont {Zamponi}}]{charbonneauExactTheoryDense2014}%
  \BibitemOpen
  \bibfield  {author} {\bibinfo {author} {\bibfnamefont {P.}~\bibnamefont
  {Charbonneau}}, \bibinfo {author} {\bibfnamefont {J.}~\bibnamefont
  {Kurchan}}, \bibinfo {author} {\bibfnamefont {G.}~\bibnamefont {Parisi}},
  \bibinfo {author} {\bibfnamefont {P.}~\bibnamefont {Urbani}},\ and\ \bibinfo
  {author} {\bibfnamefont {F.}~\bibnamefont {Zamponi}},\ }\bibfield  {title}
  {\bibinfo {title} {Exact theory of dense amorphous hard spheres in high
  dimension. {{III}}. {{The}} full replica symmetry breaking solution},\ }\href
  {https://doi.org/10.1088/1742-5468/2014/10/P10009} {\bibfield  {journal}
  {\bibinfo  {journal} {Journal of Statistical Mechanics: Theory and
  Experiment}\ }\textbf {\bibinfo {volume} {2014}},\ \bibinfo {pages} {P10009}
  (\bibinfo {year} {2014}{\natexlab{b}})}\BibitemShut {NoStop}%
\bibitem [{\citenamefont {Charbonneau}\ \emph {et~al.}(2017)\citenamefont
  {Charbonneau}, \citenamefont {Kurchan}, \citenamefont {Parisi}, \citenamefont
  {Urbani},\ and\ \citenamefont
  {Zamponi}}]{charbonneauGlassJammingTransitions2017}%
  \BibitemOpen
  \bibfield  {author} {\bibinfo {author} {\bibfnamefont {P.}~\bibnamefont
  {Charbonneau}}, \bibinfo {author} {\bibfnamefont {J.}~\bibnamefont
  {Kurchan}}, \bibinfo {author} {\bibfnamefont {G.}~\bibnamefont {Parisi}},
  \bibinfo {author} {\bibfnamefont {P.}~\bibnamefont {Urbani}},\ and\ \bibinfo
  {author} {\bibfnamefont {F.}~\bibnamefont {Zamponi}},\ }\bibfield  {title}
  {\bibinfo {title} {Glass and {{Jamming Transitions}}: {{From Exact Results}}
  to {{Finite}}-{{Dimensional Descriptions}}},\ }\href
  {https://doi.org/10.1146/annurev-conmatphys-031016-025334} {\bibfield
  {journal} {\bibinfo  {journal} {Annual Review of Condensed Matter Physics}\
  }\textbf {\bibinfo {volume} {8}},\ \bibinfo {pages} {265--288} (\bibinfo
  {year} {2017})}\BibitemShut {NoStop}%
\bibitem [{\citenamefont {Parisi}\ \emph {et~al.}(2020)\citenamefont {Parisi},
  \citenamefont {Urbani},\ and\ \citenamefont
  {Zamponi}}]{puz_TheorySimpleGlasses2020}%
  \BibitemOpen
  \bibfield  {author} {\bibinfo {author} {\bibfnamefont {G.}~\bibnamefont
  {Parisi}}, \bibinfo {author} {\bibfnamefont {P.}~\bibnamefont {Urbani}},\
  and\ \bibinfo {author} {\bibfnamefont {F.}~\bibnamefont {Zamponi}},\ }\href
  {https://doi.org/10.1017/9781108120494} {\emph {\bibinfo {title} {Theory of
  {{Simple Glasses}}: {{Exact Solutions}} in {{Infinite Dimensions}}}}}\
  (\bibinfo  {publisher} {{Cambridge University Press}},\ \bibinfo {address}
  {{Cambridge}},\ \bibinfo {year} {2020})\BibitemShut {NoStop}%
\bibitem [{\citenamefont {Berthier}\ \emph {et~al.}(2019)\citenamefont
  {Berthier}, \citenamefont {Biroli}, \citenamefont {Charbonneau},
  \citenamefont {Corwin}, \citenamefont {Franz},\ and\ \citenamefont
  {Zamponi}}]{berthierGardnerPhysicsAmorphous2019}%
  \BibitemOpen
  \bibfield  {author} {\bibinfo {author} {\bibfnamefont {L.}~\bibnamefont
  {Berthier}}, \bibinfo {author} {\bibfnamefont {G.}~\bibnamefont {Biroli}},
  \bibinfo {author} {\bibfnamefont {P.}~\bibnamefont {Charbonneau}}, \bibinfo
  {author} {\bibfnamefont {E.~I.}\ \bibnamefont {Corwin}}, \bibinfo {author}
  {\bibfnamefont {S.}~\bibnamefont {Franz}},\ and\ \bibinfo {author}
  {\bibfnamefont {F.}~\bibnamefont {Zamponi}},\ }\bibfield  {title} {\bibinfo
  {title} {Gardner physics in amorphous solids and beyond},\ }\href
  {https://doi.org/10.1063/1.5097175} {\bibfield  {journal} {\bibinfo
  {journal} {The Journal of Chemical Physics}\ }\textbf {\bibinfo {volume}
  {151}},\ \bibinfo {pages} {010901} (\bibinfo {year} {2019})}\BibitemShut
  {NoStop}%
\bibitem [{\citenamefont {Franz}\ and\ \citenamefont
  {Parisi}(2016)}]{franz2016simplest}%
  \BibitemOpen
  \bibfield  {author} {\bibinfo {author} {\bibfnamefont {S.}~\bibnamefont
  {Franz}}\ and\ \bibinfo {author} {\bibfnamefont {G.}~\bibnamefont {Parisi}},\
  }\bibfield  {title} {\bibinfo {title} {The simplest model of jamming},\
  }\href {https://doi.org/10.1088/1751-8113/49/14/145001} {\bibfield  {journal}
  {\bibinfo  {journal} {Journal of Physics A: Mathematical and Theoretical}\
  }\textbf {\bibinfo {volume} {49}},\ \bibinfo {pages} {145001} (\bibinfo
  {year} {2016})}\BibitemShut {NoStop}%
\bibitem [{\citenamefont {Franz}\ \emph {et~al.}(2017)\citenamefont {Franz},
  \citenamefont {Parisi}, \citenamefont {Sevelev}, \citenamefont {Urbani},\
  and\ \citenamefont {Zamponi}}]{franzUniversalitySATUNSATJamming2017}%
  \BibitemOpen
  \bibfield  {author} {\bibinfo {author} {\bibfnamefont {S.}~\bibnamefont
  {Franz}}, \bibinfo {author} {\bibfnamefont {G.}~\bibnamefont {Parisi}},
  \bibinfo {author} {\bibfnamefont {M.}~\bibnamefont {Sevelev}}, \bibinfo
  {author} {\bibfnamefont {P.}~\bibnamefont {Urbani}},\ and\ \bibinfo {author}
  {\bibfnamefont {F.}~\bibnamefont {Zamponi}},\ }\bibfield  {title} {\bibinfo
  {title} {Universality of the {{SAT}}-{{UNSAT}} (jamming) threshold in
  non-convex continuous constraint satisfaction problems},\ }\href
  {https://doi.org/10.21468/SciPostPhys.2.3.019} {\bibfield  {journal}
  {\bibinfo  {journal} {SciPost Physics}\ }\textbf {\bibinfo {volume} {2}},\
  \bibinfo {pages} {019} (\bibinfo {year} {2017})}\BibitemShut {NoStop}%
\bibitem [{\citenamefont {Geiger}\ \emph {et~al.}(2019)\citenamefont {Geiger},
  \citenamefont {Spigler}, \citenamefont {{d'Ascoli}}, \citenamefont {Sagun},
  \citenamefont {{Baity-Jesi}}, \citenamefont {Biroli},\ and\ \citenamefont
  {Wyart}}]{geigerJammingTransitionParadigm2019}%
  \BibitemOpen
  \bibfield  {author} {\bibinfo {author} {\bibfnamefont {M.}~\bibnamefont
  {Geiger}}, \bibinfo {author} {\bibfnamefont {S.}~\bibnamefont {Spigler}},
  \bibinfo {author} {\bibfnamefont {S.}~\bibnamefont {{d'Ascoli}}}, \bibinfo
  {author} {\bibfnamefont {L.}~\bibnamefont {Sagun}}, \bibinfo {author}
  {\bibfnamefont {M.}~\bibnamefont {{Baity-Jesi}}}, \bibinfo {author}
  {\bibfnamefont {G.}~\bibnamefont {Biroli}},\ and\ \bibinfo {author}
  {\bibfnamefont {M.}~\bibnamefont {Wyart}},\ }\bibfield  {title} {\bibinfo
  {title} {Jamming transition as a paradigm to understand the loss landscape of
  deep neural networks},\ }\href {https://doi.org/10.1103/PhysRevE.100.012115}
  {\bibfield  {journal} {\bibinfo  {journal} {Physical Review E}\ }\textbf
  {\bibinfo {volume} {100}},\ \bibinfo {pages} {012115} (\bibinfo {year}
  {2019})},\ \Eprint {https://arxiv.org/abs/1809.09349} {1809.09349}
  \BibitemShut {NoStop}%
\bibitem [{\citenamefont {Spigler}\ \emph {et~al.}(2019)\citenamefont
  {Spigler}, \citenamefont {Geiger}, \citenamefont {{d'Ascoli}}, \citenamefont
  {Sagun}, \citenamefont {Biroli},\ and\ \citenamefont
  {Wyart}}]{spiglerJammingTransitionOverparametrization2019}%
  \BibitemOpen
  \bibfield  {author} {\bibinfo {author} {\bibfnamefont {S.}~\bibnamefont
  {Spigler}}, \bibinfo {author} {\bibfnamefont {M.}~\bibnamefont {Geiger}},
  \bibinfo {author} {\bibfnamefont {S.}~\bibnamefont {{d'Ascoli}}}, \bibinfo
  {author} {\bibfnamefont {L.}~\bibnamefont {Sagun}}, \bibinfo {author}
  {\bibfnamefont {G.}~\bibnamefont {Biroli}},\ and\ \bibinfo {author}
  {\bibfnamefont {M.}~\bibnamefont {Wyart}},\ }\bibfield  {title} {\bibinfo
  {title} {A jamming transition from under- to over-parametrization affects
  generalization in deep learning},\ }\href
  {https://doi.org/10.1088/1751-8121/ab4c8b} {\bibfield  {journal} {\bibinfo
  {journal} {Journal of Physics A: Mathematical and Theoretical}\ }\textbf
  {\bibinfo {volume} {52}},\ \bibinfo {pages} {474001} (\bibinfo {year}
  {2019})}\BibitemShut {NoStop}%
\bibitem [{\citenamefont {Antenucci}\ \emph {et~al.}(2019)\citenamefont
  {Antenucci}, \citenamefont {Franz}, \citenamefont {Urbani},\ and\
  \citenamefont {Zdeborov{\'a}}}]{antenucciGlassyNatureHard2019}%
  \BibitemOpen
  \bibfield  {author} {\bibinfo {author} {\bibfnamefont {F.}~\bibnamefont
  {Antenucci}}, \bibinfo {author} {\bibfnamefont {S.}~\bibnamefont {Franz}},
  \bibinfo {author} {\bibfnamefont {P.}~\bibnamefont {Urbani}},\ and\ \bibinfo
  {author} {\bibfnamefont {L.}~\bibnamefont {Zdeborov{\'a}}},\ }\bibfield
  {title} {\bibinfo {title} {Glassy {{Nature}} of the {{Hard Phase}} in
  {{Inference Problems}}},\ }\href {https://doi.org/10.1103/PhysRevX.9.011020}
  {\bibfield  {journal} {\bibinfo  {journal} {Physical Review X}\ }\textbf
  {\bibinfo {volume} {9}},\ \bibinfo {pages} {011020} (\bibinfo {year}
  {2019})}\BibitemShut {NoStop}%
\bibitem [{\citenamefont {Franz}\ \emph {et~al.}(2015)\citenamefont {Franz},
  \citenamefont {Parisi}, \citenamefont {Urbani},\ and\ \citenamefont
  {Zamponi}}]{franzUniversalSpectrumNormal2015}%
  \BibitemOpen
  \bibfield  {author} {\bibinfo {author} {\bibfnamefont {S.}~\bibnamefont
  {Franz}}, \bibinfo {author} {\bibfnamefont {G.}~\bibnamefont {Parisi}},
  \bibinfo {author} {\bibfnamefont {P.}~\bibnamefont {Urbani}},\ and\ \bibinfo
  {author} {\bibfnamefont {F.}~\bibnamefont {Zamponi}},\ }\bibfield  {title}
  {\bibinfo {title} {Universal spectrum of normal modes in low-temperature
  glasses},\ }\href {https://doi.org/10.1073/pnas.1511134112} {\bibfield
  {journal} {\bibinfo  {journal} {Proceedings of the National Academy of
  Sciences}\ }\textbf {\bibinfo {volume} {112}},\ \bibinfo {pages}
  {14539--14544} (\bibinfo {year} {2015})}\BibitemShut {NoStop}%
\bibitem [{\citenamefont {Franz}\ \emph
  {et~al.}(2019{\natexlab{a}})\citenamefont {Franz}, \citenamefont {Sclocchi},\
  and\ \citenamefont {Urbani}}]{franz2019critical}%
  \BibitemOpen
  \bibfield  {author} {\bibinfo {author} {\bibfnamefont {S.}~\bibnamefont
  {Franz}}, \bibinfo {author} {\bibfnamefont {A.}~\bibnamefont {Sclocchi}},\
  and\ \bibinfo {author} {\bibfnamefont {P.}~\bibnamefont {Urbani}},\
  }\bibfield  {title} {\bibinfo {title} {Critical jammed phase of the linear
  perceptron},\ }\href {https://doi.org/10.1103/PhysRevLett.123.115702}
  {\bibfield  {journal} {\bibinfo  {journal} {Physical Review Letters}\
  }\textbf {\bibinfo {volume} {123}},\ \bibinfo {pages} {115702} (\bibinfo
  {year} {2019}{\natexlab{a}})}\BibitemShut {NoStop}%
\bibitem [{\citenamefont {Franz}\ \emph
  {et~al.}(2019{\natexlab{b}})\citenamefont {Franz}, \citenamefont {Maimbourg},
  \citenamefont {Parisi},\ and\ \citenamefont {Scardicchio}}]{franz2019impact}%
  \BibitemOpen
  \bibfield  {author} {\bibinfo {author} {\bibfnamefont {S.}~\bibnamefont
  {Franz}}, \bibinfo {author} {\bibfnamefont {T.}~\bibnamefont {Maimbourg}},
  \bibinfo {author} {\bibfnamefont {G.}~\bibnamefont {Parisi}},\ and\ \bibinfo
  {author} {\bibfnamefont {A.}~\bibnamefont {Scardicchio}},\ }\bibfield
  {title} {\bibinfo {title} {Impact of jamming criticality on low-temperature
  anomalies in structural glasses},\ }\href
  {https://doi.org/10.1073/pnas.1820360116} {\bibfield  {journal} {\bibinfo
  {journal} {Proceedings of the National Academy of Sciences}\ }\textbf
  {\bibinfo {volume} {116}},\ \bibinfo {pages} {13768--13773} (\bibinfo {year}
  {2019}{\natexlab{b}})}\BibitemShut {NoStop}%
\bibitem [{\citenamefont {Artiaco}\ \emph {et~al.}(2021)\citenamefont
  {Artiaco}, \citenamefont {Balducci}, \citenamefont {Parisi},\ and\
  \citenamefont {Scardicchio}}]{artiaco2021quantum}%
  \BibitemOpen
  \bibfield  {author} {\bibinfo {author} {\bibfnamefont {C.}~\bibnamefont
  {Artiaco}}, \bibinfo {author} {\bibfnamefont {F.}~\bibnamefont {Balducci}},
  \bibinfo {author} {\bibfnamefont {G.}~\bibnamefont {Parisi}},\ and\ \bibinfo
  {author} {\bibfnamefont {A.}~\bibnamefont {Scardicchio}},\ }\bibfield
  {title} {\bibinfo {title} {Quantum jamming: Critical properties of a quantum
  mechanical perceptron},\ }\href
  {https://doi.org/10.1103/PhysRevA.103.L040203} {\bibfield  {journal}
  {\bibinfo  {journal} {Physical Review A}\ }\textbf {\bibinfo {volume}
  {103}},\ \bibinfo {pages} {L040203} (\bibinfo {year} {2021})}\BibitemShut
  {NoStop}%
\bibitem [{\citenamefont {Charbonneau}\ \emph
  {et~al.}(2015{\natexlab{a}})\citenamefont {Charbonneau}, \citenamefont
  {Corwin}, \citenamefont {Parisi},\ and\ \citenamefont
  {Zamponi}}]{charbonneauJammingCriticalityRevealed2015}%
  \BibitemOpen
  \bibfield  {author} {\bibinfo {author} {\bibfnamefont {P.}~\bibnamefont
  {Charbonneau}}, \bibinfo {author} {\bibfnamefont {E.~I.}\ \bibnamefont
  {Corwin}}, \bibinfo {author} {\bibfnamefont {G.}~\bibnamefont {Parisi}},\
  and\ \bibinfo {author} {\bibfnamefont {F.}~\bibnamefont {Zamponi}},\
  }\bibfield  {title} {\bibinfo {title} {Jamming {{Criticality Revealed}} by
  {{Removing Localized Buckling Excitations}}},\ }\href
  {https://doi.org/10.1103/PhysRevLett.114.125504} {\bibfield  {journal}
  {\bibinfo  {journal} {Physical Review Letters}\ }\textbf {\bibinfo {volume}
  {114}},\ \bibinfo {pages} {125504} (\bibinfo {year}
  {2015}{\natexlab{a}})}\BibitemShut {NoStop}%
\bibitem [{\citenamefont {Charbonneau}\ \emph {et~al.}(2012)\citenamefont
  {Charbonneau}, \citenamefont {Corwin}, \citenamefont {Parisi},\ and\
  \citenamefont {Zamponi}}]{charbonneauUniversalMicrostructureMechanical2012}%
  \BibitemOpen
  \bibfield  {author} {\bibinfo {author} {\bibfnamefont {P.}~\bibnamefont
  {Charbonneau}}, \bibinfo {author} {\bibfnamefont {E.~I.}\ \bibnamefont
  {Corwin}}, \bibinfo {author} {\bibfnamefont {G.}~\bibnamefont {Parisi}},\
  and\ \bibinfo {author} {\bibfnamefont {F.}~\bibnamefont {Zamponi}},\
  }\bibfield  {title} {\bibinfo {title} {Universal {{Microstructure}} and
  {{Mechanical Stability}} of {{Jammed Packings}}},\ }\href
  {https://doi.org/10.1103/PhysRevLett.109.205501} {\bibfield  {journal}
  {\bibinfo  {journal} {Physical Review Letters}\ }\textbf {\bibinfo {volume}
  {109}},\ \bibinfo {pages} {205501} (\bibinfo {year} {2012})}\BibitemShut
  {NoStop}%
\bibitem [{\citenamefont {Dennis}\ and\ \citenamefont
  {Corwin}(2020)}]{dennisJammingEnergyLandscape2020}%
  \BibitemOpen
  \bibfield  {author} {\bibinfo {author} {\bibfnamefont {R.~C.}\ \bibnamefont
  {Dennis}}\ and\ \bibinfo {author} {\bibfnamefont {E.~I.}\ \bibnamefont
  {Corwin}},\ }\bibfield  {title} {\bibinfo {title} {Jamming {{Energy
  Landscape}} is {{Hierarchical}} and {{Ultrametric}}},\ }\href
  {https://doi.org/10.1103/PhysRevLett.124.078002} {\bibfield  {journal}
  {\bibinfo  {journal} {Physical Review Letters}\ }\textbf {\bibinfo {volume}
  {124}},\ \bibinfo {pages} {078002} (\bibinfo {year} {2020})}\BibitemShut
  {NoStop}%
\bibitem [{\citenamefont {DeGiuli}\ \emph {et~al.}(2014)\citenamefont
  {DeGiuli}, \citenamefont {Lerner}, \citenamefont {Brito},\ and\ \citenamefont
  {Wyart}}]{degiuliForceDistributionAffects2014}%
  \BibitemOpen
  \bibfield  {author} {\bibinfo {author} {\bibfnamefont {E.}~\bibnamefont
  {DeGiuli}}, \bibinfo {author} {\bibfnamefont {E.}~\bibnamefont {Lerner}},
  \bibinfo {author} {\bibfnamefont {C.}~\bibnamefont {Brito}},\ and\ \bibinfo
  {author} {\bibfnamefont {M.}~\bibnamefont {Wyart}},\ }\bibfield  {title}
  {\bibinfo {title} {Force distribution affects vibrational properties in
  hard-sphere glasses},\ }\href {https://doi.org/10.1073/pnas.1415298111}
  {\bibfield  {journal} {\bibinfo  {journal} {Proceedings of the National
  Academy of Sciences}\ }\textbf {\bibinfo {volume} {111}},\ \bibinfo {pages}
  {17054--17059} (\bibinfo {year} {2014})}\BibitemShut {NoStop}%
\bibitem [{\citenamefont {Lerner}\ \emph
  {et~al.}(2013{\natexlab{a}})\citenamefont {Lerner}, \citenamefont
  {D{\"u}ring},\ and\ \citenamefont
  {Wyart}}]{lernerLowenergyNonlinearExcitations2013}%
  \BibitemOpen
  \bibfield  {author} {\bibinfo {author} {\bibfnamefont {E.}~\bibnamefont
  {Lerner}}, \bibinfo {author} {\bibfnamefont {G.}~\bibnamefont {D{\"u}ring}},\
  and\ \bibinfo {author} {\bibfnamefont {M.}~\bibnamefont {Wyart}},\ }\bibfield
   {title} {\bibinfo {title} {Low-energy non-linear excitations in sphere
  packings},\ }\href {https://doi.org/10.1039/C3SM50515D} {\bibfield  {journal}
  {\bibinfo  {journal} {Soft Matter}\ }\textbf {\bibinfo {volume} {9}},\
  \bibinfo {pages} {8252--8263} (\bibinfo {year}
  {2013}{\natexlab{a}})}\BibitemShut {NoStop}%
\bibitem [{\citenamefont {Charbonneau}\ \emph
  {et~al.}(2021{\natexlab{a}})\citenamefont {Charbonneau}, \citenamefont
  {Corwin}, \citenamefont {Dennis}, \citenamefont {D{\'i}az
  Hern{\'a}ndez~Rojas}, \citenamefont {Ikeda}, \citenamefont {Parisi},\ and\
  \citenamefont
  {{Ricci-Tersenghi}}}]{charbonneauFinitesizeEffectsMicroscopic2021}%
  \BibitemOpen
  \bibfield  {author} {\bibinfo {author} {\bibfnamefont {P.}~\bibnamefont
  {Charbonneau}}, \bibinfo {author} {\bibfnamefont {E.~I.}\ \bibnamefont
  {Corwin}}, \bibinfo {author} {\bibfnamefont {R.~C.}\ \bibnamefont {Dennis}},
  \bibinfo {author} {\bibfnamefont {R.}~\bibnamefont {D{\'i}az
  Hern{\'a}ndez~Rojas}}, \bibinfo {author} {\bibfnamefont {H.}~\bibnamefont
  {Ikeda}}, \bibinfo {author} {\bibfnamefont {G.}~\bibnamefont {Parisi}},\ and\
  \bibinfo {author} {\bibfnamefont {F.}~\bibnamefont {{Ricci-Tersenghi}}},\
  }\bibfield  {title} {\bibinfo {title} {Finite-size effects in the microscopic
  critical properties of jammed configurations: {{A}} comprehensive study of
  the effects of different types of disorder},\ }\href
  {https://doi.org/10.1103/PhysRevE.104.014102} {\bibfield  {journal} {\bibinfo
   {journal} {Physical Review E}\ }\textbf {\bibinfo {volume} {104}},\ \bibinfo
  {pages} {014102} (\bibinfo {year} {2021}{\natexlab{a}})}\BibitemShut
  {NoStop}%
\bibitem [{\citenamefont {Goodrich}\ \emph {et~al.}(2012)\citenamefont
  {Goodrich}, \citenamefont {Liu},\ and\ \citenamefont
  {Nagel}}]{goodrichFiniteSizeScalingJamming2012}%
  \BibitemOpen
  \bibfield  {author} {\bibinfo {author} {\bibfnamefont {C.~P.}\ \bibnamefont
  {Goodrich}}, \bibinfo {author} {\bibfnamefont {A.~J.}\ \bibnamefont {Liu}},\
  and\ \bibinfo {author} {\bibfnamefont {S.~R.}\ \bibnamefont {Nagel}},\
  }\bibfield  {title} {\bibinfo {title} {Finite-{{Size Scaling}} at the
  {{Jamming Transition}}},\ }\href
  {https://doi.org/10.1103/PhysRevLett.109.095704} {\bibfield  {journal}
  {\bibinfo  {journal} {Physical Review Letters}\ }\textbf {\bibinfo {volume}
  {109}},\ \bibinfo {pages} {095704} (\bibinfo {year} {2012})}\BibitemShut
  {NoStop}%
\bibitem [{\citenamefont {Goodrich}\ \emph {et~al.}(2016)\citenamefont
  {Goodrich}, \citenamefont {Liu},\ and\ \citenamefont
  {Sethna}}]{goodrichScalingAnsatzJamming2016}%
  \BibitemOpen
  \bibfield  {author} {\bibinfo {author} {\bibfnamefont {C.~P.}\ \bibnamefont
  {Goodrich}}, \bibinfo {author} {\bibfnamefont {A.~J.}\ \bibnamefont {Liu}},\
  and\ \bibinfo {author} {\bibfnamefont {J.~P.}\ \bibnamefont {Sethna}},\
  }\bibfield  {title} {\bibinfo {title} {Scaling ansatz for the jamming
  transition},\ }\href {https://doi.org/10.1073/pnas.1601858113} {\bibfield
  {journal} {\bibinfo  {journal} {Proceedings of the National Academy of
  Sciences}\ }\textbf {\bibinfo {volume} {113}},\ \bibinfo {pages} {9745--9750}
  (\bibinfo {year} {2016})}\BibitemShut {NoStop}%
\bibitem [{\citenamefont {Artiaco}\ \emph {et~al.}(2020)\citenamefont
  {Artiaco}, \citenamefont {Baldan},\ and\ \citenamefont
  {Parisi}}]{artiacoExploratoryStudyGlassy2020}%
  \BibitemOpen
  \bibfield  {author} {\bibinfo {author} {\bibfnamefont {C.}~\bibnamefont
  {Artiaco}}, \bibinfo {author} {\bibfnamefont {P.}~\bibnamefont {Baldan}},\
  and\ \bibinfo {author} {\bibfnamefont {G.}~\bibnamefont {Parisi}},\
  }\bibfield  {title} {\bibinfo {title} {Exploratory study of the glassy
  landscape near jamming},\ }\href
  {https://doi.org/10.1103/PhysRevE.101.052605} {\bibfield  {journal} {\bibinfo
   {journal} {Physical Review E}\ }\textbf {\bibinfo {volume} {101}},\ \bibinfo
  {pages} {052605} (\bibinfo {year} {2020})}\BibitemShut {NoStop}%
\bibitem [{\citenamefont {Hexner}\ \emph {et~al.}(2019)\citenamefont {Hexner},
  \citenamefont {Urbani},\ and\ \citenamefont
  {Zamponi}}]{hexnerCanLargePacking2019}%
  \BibitemOpen
  \bibfield  {author} {\bibinfo {author} {\bibfnamefont {D.}~\bibnamefont
  {Hexner}}, \bibinfo {author} {\bibfnamefont {P.}~\bibnamefont {Urbani}},\
  and\ \bibinfo {author} {\bibfnamefont {F.}~\bibnamefont {Zamponi}},\
  }\bibfield  {title} {\bibinfo {title} {Can a {{Large Packing}} be
  {{Assembled}} from {{Smaller Ones}}?},\ }\href
  {https://doi.org/10.1103/PhysRevLett.123.068003} {\bibfield  {journal}
  {\bibinfo  {journal} {Physical Review Letters}\ }\textbf {\bibinfo {volume}
  {123}},\ \bibinfo {pages} {068003} (\bibinfo {year} {2019})}\BibitemShut
  {NoStop}%
\bibitem [{\citenamefont {Hexner}\ \emph {et~al.}(2018)\citenamefont {Hexner},
  \citenamefont {Liu},\ and\ \citenamefont
  {Nagel}}]{hexnerTwoDivergingLength2018}%
  \BibitemOpen
  \bibfield  {author} {\bibinfo {author} {\bibfnamefont {D.}~\bibnamefont
  {Hexner}}, \bibinfo {author} {\bibfnamefont {A.~J.}\ \bibnamefont {Liu}},\
  and\ \bibinfo {author} {\bibfnamefont {S.~R.}\ \bibnamefont {Nagel}},\
  }\bibfield  {title} {\bibinfo {title} {Two {{Diverging Length Scales}} in the
  {{Structure}} of {{Jammed Packings}}},\ }\href
  {https://doi.org/10.1103/PhysRevLett.121.115501} {\bibfield  {journal}
  {\bibinfo  {journal} {Physical Review Letters}\ }\textbf {\bibinfo {volume}
  {121}},\ \bibinfo {pages} {115501} (\bibinfo {year} {2018})}\BibitemShut
  {NoStop}%
\bibitem [{\citenamefont {Arceri}\ and\ \citenamefont
  {Corwin}(2020)}]{arceriVibrationalPropertiesHard2020}%
  \BibitemOpen
  \bibfield  {author} {\bibinfo {author} {\bibfnamefont {F.}~\bibnamefont
  {Arceri}}\ and\ \bibinfo {author} {\bibfnamefont {E.~I.}\ \bibnamefont
  {Corwin}},\ }\bibfield  {title} {\bibinfo {title} {Vibrational {{Properties}}
  of {{Hard}} and {{Soft Spheres Are Unified}} at {{Jamming}}},\ }\href
  {https://doi.org/10.1103/PhysRevLett.124.238002} {\bibfield  {journal}
  {\bibinfo  {journal} {Physical Review Letters}\ }\textbf {\bibinfo {volume}
  {124}},\ \bibinfo {pages} {238002} (\bibinfo {year} {2020})}\BibitemShut
  {NoStop}%
\bibitem [{\citenamefont {Hagh}\ \emph {et~al.}(2019)\citenamefont {Hagh},
  \citenamefont {Corwin}, \citenamefont {Stephenson},\ and\ \citenamefont
  {Thorpe}}]{haghBroaderViewJamming2019}%
  \BibitemOpen
  \bibfield  {author} {\bibinfo {author} {\bibfnamefont {V.~F.}\ \bibnamefont
  {Hagh}}, \bibinfo {author} {\bibfnamefont {E.~I.}\ \bibnamefont {Corwin}},
  \bibinfo {author} {\bibfnamefont {K.}~\bibnamefont {Stephenson}},\ and\
  \bibinfo {author} {\bibfnamefont {M.~F.}\ \bibnamefont {Thorpe}},\ }\bibfield
   {title} {\bibinfo {title} {A broader view on jamming: From spring networks
  to circle packings},\ }\href {https://doi.org/10.1039/C8SM01768A} {\bibfield
  {journal} {\bibinfo  {journal} {Soft Matter}\ }\textbf {\bibinfo {volume}
  {15}},\ \bibinfo {pages} {3076--3084} (\bibinfo {year} {2019})}\BibitemShut
  {NoStop}%
\bibitem [{\citenamefont {Coulais}\ \emph {et~al.}(2014)\citenamefont
  {Coulais}, \citenamefont {Behringer},\ and\ \citenamefont
  {Dauchot}}]{coulaisHowIdealJamming2014}%
  \BibitemOpen
  \bibfield  {author} {\bibinfo {author} {\bibfnamefont {C.}~\bibnamefont
  {Coulais}}, \bibinfo {author} {\bibfnamefont {R.~P.}\ \bibnamefont
  {Behringer}},\ and\ \bibinfo {author} {\bibfnamefont {O.}~\bibnamefont
  {Dauchot}},\ }\bibfield  {title} {\bibinfo {title} {How the ideal jamming
  point illuminates the world of granular media},\ }\href
  {https://doi.org/10.1039/C3SM51231B} {\bibfield  {journal} {\bibinfo
  {journal} {Soft Matter}\ }\textbf {\bibinfo {volume} {10}},\ \bibinfo {pages}
  {1519--1536} (\bibinfo {year} {2014})}\BibitemShut {NoStop}%
\bibitem [{\citenamefont {Dauchot}\ \emph {et~al.}(2005)\citenamefont
  {Dauchot}, \citenamefont {Marty},\ and\ \citenamefont
  {Biroli}}]{dauchotDynamicalHeterogeneityClose2005}%
  \BibitemOpen
  \bibfield  {author} {\bibinfo {author} {\bibfnamefont {O.}~\bibnamefont
  {Dauchot}}, \bibinfo {author} {\bibfnamefont {G.}~\bibnamefont {Marty}},\
  and\ \bibinfo {author} {\bibfnamefont {G.}~\bibnamefont {Biroli}},\
  }\bibfield  {title} {\bibinfo {title} {Dynamical {{Heterogeneity Close}} to
  the {{Jamming Transition}} in a {{Sheared Granular Material}}},\ }\href
  {https://doi.org/10.1103/PhysRevLett.95.265701} {\bibfield  {journal}
  {\bibinfo  {journal} {Physical Review Letters}\ }\textbf {\bibinfo {volume}
  {95}},\ \bibinfo {pages} {265701} (\bibinfo {year} {2005})}\BibitemShut
  {NoStop}%
\bibitem [{\citenamefont {Lechenault}\ \emph {et~al.}(2008)\citenamefont
  {Lechenault}, \citenamefont {Dauchot}, \citenamefont {Biroli},\ and\
  \citenamefont {Bouchaud}}]{lechenaultCriticalScalingHeterogeneous2008}%
  \BibitemOpen
  \bibfield  {author} {\bibinfo {author} {\bibfnamefont {F.}~\bibnamefont
  {Lechenault}}, \bibinfo {author} {\bibfnamefont {O.}~\bibnamefont {Dauchot}},
  \bibinfo {author} {\bibfnamefont {G.}~\bibnamefont {Biroli}},\ and\ \bibinfo
  {author} {\bibfnamefont {J.~P.}\ \bibnamefont {Bouchaud}},\ }\bibfield
  {title} {\bibinfo {title} {Critical scaling and heterogeneous superdiffusion
  across the jamming/rigidity transition of a granular glass},\ }\href
  {https://doi.org/10.1209/0295-5075/83/46003} {\bibfield  {journal} {\bibinfo
  {journal} {EPL (Europhysics Letters)}\ }\textbf {\bibinfo {volume} {83}},\
  \bibinfo {pages} {46003} (\bibinfo {year} {2008})}\BibitemShut {NoStop}%
\bibitem [{\citenamefont {Seguin}\ and\ \citenamefont
  {Dauchot}(2016)}]{seguinExperimentalEvidenceGardner2016}%
  \BibitemOpen
  \bibfield  {author} {\bibinfo {author} {\bibfnamefont {A.}~\bibnamefont
  {Seguin}}\ and\ \bibinfo {author} {\bibfnamefont {O.}~\bibnamefont
  {Dauchot}},\ }\bibfield  {title} {\bibinfo {title} {Experimental {{Evidence}}
  of the {{Gardner Phase}} in a {{Granular Glass}}},\ }\href
  {https://doi.org/10.1103/PhysRevLett.117.228001} {\bibfield  {journal}
  {\bibinfo  {journal} {Physical Review Letters}\ }\textbf {\bibinfo {volume}
  {117}},\ \bibinfo {pages} {228001} (\bibinfo {year} {2016})}\BibitemShut
  {NoStop}%
\bibitem [{\citenamefont {Wang}\ \emph {et~al.}(2022)\citenamefont {Wang},
  \citenamefont {Shang}, \citenamefont {Jin},\ and\ \citenamefont
  {Zhang}}]{wangExperimentalObservationsMarginal2022}%
  \BibitemOpen
  \bibfield  {author} {\bibinfo {author} {\bibfnamefont {Y.}~\bibnamefont
  {Wang}}, \bibinfo {author} {\bibfnamefont {J.}~\bibnamefont {Shang}},
  \bibinfo {author} {\bibfnamefont {Y.}~\bibnamefont {Jin}},\ and\ \bibinfo
  {author} {\bibfnamefont {J.}~\bibnamefont {Zhang}},\ }\bibfield  {title}
  {\bibinfo {title} {Experimental observations of marginal criticality in
  granular materials},\ }\href {https://doi.org/10.1073/pnas.2204879119}
  {\bibfield  {journal} {\bibinfo  {journal} {Proceedings of the National
  Academy of Sciences}\ }\textbf {\bibinfo {volume} {119}},\ \bibinfo {pages}
  {e2204879119} (\bibinfo {year} {2022})}\BibitemShut {NoStop}%
\bibitem [{\citenamefont {Aste}\ \emph {et~al.}(2005)\citenamefont {Aste},
  \citenamefont {Saadatfar},\ and\ \citenamefont
  {Senden}}]{asteGeometricalStructureDisordered2005}%
  \BibitemOpen
  \bibfield  {author} {\bibinfo {author} {\bibfnamefont {T.}~\bibnamefont
  {Aste}}, \bibinfo {author} {\bibfnamefont {M.}~\bibnamefont {Saadatfar}},\
  and\ \bibinfo {author} {\bibfnamefont {T.~J.}\ \bibnamefont {Senden}},\
  }\bibfield  {title} {\bibinfo {title} {Geometrical structure of disordered
  sphere packings},\ }\href {https://doi.org/10.1103/PhysRevE.71.061302}
  {\bibfield  {journal} {\bibinfo  {journal} {Physical Review E}\ }\textbf
  {\bibinfo {volume} {71}},\ \bibinfo {pages} {061302} (\bibinfo {year}
  {2005})}\BibitemShut {NoStop}%
\bibitem [{\citenamefont {Aste}(2006)}]{asteVolumeFluctuationsGeometrical2006}%
  \BibitemOpen
  \bibfield  {author} {\bibinfo {author} {\bibfnamefont {T.}~\bibnamefont
  {Aste}},\ }\bibfield  {title} {\bibinfo {title} {Volume {{Fluctuations}} and
  {{Geometrical Constraints}} in {{Granular Packs}}},\ }\href
  {https://doi.org/10.1103/PhysRevLett.96.018002} {\bibfield  {journal}
  {\bibinfo  {journal} {Physical Review Letters}\ }\textbf {\bibinfo {volume}
  {96}},\ \bibinfo {pages} {018002} (\bibinfo {year} {2006})}\BibitemShut
  {NoStop}%
\bibitem [{\citenamefont {Bitzek}\ \emph {et~al.}(2006)\citenamefont {Bitzek},
  \citenamefont {Koskinen}, \citenamefont {G{\"a}hler}, \citenamefont
  {Moseler},\ and\ \citenamefont {Gumbsch}}]{FIRE}%
  \BibitemOpen
  \bibfield  {author} {\bibinfo {author} {\bibfnamefont {E.}~\bibnamefont
  {Bitzek}}, \bibinfo {author} {\bibfnamefont {P.}~\bibnamefont {Koskinen}},
  \bibinfo {author} {\bibfnamefont {F.}~\bibnamefont {G{\"a}hler}}, \bibinfo
  {author} {\bibfnamefont {M.}~\bibnamefont {Moseler}},\ and\ \bibinfo {author}
  {\bibfnamefont {P.}~\bibnamefont {Gumbsch}},\ }\bibfield  {title} {\bibinfo
  {title} {Structural {{Relaxation Made Simple}}},\ }\href
  {https://doi.org/10.1103/PhysRevLett.97.170201} {\bibfield  {journal}
  {\bibinfo  {journal} {Physical Review Letters}\ }\textbf {\bibinfo {volume}
  {97}},\ \bibinfo {pages} {170201} (\bibinfo {year} {2006})}\BibitemShut
  {NoStop}%
\bibitem [{\citenamefont {Morse}\ and\ \citenamefont
  {Corwin}(2014)}]{morseGeometricSignaturesJamming2014}%
  \BibitemOpen
  \bibfield  {author} {\bibinfo {author} {\bibfnamefont {P.~K.}\ \bibnamefont
  {Morse}}\ and\ \bibinfo {author} {\bibfnamefont {E.~I.}\ \bibnamefont
  {Corwin}},\ }\bibfield  {title} {\bibinfo {title} {Geometric signatures of
  jamming in the mechanical vacuum},\ }\href
  {https://doi.org/10.1103/PhysRevLett.112.115701} {\bibfield  {journal}
  {\bibinfo  {journal} {Physical Review Letters}\ }\textbf {\bibinfo {volume}
  {112}},\ \bibinfo {pages} {115701} (\bibinfo {year} {2014})}\BibitemShut
  {NoStop}%
\bibitem [{\citenamefont {Charbonneau}\ \emph {et~al.}(2016)\citenamefont
  {Charbonneau}, \citenamefont {Corwin}, \citenamefont {Parisi}, \citenamefont
  {Poncet},\ and\ \citenamefont
  {Zamponi}}]{charbonneauUniversalNonDebyeScaling2016}%
  \BibitemOpen
  \bibfield  {author} {\bibinfo {author} {\bibfnamefont {P.}~\bibnamefont
  {Charbonneau}}, \bibinfo {author} {\bibfnamefont {E.~I.}\ \bibnamefont
  {Corwin}}, \bibinfo {author} {\bibfnamefont {G.}~\bibnamefont {Parisi}},
  \bibinfo {author} {\bibfnamefont {A.}~\bibnamefont {Poncet}},\ and\ \bibinfo
  {author} {\bibfnamefont {F.}~\bibnamefont {Zamponi}},\ }\bibfield  {title}
  {\bibinfo {title} {Universal {{Non}}-{{Debye Scaling}} in the {{Density}} of
  {{States}} of {{Amorphous Solids}}},\ }\href
  {https://doi.org/10.1103/PhysRevLett.117.045503} {\bibfield  {journal}
  {\bibinfo  {journal} {Physical Review Letters}\ }\textbf {\bibinfo {volume}
  {117}},\ \bibinfo {pages} {045503} (\bibinfo {year} {2016})}\BibitemShut
  {NoStop}%
\bibitem [{\citenamefont {Morse}\ and\ \citenamefont
  {Corwin}(2017)}]{morseEchoesGlassTransition2017}%
  \BibitemOpen
  \bibfield  {author} {\bibinfo {author} {\bibfnamefont {P.~K.}\ \bibnamefont
  {Morse}}\ and\ \bibinfo {author} {\bibfnamefont {E.~I.}\ \bibnamefont
  {Corwin}},\ }\bibfield  {title} {\bibinfo {title} {Echoes of the {{Glass
  Transition}} in {{Athermal Soft Spheres}}},\ }\href
  {https://doi.org/10.1103/PhysRevLett.119.118003} {\bibfield  {journal}
  {\bibinfo  {journal} {Physical Review Letters}\ }\textbf {\bibinfo {volume}
  {119}},\ \bibinfo {pages} {118003} (\bibinfo {year} {2017})}\BibitemShut
  {NoStop}%
\bibitem [{\citenamefont {Charbonneau}\ and\ \citenamefont
  {Morse}(2021)}]{charbonneauMemoryFormationJammed2021}%
  \BibitemOpen
  \bibfield  {author} {\bibinfo {author} {\bibfnamefont {P.}~\bibnamefont
  {Charbonneau}}\ and\ \bibinfo {author} {\bibfnamefont {P.~K.}\ \bibnamefont
  {Morse}},\ }\bibfield  {title} {\bibinfo {title} {Memory {{Formation}} in
  {{Jammed Hard Spheres}}},\ }\href
  {https://doi.org/10.1103/PhysRevLett.126.088001} {\bibfield  {journal}
  {\bibinfo  {journal} {Physical Review Letters}\ }\textbf {\bibinfo {volume}
  {126}},\ \bibinfo {pages} {088001} (\bibinfo {year} {2021})}\BibitemShut
  {NoStop}%
\bibitem [{\citenamefont {Lubachevsky}\ and\ \citenamefont
  {Stillinger}(1990)}]{lubachevskyGeometricPropertiesRandom1990}%
  \BibitemOpen
  \bibfield  {author} {\bibinfo {author} {\bibfnamefont {B.~D.}\ \bibnamefont
  {Lubachevsky}}\ and\ \bibinfo {author} {\bibfnamefont {F.~H.}\ \bibnamefont
  {Stillinger}},\ }\bibfield  {title} {\bibinfo {title} {Geometric properties
  of random disk packings},\ }\href {https://doi.org/10.1007/BF01025983}
  {\bibfield  {journal} {\bibinfo  {journal} {Journal of Statistical Physics}\
  }\textbf {\bibinfo {volume} {60}},\ \bibinfo {pages} {561--583} (\bibinfo
  {year} {1990})}\BibitemShut {NoStop}%
\bibitem [{\citenamefont {Skoge}\ \emph {et~al.}(2006)\citenamefont {Skoge},
  \citenamefont {Donev}, \citenamefont {Stillinger},\ and\ \citenamefont
  {Torquato}}]{md-code}%
  \BibitemOpen
  \bibfield  {author} {\bibinfo {author} {\bibfnamefont {M.}~\bibnamefont
  {Skoge}}, \bibinfo {author} {\bibfnamefont {A.}~\bibnamefont {Donev}},
  \bibinfo {author} {\bibfnamefont {F.~H.}\ \bibnamefont {Stillinger}},\ and\
  \bibinfo {author} {\bibfnamefont {S.}~\bibnamefont {Torquato}},\ }\bibfield
  {title} {\bibinfo {title} {Packing hyperspheres in high-dimensional
  {{Euclidean}} spaces},\ }\href {https://doi.org/10.1103/PhysRevE.74.041127}
  {\bibfield  {journal} {\bibinfo  {journal} {Physical Review E}\ }\textbf
  {\bibinfo {volume} {74}},\ \bibinfo {pages} {041127} (\bibinfo {year}
  {2006})}\BibitemShut {NoStop}%
\bibitem [{\citenamefont {Torquato}\ \emph {et~al.}(2000)\citenamefont
  {Torquato}, \citenamefont {Truskett},\ and\ \citenamefont
  {Debenedetti}}]{torquatoRandomClosePacking2000}%
  \BibitemOpen
  \bibfield  {author} {\bibinfo {author} {\bibfnamefont {S.}~\bibnamefont
  {Torquato}}, \bibinfo {author} {\bibfnamefont {T.~M.}\ \bibnamefont
  {Truskett}},\ and\ \bibinfo {author} {\bibfnamefont {P.~G.}\ \bibnamefont
  {Debenedetti}},\ }\bibfield  {title} {\bibinfo {title} {Is {{Random Close
  Packing}} of {{Spheres Well Defined}}?},\ }\href
  {https://doi.org/10.1103/PhysRevLett.84.2064} {\bibfield  {journal} {\bibinfo
   {journal} {Physical Review Letters}\ }\textbf {\bibinfo {volume} {84}},\
  \bibinfo {pages} {2064--2067} (\bibinfo {year} {2000})}\BibitemShut {NoStop}%
\bibitem [{\citenamefont {Zhang}\ \emph {et~al.}(2014)\citenamefont {Zhang},
  \citenamefont {Smith}, \citenamefont {Wang}, \citenamefont {Liu},
  \citenamefont {Schroers}, \citenamefont {Shattuck},\ and\ \citenamefont
  {O'Hern}}]{zhangConnectionPackingEfficiency2014}%
  \BibitemOpen
  \bibfield  {author} {\bibinfo {author} {\bibfnamefont {K.}~\bibnamefont
  {Zhang}}, \bibinfo {author} {\bibfnamefont {W.~W.}\ \bibnamefont {Smith}},
  \bibinfo {author} {\bibfnamefont {M.}~\bibnamefont {Wang}}, \bibinfo {author}
  {\bibfnamefont {Y.}~\bibnamefont {Liu}}, \bibinfo {author} {\bibfnamefont
  {J.}~\bibnamefont {Schroers}}, \bibinfo {author} {\bibfnamefont {M.~D.}\
  \bibnamefont {Shattuck}},\ and\ \bibinfo {author} {\bibfnamefont {C.~S.}\
  \bibnamefont {O'Hern}},\ }\bibfield  {title} {\bibinfo {title} {Connection
  between the packing efficiency of binary hard spheres and the glass-forming
  ability of bulk metallic glasses},\ }\href
  {https://doi.org/10.1103/PhysRevE.90.032311} {\bibfield  {journal} {\bibinfo
  {journal} {Physical Review E}\ }\textbf {\bibinfo {volume} {90}},\ \bibinfo
  {pages} {032311} (\bibinfo {year} {2014})}\BibitemShut {NoStop}%
\bibitem [{\citenamefont {Lerner}\ \emph
  {et~al.}(2013{\natexlab{b}})\citenamefont {Lerner}, \citenamefont
  {D{\"u}ring},\ and\ \citenamefont
  {Wyart}}]{lernerSimulationsDrivenOverdamped2013}%
  \BibitemOpen
  \bibfield  {author} {\bibinfo {author} {\bibfnamefont {E.}~\bibnamefont
  {Lerner}}, \bibinfo {author} {\bibfnamefont {G.}~\bibnamefont {D{\"u}ring}},\
  and\ \bibinfo {author} {\bibfnamefont {M.}~\bibnamefont {Wyart}},\ }\bibfield
   {title} {\bibinfo {title} {Simulations of driven overdamped frictionless
  hard spheres},\ }\href {https://doi.org/10.1016/j.cpc.2012.10.020} {\bibfield
   {journal} {\bibinfo  {journal} {Computer Physics Communications}\ }\textbf
  {\bibinfo {volume} {184}},\ \bibinfo {pages} {628--637} (\bibinfo {year}
  {2013}{\natexlab{b}})}\BibitemShut {NoStop}%
\bibitem [{\citenamefont {Brito}\ and\ \citenamefont
  {Wyart}(2006)}]{britoRigidityHardsphereGlass2006}%
  \BibitemOpen
  \bibfield  {author} {\bibinfo {author} {\bibfnamefont {C.}~\bibnamefont
  {Brito}}\ and\ \bibinfo {author} {\bibfnamefont {M.}~\bibnamefont {Wyart}},\
  }\bibfield  {title} {\bibinfo {title} {On the rigidity of a hard-sphere glass
  near random close packing},\ }\href
  {https://doi.org/10.1209/epl/i2006-10238-x} {\bibfield  {journal} {\bibinfo
  {journal} {EPL (Europhysics Letters)}\ }\textbf {\bibinfo {volume} {76}},\
  \bibinfo {pages} {149} (\bibinfo {year} {2006})}\BibitemShut {NoStop}%
\bibitem [{\citenamefont {Brito}\ and\ \citenamefont
  {Wyart}(2009)}]{britoGeometricInterpretationPrevitrification2009}%
  \BibitemOpen
  \bibfield  {author} {\bibinfo {author} {\bibfnamefont {C.}~\bibnamefont
  {Brito}}\ and\ \bibinfo {author} {\bibfnamefont {M.}~\bibnamefont {Wyart}},\
  }\bibfield  {title} {\bibinfo {title} {Geometric interpretation of
  previtrification in hard sphere liquids},\ }\href
  {https://doi.org/10.1063/1.3157261} {\bibfield  {journal} {\bibinfo
  {journal} {The Journal of Chemical Physics}\ }\textbf {\bibinfo {volume}
  {131}},\ \bibinfo {pages} {024504} (\bibinfo {year} {2009})}\BibitemShut
  {NoStop}%
\bibitem [{\citenamefont {Henkes}\ \emph {et~al.}(2012)\citenamefont {Henkes},
  \citenamefont {Brito},\ and\ \citenamefont
  {Dauchot}}]{henkesExtractingVibrationalModes2012}%
  \BibitemOpen
  \bibfield  {author} {\bibinfo {author} {\bibfnamefont {S.}~\bibnamefont
  {Henkes}}, \bibinfo {author} {\bibfnamefont {C.}~\bibnamefont {Brito}},\ and\
  \bibinfo {author} {\bibfnamefont {O.}~\bibnamefont {Dauchot}},\ }\bibfield
  {title} {\bibinfo {title} {Extracting vibrational modes from fluctuations: A
  pedagogical discussion},\ }\href {https://doi.org/10.1039/C2SM07445A}
  {\bibfield  {journal} {\bibinfo  {journal} {Soft Matter}\ }\textbf {\bibinfo
  {volume} {8}},\ \bibinfo {pages} {6092--6109} (\bibinfo {year}
  {2012})}\BibitemShut {NoStop}%
\bibitem [{\citenamefont {Altieri}\ \emph {et~al.}(2016)\citenamefont
  {Altieri}, \citenamefont {Franz},\ and\ \citenamefont
  {Parisi}}]{altieriJammingTransitionHigh2016}%
  \BibitemOpen
  \bibfield  {author} {\bibinfo {author} {\bibfnamefont {A.}~\bibnamefont
  {Altieri}}, \bibinfo {author} {\bibfnamefont {S.}~\bibnamefont {Franz}},\
  and\ \bibinfo {author} {\bibfnamefont {G.}~\bibnamefont {Parisi}},\
  }\bibfield  {title} {\bibinfo {title} {The jamming transition in high
  dimension: An analytical study of the {{TAP}} equations and the effective
  thermodynamic potential},\ }\href
  {https://doi.org/10.1088/1742-5468/2016/09/093301} {\bibfield  {journal}
  {\bibinfo  {journal} {Journal of Statistical Mechanics: Theory and
  Experiment}\ }\textbf {\bibinfo {volume} {2016}},\ \bibinfo {pages} {093301}
  (\bibinfo {year} {2016})}\BibitemShut {NoStop}%
\bibitem [{\citenamefont {Frenkel}(2015)}]{frenkelOrderEntropy2015}%
  \BibitemOpen
  \bibfield  {author} {\bibinfo {author} {\bibfnamefont {D.}~\bibnamefont
  {Frenkel}},\ }\bibfield  {title} {\bibinfo {title} {Order through entropy},\
  }\href {https://doi.org/10.1038/nmat4178} {\bibfield  {journal} {\bibinfo
  {journal} {Nature Materials}\ }\textbf {\bibinfo {volume} {14}},\ \bibinfo
  {pages} {9--12} (\bibinfo {year} {2015})}\BibitemShut {NoStop}%
\bibitem [{\citenamefont {Bezanson}\ \emph {et~al.}(2017)\citenamefont
  {Bezanson}, \citenamefont {Edelman}, \citenamefont {Karpinski},\ and\
  \citenamefont {Shah}}]{bezansonJuliaFreshApproach2017}%
  \BibitemOpen
  \bibfield  {author} {\bibinfo {author} {\bibfnamefont {J.}~\bibnamefont
  {Bezanson}}, \bibinfo {author} {\bibfnamefont {A.}~\bibnamefont {Edelman}},
  \bibinfo {author} {\bibfnamefont {S.}~\bibnamefont {Karpinski}},\ and\
  \bibinfo {author} {\bibfnamefont {V.~B.}\ \bibnamefont {Shah}},\ }\bibfield
  {title} {\bibinfo {title} {Julia: A {Fresh Approach} to {Numerical
  Computing}},\ }\href {https://doi.org/10.1137/141000671} {\bibfield
  {journal} {\bibinfo  {journal} {SIAM Review}\ }\textbf {\bibinfo {volume}
  {59}},\ \bibinfo {pages} {65--98} (\bibinfo {year} {2017})}\BibitemShut
  {NoStop}%
\bibitem [{cod()}]{code2022github}%
  \BibitemOpen
  \href@noop {} {\bibinfo {title} {Our implementation of the {CALiPPSO}
  algorithm}},\ \bibinfo {howpublished}
  {\url{https://github.com/rdhr/CALiPPSO.jl}}\BibitemShut {NoStop}%
\bibitem [{\citenamefont {D{\'i}az Hern{\'a}ndez~Rojas}\ \emph
  {et~al.}(2021)\citenamefont {D{\'i}az Hern{\'a}ndez~Rojas}, \citenamefont
  {Parisi},\ and\ \citenamefont
  {{Ricci-Tersenghi}}}]{diazhernandezrojasInferringParticlewiseDynamics2021}%
  \BibitemOpen
  \bibfield  {author} {\bibinfo {author} {\bibfnamefont {R.}~\bibnamefont
  {D{\'i}az Hern{\'a}ndez~Rojas}}, \bibinfo {author} {\bibfnamefont
  {G.}~\bibnamefont {Parisi}},\ and\ \bibinfo {author} {\bibfnamefont
  {F.}~\bibnamefont {{Ricci-Tersenghi}}},\ }\bibfield  {title} {\bibinfo
  {title} {Inferring the particle-wise dynamics of amorphous solids from the
  local structure at the jamming point},\ }\href
  {https://doi.org/10.1039/C9SM02283J} {\bibfield  {journal} {\bibinfo
  {journal} {Soft Matter}\ }\textbf {\bibinfo {volume} {17}},\ \bibinfo {pages}
  {1056--1083} (\bibinfo {year} {2021})}\BibitemShut {NoStop}%
\bibitem [{\citenamefont {Torquato}\ and\ \citenamefont
  {Jiao}(2010)}]{torquatoRobustAlgorithmGenerate2010}%
  \BibitemOpen
  \bibfield  {author} {\bibinfo {author} {\bibfnamefont {S.}~\bibnamefont
  {Torquato}}\ and\ \bibinfo {author} {\bibfnamefont {Y.}~\bibnamefont
  {Jiao}},\ }\bibfield  {title} {\bibinfo {title} {Robust algorithm to generate
  a diverse class of dense disordered and ordered sphere packings via linear
  programming},\ }\href {https://doi.org/10.1103/PhysRevE.82.061302} {\bibfield
   {journal} {\bibinfo  {journal} {Physical Review E}\ }\textbf {\bibinfo
  {volume} {82}},\ \bibinfo {pages} {061302} (\bibinfo {year}
  {2010})}\BibitemShut {NoStop}%
\bibitem [{\citenamefont {Krabbenhoft}\ \emph {et~al.}(2012)\citenamefont
  {Krabbenhoft}, \citenamefont {Lyamin}, \citenamefont {Huang},\ and\
  \citenamefont {{Vicente da Silva}}}]{krabbenhoftGranularContactDynamics2012}%
  \BibitemOpen
  \bibfield  {author} {\bibinfo {author} {\bibfnamefont {K.}~\bibnamefont
  {Krabbenhoft}}, \bibinfo {author} {\bibfnamefont {A.~V.}\ \bibnamefont
  {Lyamin}}, \bibinfo {author} {\bibfnamefont {J.}~\bibnamefont {Huang}},\ and\
  \bibinfo {author} {\bibfnamefont {M.}~\bibnamefont {{Vicente da Silva}}},\
  }\bibfield  {title} {\bibinfo {title} {Granular contact dynamics using
  mathematical programming methods},\ }\href
  {https://doi.org/10.1016/j.compgeo.2012.02.006} {\bibfield  {journal}
  {\bibinfo  {journal} {Computers and Geotechnics}\ }\textbf {\bibinfo {volume}
  {43}},\ \bibinfo {pages} {165--176} (\bibinfo {year} {2012})}\BibitemShut
  {NoStop}%
\bibitem [{\citenamefont {Donev}\ \emph
  {et~al.}(2004{\natexlab{a}})\citenamefont {Donev}, \citenamefont {Torquato},
  \citenamefont {Stillinger},\ and\ \citenamefont
  {Connelly}}]{donevLinearProgrammingAlgorithm2004}%
  \BibitemOpen
  \bibfield  {author} {\bibinfo {author} {\bibfnamefont {A.}~\bibnamefont
  {Donev}}, \bibinfo {author} {\bibfnamefont {S.}~\bibnamefont {Torquato}},
  \bibinfo {author} {\bibfnamefont {F.~H.}\ \bibnamefont {Stillinger}},\ and\
  \bibinfo {author} {\bibfnamefont {R.}~\bibnamefont {Connelly}},\ }\bibfield
  {title} {\bibinfo {title} {A linear programming algorithm to test for jamming
  in hard-sphere packings},\ }\href {https://doi.org/10.1016/j.jcp.2003.11.022}
  {\bibfield  {journal} {\bibinfo  {journal} {Journal of Computational
  Physics}\ }\textbf {\bibinfo {volume} {197}},\ \bibinfo {pages} {139--166}
  (\bibinfo {year} {2004}{\natexlab{a}})}\BibitemShut {NoStop}%
\bibitem [{\citenamefont {Hopkins}\ \emph {et~al.}(2011)\citenamefont
  {Hopkins}, \citenamefont {Jiao}, \citenamefont {Stillinger},\ and\
  \citenamefont {Torquato}}]{hopkinsPhaseDiagramStructural2011}%
  \BibitemOpen
  \bibfield  {author} {\bibinfo {author} {\bibfnamefont {A.~B.}\ \bibnamefont
  {Hopkins}}, \bibinfo {author} {\bibfnamefont {Y.}~\bibnamefont {Jiao}},
  \bibinfo {author} {\bibfnamefont {F.~H.}\ \bibnamefont {Stillinger}},\ and\
  \bibinfo {author} {\bibfnamefont {S.}~\bibnamefont {Torquato}},\ }\bibfield
  {title} {\bibinfo {title} {Phase {{Diagram}} and {{Structural Diversity}} of
  the {{Densest Binary Sphere Packings}}},\ }\href
  {https://doi.org/10.1103/PhysRevLett.107.125501} {\bibfield  {journal}
  {\bibinfo  {journal} {Physical Review Letters}\ }\textbf {\bibinfo {volume}
  {107}},\ \bibinfo {pages} {125501} (\bibinfo {year} {2011})}\BibitemShut
  {NoStop}%
\bibitem [{\citenamefont {Hopkins}\ \emph {et~al.}(2012)\citenamefont
  {Hopkins}, \citenamefont {Stillinger},\ and\ \citenamefont
  {Torquato}}]{hopkinsDensestBinarySphere2012}%
  \BibitemOpen
  \bibfield  {author} {\bibinfo {author} {\bibfnamefont {A.~B.}\ \bibnamefont
  {Hopkins}}, \bibinfo {author} {\bibfnamefont {F.~H.}\ \bibnamefont
  {Stillinger}},\ and\ \bibinfo {author} {\bibfnamefont {S.}~\bibnamefont
  {Torquato}},\ }\bibfield  {title} {\bibinfo {title} {Densest binary sphere
  packings},\ }\href {https://doi.org/10.1103/PhysRevE.85.021130} {\bibfield
  {journal} {\bibinfo  {journal} {Physical Review E}\ }\textbf {\bibinfo
  {volume} {85}},\ \bibinfo {pages} {021130} (\bibinfo {year}
  {2012})}\BibitemShut {NoStop}%
\bibitem [{\citenamefont {Hopkins}\ \emph {et~al.}(2013)\citenamefont
  {Hopkins}, \citenamefont {Stillinger},\ and\ \citenamefont
  {Torquato}}]{hopkinsDisorderedStrictlyJammed2013}%
  \BibitemOpen
  \bibfield  {author} {\bibinfo {author} {\bibfnamefont {A.~B.}\ \bibnamefont
  {Hopkins}}, \bibinfo {author} {\bibfnamefont {F.~H.}\ \bibnamefont
  {Stillinger}},\ and\ \bibinfo {author} {\bibfnamefont {S.}~\bibnamefont
  {Torquato}},\ }\bibfield  {title} {\bibinfo {title} {Disordered strictly
  jammed binary sphere packings attain an anomalously large range of
  densities},\ }\href {https://doi.org/10.1103/PhysRevE.88.022205} {\bibfield
  {journal} {\bibinfo  {journal} {Physical Review E}\ }\textbf {\bibinfo
  {volume} {88}},\ \bibinfo {pages} {022205} (\bibinfo {year}
  {2013})}\BibitemShut {NoStop}%
\bibitem [{\citenamefont {Jiao}\ \emph {et~al.}(2011)\citenamefont {Jiao},
  \citenamefont {Stillinger},\ and\ \citenamefont
  {Torquato}}]{jiaoNonuniversalityDensityDisorder2011}%
  \BibitemOpen
  \bibfield  {author} {\bibinfo {author} {\bibfnamefont {Y.}~\bibnamefont
  {Jiao}}, \bibinfo {author} {\bibfnamefont {F.~H.}\ \bibnamefont
  {Stillinger}},\ and\ \bibinfo {author} {\bibfnamefont {S.}~\bibnamefont
  {Torquato}},\ }\bibfield  {title} {\bibinfo {title} {Nonuniversality of
  density and disorder in jammed sphere packings},\ }\href
  {https://doi.org/10.1063/1.3524489} {\bibfield  {journal} {\bibinfo
  {journal} {Journal of Applied Physics}\ }\textbf {\bibinfo {volume} {109}},\
  \bibinfo {pages} {013508} (\bibinfo {year} {2011})}\BibitemShut {NoStop}%
\bibitem [{\citenamefont {Roux}(2000)}]{rouxGeometricOriginMechanical2000}%
  \BibitemOpen
  \bibfield  {author} {\bibinfo {author} {\bibfnamefont {J.-N.}\ \bibnamefont
  {Roux}},\ }\bibfield  {title} {\bibinfo {title} {Geometric origin of
  mechanical properties of granular materials},\ }\href
  {https://doi.org/10.1103/PhysRevE.61.6802} {\bibfield  {journal} {\bibinfo
  {journal} {Physical Review E}\ }\textbf {\bibinfo {volume} {61}},\ \bibinfo
  {pages} {6802--6836} (\bibinfo {year} {2000})}\BibitemShut {NoStop}%
\bibitem [{\citenamefont {Luenberger}\ and\ \citenamefont
  {Ye}(2016)}]{luenbergerLinearNonlinearProgramming2016}%
  \BibitemOpen
  \bibfield  {author} {\bibinfo {author} {\bibfnamefont {D.~G.}\ \bibnamefont
  {Luenberger}}\ and\ \bibinfo {author} {\bibfnamefont {Y.}~\bibnamefont
  {Ye}},\ }\href {https://doi.org/10.1007/978-3-319-18842-3} {\emph {\bibinfo
  {title} {Linear and {{Nonlinear Programming}}}}},\ \bibinfo {series}
  {International {{Series}} in {{Operations Research}} \& {{Management
  Science}}}, Vol.\ \bibinfo {volume} {228}\ (\bibinfo  {publisher} {{Springer
  International Publishing}},\ \bibinfo {address} {{Cham}},\ \bibinfo {year}
  {2016})\BibitemShut {NoStop}%
\bibitem [{\citenamefont {Boyd}\ and\ \citenamefont
  {Vandenberghe}(2004)}]{boydConvexOptimization2004}%
  \BibitemOpen
  \bibfield  {author} {\bibinfo {author} {\bibfnamefont {S.~P.}\ \bibnamefont
  {Boyd}}\ and\ \bibinfo {author} {\bibfnamefont {L.}~\bibnamefont
  {Vandenberghe}},\ }\href@noop {} {\emph {\bibinfo {title} {Convex
  Optimization}}}\ (\bibinfo  {publisher} {{Cambridge University Press}},\
  \bibinfo {address} {{Cambridge, UK; New York}},\ \bibinfo {year}
  {2004})\BibitemShut {NoStop}%
\bibitem [{\citenamefont {Conway}\ and\ \citenamefont
  {Sloane}(2013)}]{conwaySpherePackingsLattices2013}%
  \BibitemOpen
  \bibfield  {author} {\bibinfo {author} {\bibfnamefont {J.~H.}\ \bibnamefont
  {Conway}}\ and\ \bibinfo {author} {\bibfnamefont {N.~J.~A.}\ \bibnamefont
  {Sloane}},\ }\href@noop {} {\emph {\bibinfo {title} {Sphere {{Packings}},
  {{Lattices}} and {{Groups}}}}}\ (\bibinfo  {publisher} {{Springer Science \&
  Business Media}},\ \bibinfo {year} {2013})\BibitemShut {NoStop}%
\bibitem [{\citenamefont {Charbonneau}\ \emph
  {et~al.}(2021{\natexlab{b}})\citenamefont {Charbonneau}, \citenamefont
  {Morse}, \citenamefont {Perkins},\ and\ \citenamefont
  {Zamponi}}]{charbonneauThreeSimpleScenarios2021}%
  \BibitemOpen
  \bibfield  {author} {\bibinfo {author} {\bibfnamefont {P.}~\bibnamefont
  {Charbonneau}}, \bibinfo {author} {\bibfnamefont {P.~K.}\ \bibnamefont
  {Morse}}, \bibinfo {author} {\bibfnamefont {W.}~\bibnamefont {Perkins}},\
  and\ \bibinfo {author} {\bibfnamefont {F.}~\bibnamefont {Zamponi}},\
  }\bibfield  {title} {\bibinfo {title} {Three simple scenarios for
  high-dimensional sphere packings},\ }\href
  {https://doi.org/10.1103/PhysRevE.104.064612} {\bibfield  {journal} {\bibinfo
   {journal} {Physical Review E}\ }\textbf {\bibinfo {volume} {104}},\ \bibinfo
  {pages} {064612} (\bibinfo {year} {2021}{\natexlab{b}})}\BibitemShut
  {NoStop}%
\bibitem [{\citenamefont
  {Blumenfeld}(2021)}]{blumenfeldDisorderCriterionExplicit2021}%
  \BibitemOpen
  \bibfield  {author} {\bibinfo {author} {\bibfnamefont {R.}~\bibnamefont
  {Blumenfeld}},\ }\bibfield  {title} {\bibinfo {title} {Disorder {{Criterion}}
  and {{Explicit Solution}} for the {{Disc Random Packing Problem}}},\ }\href
  {https://doi.org/10.1103/PhysRevLett.127.118002} {\bibfield  {journal}
  {\bibinfo  {journal} {Physical Review Letters}\ }\textbf {\bibinfo {volume}
  {127}},\ \bibinfo {pages} {118002} (\bibinfo {year} {2021})}\BibitemShut
  {NoStop}%
\bibitem [{\citenamefont {Hinrichsen}\ \emph {et~al.}(1990)\citenamefont
  {Hinrichsen}, \citenamefont {Feder},\ and\ \citenamefont
  {J{\o}ssang}}]{hinrichsenRandomPackingDisks1990}%
  \BibitemOpen
  \bibfield  {author} {\bibinfo {author} {\bibfnamefont {E.~L.}\ \bibnamefont
  {Hinrichsen}}, \bibinfo {author} {\bibfnamefont {J.}~\bibnamefont {Feder}},\
  and\ \bibinfo {author} {\bibfnamefont {T.}~\bibnamefont {J{\o}ssang}},\
  }\bibfield  {title} {\bibinfo {title} {Random packing of disks in two
  dimensions},\ }\href {https://doi.org/10.1103/PhysRevA.41.4199} {\bibfield
  {journal} {\bibinfo  {journal} {Physical Review A}\ }\textbf {\bibinfo
  {volume} {41}},\ \bibinfo {pages} {4199--4209} (\bibinfo {year}
  {1990})}\BibitemShut {NoStop}%
\bibitem [{\citenamefont {Donev}\ \emph
  {et~al.}(2005{\natexlab{a}})\citenamefont {Donev}, \citenamefont {Torquato},\
  and\ \citenamefont {Stillinger}}]{donevPairCorrelationFunction2005}%
  \BibitemOpen
  \bibfield  {author} {\bibinfo {author} {\bibfnamefont {A.}~\bibnamefont
  {Donev}}, \bibinfo {author} {\bibfnamefont {S.}~\bibnamefont {Torquato}},\
  and\ \bibinfo {author} {\bibfnamefont {F.~H.}\ \bibnamefont {Stillinger}},\
  }\bibfield  {title} {\bibinfo {title} {Pair correlation function
  characteristics of nearly jammed disordered and ordered hard-sphere
  packings},\ }\href {https://doi.org/10.1103/PhysRevE.71.011105} {\bibfield
  {journal} {\bibinfo  {journal} {Physical Review E}\ }\textbf {\bibinfo
  {volume} {71}},\ \bibinfo {pages} {011105} (\bibinfo {year}
  {2005}{\natexlab{a}})}\BibitemShut {NoStop}%
\bibitem [{\citenamefont {Donev}\ \emph
  {et~al.}(2004{\natexlab{b}})\citenamefont {Donev}, \citenamefont {Torquato},
  \citenamefont {Stillinger},\ and\ \citenamefont
  {Connelly}}]{donevCommentJammingZero2004}%
  \BibitemOpen
  \bibfield  {author} {\bibinfo {author} {\bibfnamefont {A.}~\bibnamefont
  {Donev}}, \bibinfo {author} {\bibfnamefont {S.}~\bibnamefont {Torquato}},
  \bibinfo {author} {\bibfnamefont {F.~H.}\ \bibnamefont {Stillinger}},\ and\
  \bibinfo {author} {\bibfnamefont {R.}~\bibnamefont {Connelly}},\ }\bibfield
  {title} {\bibinfo {title} {Comment on ``{{Jamming}} at zero temperature and
  zero applied stress: {{The}} epitome of disorder''},\ }\href
  {https://doi.org/10.1103/PhysRevE.70.043301} {\bibfield  {journal} {\bibinfo
  {journal} {Physical Review E}\ }\textbf {\bibinfo {volume} {70}},\ \bibinfo
  {pages} {043301} (\bibinfo {year} {2004}{\natexlab{b}})}\BibitemShut
  {NoStop}%
\bibitem [{\citenamefont {O'Hern}\ \emph {et~al.}(2004)\citenamefont {O'Hern},
  \citenamefont {Silbert}, \citenamefont {Liu},\ and\ \citenamefont
  {Nagel}}]{ohernReplyCommentJamming2004}%
  \BibitemOpen
  \bibfield  {author} {\bibinfo {author} {\bibfnamefont {C.~S.}\ \bibnamefont
  {O'Hern}}, \bibinfo {author} {\bibfnamefont {L.~E.}\ \bibnamefont {Silbert}},
  \bibinfo {author} {\bibfnamefont {A.~J.}\ \bibnamefont {Liu}},\ and\ \bibinfo
  {author} {\bibfnamefont {S.~R.}\ \bibnamefont {Nagel}},\ }\bibfield  {title}
  {\bibinfo {title} {Reply to ``{{Comment}} on `{{Jamming}} at zero temperature
  and zero applied stress: The epitome of disorder' ''},\ }\href
  {https://doi.org/10.1103/PhysRevE.70.043302} {\bibfield  {journal} {\bibinfo
  {journal} {Physical Review E}\ }\textbf {\bibinfo {volume} {70}},\ \bibinfo
  {pages} {043302} (\bibinfo {year} {2004})}\BibitemShut {NoStop}%
\bibitem [{\citenamefont {Donev}\ \emph
  {et~al.}(2005{\natexlab{b}})\citenamefont {Donev}, \citenamefont {Torquato},\
  and\ \citenamefont {Stillinger}}]{donevNeighborListCollisiondriven2005}%
  \BibitemOpen
  \bibfield  {author} {\bibinfo {author} {\bibfnamefont {A.}~\bibnamefont
  {Donev}}, \bibinfo {author} {\bibfnamefont {S.}~\bibnamefont {Torquato}},\
  and\ \bibinfo {author} {\bibfnamefont {F.~H.}\ \bibnamefont {Stillinger}},\
  }\bibfield  {title} {\bibinfo {title} {Neighbor list collision-driven
  molecular dynamics simulation for nonspherical hard particles. {{I}}.
  {{Algorithmic}} details},\ }\href {https://doi.org/10.1016/j.jcp.2004.08.014}
  {\bibfield  {journal} {\bibinfo  {journal} {Journal of Computational
  Physics}\ }\textbf {\bibinfo {volume} {202}},\ \bibinfo {pages} {737--764}
  (\bibinfo {year} {2005}{\natexlab{b}})}\BibitemShut {NoStop}%
\bibitem [{\citenamefont {Dunning}\ \emph {et~al.}(2017)\citenamefont
  {Dunning}, \citenamefont {Huchette},\ and\ \citenamefont {Lubin}}]{jump}%
  \BibitemOpen
  \bibfield  {author} {\bibinfo {author} {\bibfnamefont {I.}~\bibnamefont
  {Dunning}}, \bibinfo {author} {\bibfnamefont {J.}~\bibnamefont {Huchette}},\
  and\ \bibinfo {author} {\bibfnamefont {M.}~\bibnamefont {Lubin}},\ }\bibfield
   {title} {\bibinfo {title} {Jump: A modeling language for mathematical
  optimization},\ }\href {https://doi.org/10.1137/15M1020575} {\bibfield
  {journal} {\bibinfo  {journal} {SIAM Review}\ }\textbf {\bibinfo {volume}
  {59}},\ \bibinfo {pages} {295--320} (\bibinfo {year} {2017})}\BibitemShut
  {NoStop}%
\bibitem [{\citenamefont {{Gurobi Optimization, LLC}}(2021)}]{gurobi}%
  \BibitemOpen
  \bibfield  {author} {\bibinfo {author} {\bibnamefont {{Gurobi Optimization,
  LLC}}},\ }\href {https://www.gurobi.com} {\bibinfo {title} {{Gurobi Optimizer
  Reference Manual}}} (\bibinfo {year} {2021})\BibitemShut {NoStop}%
\bibitem [{\citenamefont {Huangfu}\ and\ \citenamefont {Hall}(2018)}]{highs}%
  \BibitemOpen
  \bibfield  {author} {\bibinfo {author} {\bibfnamefont {Q.}~\bibnamefont
  {Huangfu}}\ and\ \bibinfo {author} {\bibfnamefont {J.~A.~J.}\ \bibnamefont
  {Hall}},\ }\bibfield  {title} {\bibinfo {title} {Parallelizing the dual
  revised simplex method},\ }\href {https://doi.org/10.1007/s12532-017-0130-5}
  {\bibfield  {journal} {\bibinfo  {journal} {Mathematical Programming
  Computation}\ }\textbf {\bibinfo {volume} {10}},\ \bibinfo {pages} {119--142}
  (\bibinfo {year} {2018})}\BibitemShut {NoStop}%
\bibitem [{\citenamefont {Makhorin}(2008)}]{glpk}%
  \BibitemOpen
  \bibfield  {author} {\bibinfo {author} {\bibfnamefont {A.}~\bibnamefont
  {Makhorin}},\ }\href {http://www. gnu. org/s/glpk/glpk. html} {\bibinfo
  {title} {{GLPK} ({GNU} linear programming kit)}} (\bibinfo {year}
  {2008})\BibitemShut {NoStop}%
\bibitem [{\citenamefont {Stukowski}(2009)}]{ovito}%
  \BibitemOpen
  \bibfield  {author} {\bibinfo {author} {\bibfnamefont {A.}~\bibnamefont
  {Stukowski}},\ }\bibfield  {title} {\bibinfo {title} {Visualization and
  analysis of atomistic simulation data with {{OVITO}}\textendash the {{Open
  Visualization Tool}}},\ }\href
  {https://doi.org/10.1088/0965-0393/18/1/015012} {\bibfield  {journal}
  {\bibinfo  {journal} {Modelling and Simulation in Materials Science and
  Engineering}\ }\textbf {\bibinfo {volume} {18}},\ \bibinfo {pages} {015012}
  (\bibinfo {year} {2009})}\BibitemShut {NoStop}%
\bibitem [{Note1()}]{Note1}%
  \BibitemOpen
  \bibinfo {note} {Note that Eq.~\protect \textup {\hbox {\mathsurround \z@
  \protect \normalfont (\ignorespaces \ref {eq:hs-glass-eos}\unskip
  \@@italiccorr )}} has been derived from a free volume analysis~\cite
  {salsburgEquationStateClassical1962}. Therefore, it does not correspond to
  the true thermodynamic equation of state~\cite
  {puz_TheorySimpleGlasses2020,charbonneauGlassJammingTransitions2017} (see
  also Fig.~\ref {fig:hs-eos-liquid-and-glass} in App.~\ref {app:further-md}
  for more details).}\BibitemShut {Stop}%
\bibitem [{\citenamefont {Berthier}\ \emph
  {et~al.}(2016{\natexlab{a}})\citenamefont {Berthier}, \citenamefont
  {Charbonneau}, \citenamefont {Jin}, \citenamefont {Parisi}, \citenamefont
  {Seoane},\ and\ \citenamefont
  {Zamponi}}]{berthierGrowingTimescalesLengthscales2016}%
  \BibitemOpen
  \bibfield  {author} {\bibinfo {author} {\bibfnamefont {L.}~\bibnamefont
  {Berthier}}, \bibinfo {author} {\bibfnamefont {P.}~\bibnamefont
  {Charbonneau}}, \bibinfo {author} {\bibfnamefont {Y.}~\bibnamefont {Jin}},
  \bibinfo {author} {\bibfnamefont {G.}~\bibnamefont {Parisi}}, \bibinfo
  {author} {\bibfnamefont {B.}~\bibnamefont {Seoane}},\ and\ \bibinfo {author}
  {\bibfnamefont {F.}~\bibnamefont {Zamponi}},\ }\bibfield  {title} {\bibinfo
  {title} {Growing timescales and lengthscales characterizing vibrations of
  amorphous solids},\ }\href {https://doi.org/10.1073/pnas.1607730113}
  {\bibfield  {journal} {\bibinfo  {journal} {Proceedings of the National
  Academy of Sciences}\ }\textbf {\bibinfo {volume} {113}},\ \bibinfo {pages}
  {8397--8401} (\bibinfo {year} {2016}{\natexlab{a}})}\BibitemShut {NoStop}%
\bibitem [{\citenamefont {Salsburg}\ and\ \citenamefont
  {Wood}(1962)}]{salsburgEquationStateClassical1962}%
  \BibitemOpen
  \bibfield  {author} {\bibinfo {author} {\bibfnamefont {Z.~W.}\ \bibnamefont
  {Salsburg}}\ and\ \bibinfo {author} {\bibfnamefont {W.~W.}\ \bibnamefont
  {Wood}},\ }\bibfield  {title} {\bibinfo {title} {Equation of {{State}} of
  {{Classical Hard Spheres}} at {{High Density}}},\ }\href
  {https://doi.org/10.1063/1.1733163} {\bibfield  {journal} {\bibinfo
  {journal} {The Journal of Chemical Physics}\ }\textbf {\bibinfo {volume}
  {37}},\ \bibinfo {pages} {798--804} (\bibinfo {year} {1962})}\BibitemShut
  {NoStop}%
\bibitem [{\citenamefont {Charbonneau}\ \emph
  {et~al.}(2015{\natexlab{b}})\citenamefont {Charbonneau}, \citenamefont {Jin},
  \citenamefont {Parisi}, \citenamefont {Rainone}, \citenamefont {Seoane},\
  and\ \citenamefont {Zamponi}}]{charbonneauNumericalDetectionGardner2015}%
  \BibitemOpen
  \bibfield  {author} {\bibinfo {author} {\bibfnamefont {P.}~\bibnamefont
  {Charbonneau}}, \bibinfo {author} {\bibfnamefont {Y.}~\bibnamefont {Jin}},
  \bibinfo {author} {\bibfnamefont {G.}~\bibnamefont {Parisi}}, \bibinfo
  {author} {\bibfnamefont {C.}~\bibnamefont {Rainone}}, \bibinfo {author}
  {\bibfnamefont {B.}~\bibnamefont {Seoane}},\ and\ \bibinfo {author}
  {\bibfnamefont {F.}~\bibnamefont {Zamponi}},\ }\bibfield  {title} {\bibinfo
  {title} {Numerical detection of the {{Gardner}} transition in a mean-field
  glass former},\ }\href {https://doi.org/10.1103/PhysRevE.92.012316}
  {\bibfield  {journal} {\bibinfo  {journal} {Physical Review E}\ }\textbf
  {\bibinfo {volume} {92}},\ \bibinfo {pages} {012316} (\bibinfo {year}
  {2015}{\natexlab{b}})}\BibitemShut {NoStop}%
\bibitem [{Note2()}]{Note2}%
  \BibitemOpen
  \bibinfo {note} {Jamming criticality can also be clearly observed in two
  dimensional systems, provided that bidisperse or polydisperse packings are
  used. In that sense, $3d$ monodisperse systems are simpler because the random
  particles' positions are the only source of disorder.}\BibitemShut {Stop}%
\bibitem [{\citenamefont {Rainone}\ and\ \citenamefont
  {Urbani}(2016)}]{rainoneFollowingEvolutionGlassy2016}%
  \BibitemOpen
  \bibfield  {author} {\bibinfo {author} {\bibfnamefont {C.}~\bibnamefont
  {Rainone}}\ and\ \bibinfo {author} {\bibfnamefont {P.}~\bibnamefont
  {Urbani}},\ }\bibfield  {title} {\bibinfo {title} {Following the evolution of
  glassy states under external perturbations: The full replica symmetry
  breaking solution},\ }\href
  {https://doi.org/10.1088/1742-5468/2016/05/053302} {\bibfield  {journal}
  {\bibinfo  {journal} {Journal of Statistical Mechanics: Theory and
  Experiment}\ }\textbf {\bibinfo {volume} {2016}},\ \bibinfo {pages} {053302}
  (\bibinfo {year} {2016})}\BibitemShut {NoStop}%
\bibitem [{\citenamefont {Stillinger}(1999)}]{stillinger1999exponential}%
  \BibitemOpen
  \bibfield  {author} {\bibinfo {author} {\bibfnamefont {F.~H.}\ \bibnamefont
  {Stillinger}},\ }\bibfield  {title} {\bibinfo {title} {Exponential
  multiplicity of inherent structures},\ }\href
  {https://doi.org/10.1103/PhysRevE.59.48} {\bibfield  {journal} {\bibinfo
  {journal} {Physical Review E}\ }\textbf {\bibinfo {volume} {59}},\ \bibinfo
  {pages} {48} (\bibinfo {year} {1999})}\BibitemShut {NoStop}%
\bibitem [{\citenamefont {Santos}\ \emph {et~al.}(2020)\citenamefont {Santos},
  \citenamefont {Yuste},\ and\ \citenamefont {{L{\'o}pez de
  Haro}}}]{santosStructuralThermodynamicProperties2020}%
  \BibitemOpen
  \bibfield  {author} {\bibinfo {author} {\bibfnamefont {A.}~\bibnamefont
  {Santos}}, \bibinfo {author} {\bibfnamefont {S.~B.}\ \bibnamefont {Yuste}},\
  and\ \bibinfo {author} {\bibfnamefont {M.}~\bibnamefont {{L{\'o}pez de
  Haro}}},\ }\bibfield  {title} {\bibinfo {title} {Structural and thermodynamic
  properties of hard-sphere fluids},\ }\href
  {https://doi.org/10.1063/5.0023903} {\bibfield  {journal} {\bibinfo
  {journal} {The Journal of Chemical Physics}\ }\textbf {\bibinfo {volume}
  {153}},\ \bibinfo {pages} {120901} (\bibinfo {year} {2020})}\BibitemShut
  {NoStop}%
\bibitem [{\citenamefont {Berthier}\ \emph
  {et~al.}(2016{\natexlab{b}})\citenamefont {Berthier}, \citenamefont
  {Coslovich}, \citenamefont {Ninarello},\ and\ \citenamefont
  {Ozawa}}]{berthierEquilibriumSamplingHard2016}%
  \BibitemOpen
  \bibfield  {author} {\bibinfo {author} {\bibfnamefont {L.}~\bibnamefont
  {Berthier}}, \bibinfo {author} {\bibfnamefont {D.}~\bibnamefont {Coslovich}},
  \bibinfo {author} {\bibfnamefont {A.}~\bibnamefont {Ninarello}},\ and\
  \bibinfo {author} {\bibfnamefont {M.}~\bibnamefont {Ozawa}},\ }\bibfield
  {title} {\bibinfo {title} {Equilibrium {{Sampling}} of {{Hard Spheres}} up to
  the {{Jamming Density}} and {{Beyond}}},\ }\href
  {https://doi.org/10.1103/PhysRevLett.116.238002} {\bibfield  {journal}
  {\bibinfo  {journal} {Physical Review Letters}\ }\textbf {\bibinfo {volume}
  {116}},\ \bibinfo {pages} {238002} (\bibinfo {year}
  {2016}{\natexlab{b}})}\BibitemShut {NoStop}%
\bibitem [{\citenamefont {Hwang}\ and\ \citenamefont
  {Ikeda}(2020)}]{hwangForceBalanceControls2020}%
  \BibitemOpen
  \bibfield  {author} {\bibinfo {author} {\bibfnamefont {S.}~\bibnamefont
  {Hwang}}\ and\ \bibinfo {author} {\bibfnamefont {H.}~\bibnamefont {Ikeda}},\
  }\bibfield  {title} {\bibinfo {title} {Force balance controls the relaxation
  time of the gradient descent algorithm in the satisfiable phase},\ }\href
  {https://doi.org/10.1103/PhysRevE.101.052308} {\bibfield  {journal} {\bibinfo
   {journal} {Physical Review E}\ }\textbf {\bibinfo {volume} {101}},\ \bibinfo
  {pages} {052308} (\bibinfo {year} {2020})}\BibitemShut {NoStop}%
\bibitem [{\citenamefont {Nocedal}\ and\ \citenamefont
  {Wright}(2006)}]{nocedalNumericalOptimization2006}%
  \BibitemOpen
  \bibfield  {author} {\bibinfo {author} {\bibfnamefont {J.}~\bibnamefont
  {Nocedal}}\ and\ \bibinfo {author} {\bibfnamefont {S.~J.}\ \bibnamefont
  {Wright}},\ }\href@noop {} {\emph {\bibinfo {title} {Numerical
  Optimization}}},\ \bibinfo {edition} {2nd}\ ed.,\ Springer Series in
  Operations Research\ (\bibinfo  {publisher} {{Springer}},\ \bibinfo {address}
  {{New York}},\ \bibinfo {year} {2006})\BibitemShut {NoStop}%
\bibitem [{\citenamefont {Kurchan}\ \emph {et~al.}(2013)\citenamefont
  {Kurchan}, \citenamefont {Parisi}, \citenamefont {Urbani},\ and\
  \citenamefont {Zamponi}}]{kurchanExactTheoryDense2013}%
  \BibitemOpen
  \bibfield  {author} {\bibinfo {author} {\bibfnamefont {J.}~\bibnamefont
  {Kurchan}}, \bibinfo {author} {\bibfnamefont {G.}~\bibnamefont {Parisi}},
  \bibinfo {author} {\bibfnamefont {P.}~\bibnamefont {Urbani}},\ and\ \bibinfo
  {author} {\bibfnamefont {F.}~\bibnamefont {Zamponi}},\ }\bibfield  {title}
  {\bibinfo {title} {Exact {{Theory}} of {{Dense Amorphous Hard Spheres}} in
  {{High Dimension}}. {{II}}. {{The High Density Regime}} and the {{Gardner
  Transition}}},\ }\href {https://doi.org/10.1021/jp402235d} {\bibfield
  {journal} {\bibinfo  {journal} {The Journal of Physical Chemistry B}\
  }\textbf {\bibinfo {volume} {117}},\ \bibinfo {pages} {12979--12994}
  (\bibinfo {year} {2013})}\BibitemShut {NoStop}%
\bibitem [{\citenamefont {Rissone}\ \emph {et~al.}(2021)\citenamefont
  {Rissone}, \citenamefont {Corwin},\ and\ \citenamefont
  {Parisi}}]{rissoneLongRangeAnomalousDecay2021}%
  \BibitemOpen
  \bibfield  {author} {\bibinfo {author} {\bibfnamefont {P.}~\bibnamefont
  {Rissone}}, \bibinfo {author} {\bibfnamefont {E.~I.}\ \bibnamefont
  {Corwin}},\ and\ \bibinfo {author} {\bibfnamefont {G.}~\bibnamefont
  {Parisi}},\ }\bibfield  {title} {\bibinfo {title} {Long-{{Range Anomalous
  Decay}} of the {{Correlation}} in {{Jammed Packings}}},\ }\href
  {https://doi.org/10.1103/PhysRevLett.127.038001} {\bibfield  {journal}
  {\bibinfo  {journal} {Physical Review Letters}\ }\textbf {\bibinfo {volume}
  {127}},\ \bibinfo {pages} {038001} (\bibinfo {year} {2021})}\BibitemShut
  {NoStop}%
\end{thebibliography}%

\end{document}